\documentclass[useAMS,usenatbib]{mn2e}
\input psfig.tex
\usepackage{latexsym}
\usepackage{amssymb}
\usepackage{amsmath}
\usepackage{graphics}
\usepackage{graphicx}
\usepackage{natbib}
\usepackage{subfigure}

\title[Galaxy spectra with interstellar dust]{Modelling galaxy spectra in presence of interstellar dust-III. From nearby galaxies to the distant Universe}
\author[L. P. Cassar\`{a}, L. Piovan and C. Chiosi]{L. P. Cassar\`{a}$^{1,2}$\thanks{E-mail:
letizia@lambrate.inaf.it (LPC); lorenzo.piovan@gmail.com (LP); cesare.chiosi@unipd.it (CC)}, 
L. Piovan$^{2,3}$\footnotemark[1], C. Chiosi$^{2}$\footnotemark[1]\\
 $^{1}$INAF-IASF Milano, Via E. Bassini 15, 20133 Milano, Italy\\
$^{2}$Department of Physics and Astronomy, University of Padova,
       Via Marzolo 8-I, 35131, Padova, Italy\\
 $^{3}$Max-Planck-Institut f\"ur Astrophysik, Karl-Schwarzschild-Str. 1, Garching bei M\"unchen, Germany}

\begin{document}

\date{Submitted 2014 November 18}

\pagerange{\pageref{firstpage}--\pageref{lastpage}} \pubyear{2014}

\maketitle

\label{firstpage}

\begin{abstract}

Improving upon the standard evolutionary population synthesis (EPS) technique, we present spectrophotometric models of galaxies 
whose morphology goes from spherical structures to discs, properly accounting for the effect of dust in the interstellar medium (ISM).
These models enclose three main physical components: the diffuse ISM  composed by
gas and dust, the complexes of molecular clouds (MCs) where active star formation occur and the stars of any age and chemical composition. 
These models are based on robust evolutionary chemical models that provide the total amount of gas and stars present at any age and that are 
adjusted in order to match the gross properties of galaxies of different morphological type. 
We have employed the results for the properties of the ISM presented in \citet{Piovan2006a} and the single stellar populations 
calculated by \citet{Cassara2013} to derive the spectral 
energy distributions (SEDs) of galaxies going from pure bulge  to discs passing through a number of composite systems with different combinations of the 
two components. The first part of the paper is devoted to recall the technical details of the method and the basic relations driving the 
interaction between the physical components of the galaxy. Then, the main parameters are examined and their effects on the spectral energy distribution 
of three prototype galaxies are highlighted. 
We conclude analyzing the capability of our galaxy models in reproducing the SEDs of real galaxies in the Local Universe and as function of redshift.

\end{abstract}


\section{Introduction} \label{intro}

In recent years the first epochs of galaxy formation have been  continuously  pushed back in time by the discovery of  galaxy size objects at 
higher and higher redshifts,  z$\sim$4-5 \citep{Madau1996,Steidel1999}, z$\sim$6 \citep{Stanway2003,Dickinson2004} to z$\sim$10 \citep{Zheng2012,Bouwens2012,Oesch2012}, 
up to z$\sim$10-20  according to the current view  \citep{RowanRobinson2012}. Furthermore, this high redshift Universe turned out to heavily  obscured by copious amounts of dust
\citep[see for instance ][]{Shapley2001,Carilli2001,Robson2004,Wang2009,Michalowski2010b}, whose origin and composition are a matter of vivid debate 
\citep{Gall2011a,Gall2011b,Dwek2009,Draine2009,Dwek2011}.
In this context, the current paradigm is that  the interstellar dust plays an important role in galaxy formation and evolution.
Therefore, understanding  the properties of this interstellar dust and setting a physically realistic interplay  between populations of stars and 
dust are critical to determine the properties of the high-z Universe and to obtain precious clues on the fundamental question about when and how galaxies formed and evolved.

It follows from all this that  to fully exploit modern data, realistic spectro-photometric models of galaxies must include this important 
component of the interstellar medium (ISM). This has spurred unprecedented effort
in the theoretical modelling of the spectro-photometry, dynamics, and chemistry  of dusty galaxies
\citep[see for instance][]{Narayanan2010,Jonsson2010,Grassi2011,Pipino2011,Popescu2011}.
Dust absorbs the stellar radiation and re-emits it at longer wavelengths, deeply changing the shape of the observed spectral energy distributions (SEDs) 
\citep{Silva1998,Piovan2006b,Popescu2011}. 
It also strongly affects the production of molecular hydrogen and the local amount of ultraviolet (UV) radiation in galaxies, 
playing a strong role in the star formation process \citep{Yamasawa2011}.

This paper is  a sequel of the series initiated with \citet{Piovan2003,Piovan2006a,Piovan2006b} devoted to modelling the spectra of galaxies of
 different morphological type in presence of dust. In this paper  we present new results for the light emitted by galaxies of different morphological type and age.  
Our galaxy model contains three interacting components: the diffuse ISM, made of gas and dust, the large complexes of molecular clouds (MCs) in which active star formation occurs and,
 finally, the populations of stars that are no longer embedded in the dusty environment of their parental MCs. Furthermore, our model for the dust takes into account 
three components, i.e. graphite, silicates and polycyclic aromatic hydrocarbons (PAHs). We consider and adapt to our aims two prescriptions for the size distribution 
of the dust grains and two models for the emission of the dusty ISM. The final model we have adopted is an hybrid one which stems from combining the analysis 
of \citet{Guhathakurta1989} for the emission of graphite and silicates and \citet{Puget1985} for the PAH emission, and using the distribution law of \citet{Weingartner2001a} 
and the ionization model for PAHs of \citet{Weingartner2001b}. The SEDs of single stellar populations (SSPs) of different age and chemical composition, 
the building blocks of galactic spectra,  are taken from \citet{Cassara2013} who have revised the contribution by asymptotic giant branch (AGB) stars 
taking into account new models of stars in this phase by \citet{Weiss2009}
and include the effect of dust on both young stars and AGB stars. 
During the history of  a SSP there are two periods of time in which self obscuration by dust in cloud surrounding a star, causing internal 
absorption and re-emission of the light emitted by central object,  plays an important role:   young  massive stars while they are still embedded in 
their parental MCs and intermediate and low-mass AGB stars when they form their own dust shell around 
\citep[see][for more details about AGB stars]{Piovan2003,Piovan2006a,Piovan2006b,Cassara2013}.
With the aid of all this, we seek  to get clues on the spectro-photometric evolution of galaxies, in particular taking into
 account the key role played by dust in determining the spectro-photometric properties of the stellar populations,  and to set up an
 library of template model galaxies of different morphological type (from pure spheroids to pure discs and  intermediate types with the 
two components in different proportions) whose SEDs at different evolutionary time can be compared with galaxies of the Local Universe and as function of the redshift.

The strategy of the paper is as follows. The model we have adopted is shortly summarized in Sect. \ref{gal_mod} where first we define the galaxy
 components we are dealing with, i.e. bare stars, stars embedded in MC complexes, and diffuse ISM; second we outline the recipes and basic equations
 for the gas infall, chemical evolution, initial mass function (IMF) and star formation rate (SFR); third we describe how the total amounts of stars, MCs and 
ISM present in the galaxy at a certain age are distributed over the galaxy volume by means of suitable density profiles, one for each component that depend on the
 galaxy type: pure disc galaxies, pure spheroidal galaxies, and composite galaxies with both disc and bulge; finally   the galaxy volume is split in suitable 
elemental volumes  to each of which the appropriate amounts of stars, MCs and ISM are assigned. In Sect. \ref{spe_syn_ism} we explain how the SEDs of galaxies of
 different morphological type are calculated in presence of dust in the ISM. First the technical details of the method are described and basic relationships/equations 
describing the interaction between the physical components of the galaxy are presented.
In Sect. \ref{compo_dust} we shortly describe the composition of the dust. In Sect. \ref{model_param} we  list and discuss the key parameters of the chemical and 
spectro-photometric models for the three basic types of galaxy: pure discs, pure spheroids and composite systems. In Sect. \ref{SED_gal} the SED of a few
 proto-type galaxies are presented, first as it would be observed at the present time and then as a function of the age. In Sect. \ref{colours} we examine the
 present-day theoretical colours in widely used photometric systems as a function of the morphological type of the underlying galaxy and compare them with literature data. 
In Sect. \ref{colour_redshift} we present the colour evolution of the model galaxies as a function of the redshift assuming the nowadays most credited 
$\Lambda$CDM cosmological scenario  and compare the theoretical colours with the  observational data currently available in literature.
Finally, some concluding remarks are drawn in Sect. \ref{conclusions}.

\section{Models of galaxies of different morphological types}\label{gal_mod}

\subsection{Physical components of galaxy models}\label{gal_comp}

As originally proposed by
\citet{Silva1998} and then adopted in many EPS models with dust
\citep[see for instance][]{Bressan2002,Piovan2006a,Piovan2006b,Galliano2008a,Popescu2011}, the most refined  galaxy models with dust available in literature contain three physical components:

\begin{itemize}
\item[-] The diffuse ISM, composed by gas and dust; the model for the ISM adopted in this study is described in detail in \citet{Piovan2006a} and 
includes a state-of-art description of a three component ISM made by graphite, silicates and PAHs, the most widely adopted scheme for a dusty ISM.

\item[-] The large complexes of MCs where active star formation takes place. In our model we do not consider HII regions
      and nebular emission. The MCs hide very young stars; their SEDs are severely affected by the dusty environment around them and skewed toward the IR. 
			The library of dusty MCs for which a ray-tracing radiative transfer code has been adopted is presented in \citet{Piovan2006a} at varying the input parameters.
		
\item[-]  The populations of stars free from their parental MCs. They are both intermediate age AGB stars (with age from about 0.1 Gyr to 1-2 Gyr) 
that are intrinsically obscured by their own dust shells \citep{Piovan2003,Marigo2008}, and the old stars that 
shine as bare objects and whose radiation propagates through the diffuse galactic ISM. The effect of dust around AGB stars 
has been included ex-novo taking into account the new models of TP-AGB stars by \citet{Weiss2009} as described 
in \citet{Cassara2013}, where new SEDs of SSPs have been presented and tested.
\end{itemize}

\noindent

The models of galaxies are based on a robust evolutionary chemical model that, considering a detailed description for the gas infall,  
star formation law, IMF and stellar ejecta, provide the total amounts of gas and stars present 
at any age, with their chemical history \citep{Chiosi1980,Tantalo1996,Tantalo1998,Portinari1998,Portinari1999,Portinari2000,Piovan2006b}.  
These chemical models are adjusted in order to match the gross properties of galaxies of different morphological type. 
The interaction between stars and ISM in building up the total SED of a galaxy is described using a suitable spatial distribution of gas and stars. 
For each type of galaxies, a simple geometrical model is assumed. 
The following step is to distribute the total gas and star mass provided by the chemical model over the whole volume,
using suitable density profiles, according to each component and depending on the galaxy type 
(pure spheroid, pure disc, and a combination of  disc plus bulge).
The galaxy is split in suitable volume elements; each elemental volume contains the appropriate amounts of stars, MCs and 
ISM and it is at the same time source of radiation (from the 
stars inside) and absorber and emitter of radiation (from and to all other elemental volume  and the ISM in between). 
These elements are the primordial seeds to calculate the global galaxy SED.

\subsection{The star formation and chemical enrichment of galaxy models}\label{chem_mod}

The so-called {\it infall-model}, developed by \citet{Chiosi1980}, and used by many authors (\citet{Bressan1994}, \citet{Tantalo1996}, 
\citet{Tantalo1998}, \citet{Portinari1998}) characterized the star formation and chemical enrichment histories of the model galaxies. 
Originally conceived for disc galaxies, over the years it has been extended also to  the early-type ones (ETGs).
In this paragraph the main features of the infall model are presented.

Within an halo of arbitrary shape and volume, which contains Dark Matter (DM) of mass $M_{DM}$, 
the mass of the Baryonic (luminous) Matter, $M_{BM}$, evolves with time by infall of primordial
gas following to the law

\begin{equation} \label{rate_inf}
\frac{dM_{BM}\left( t\right) }{dt}=M_{BM}^0\exp \left( -\frac{t}{\tau}\right)
\end{equation}

\noindent $\tau $ is the infall time scale. The constant $M_{BM}^0$ is fixed considering that at the present
  age $t_{G}$ the mass $M_{BM} \left(t \right)$ is equal to $M_{BM}\left( t_{G}\right)$.
	The baryonic (luminous) asymptotic mass of the galaxy  becomes:

\begin{equation}
M_{BM}^{0}=\frac{M_{BM}\left( t_{G}\right) }{\tau \left[ 1-e^{\left(-t_{G}/\tau \right) }\right] }
\end{equation}

\noindent
while the time variation of the BM is

\begin{equation}
M_{BM}\left( t\right) =\frac{M_{BM}\left( t_{G}\right) }{\tau \left[
1-e^{\left( -t_{G}/\tau \right) }\right] }\left[ 1-e^{\left(
-t/\tau \right) }\right]
\end{equation}

For more details we refer to \citet{Tantalo1996,Tantalo1998}.

\textbf{Early-type galaxies and/or bulges}. Applied to an ETG or a bulge, our infall model of chemical evolution can \textit{mimic} the 
collapse of the parental proto-galaxy made of BM and DM in cosmological proportions from a very extended size to the one we see today. 
Under the self gravity both DM and BM collapse and shrink at  a suitable rate. As the gas falls and cools down into the common potential well, 
the gas density increases so that star formation begins. A central visible object is formed. Soon or later both components virialize and 
settle to an equilibrium condition, whose geometrical shape is close to a sphere.  In this case, the galaxy can be approximated by a sphere of 
DM with mass $M_{DM}$ and radius $R_{DM}$, containing inside a  luminous, spherical object of mass $M_{BM}$ and radius $R_{BM}$.  As more gas flows in, 
the more efficient star formation gets. Eventually, the whole gas content is exhausted and turned into stars, thus quenching further star formation.
 The star formation rate starts small, rises to a maximum, and then declines. Because of the more efficient chemical enrichment of the infall model, 
the initial metallicity for the bulk of star forming activity is significantly different from zero. The radius of the stellar component of a 
galaxy will grow with the mass of it according to the law of virial equilibrium. It must be underlined that the collapse of the proto-galactic 
cloud cannot be modelled in a realistic dynamical way using a traditional chemical code: to simulate the gas dynamics other technique must be used
\citep[see for instance][]{Chiosi2002,Springel2005,Merlin2007,Gibson2007,Merlin2010,Merlin2012}. However, these models require lots of computational 
time and do not allow us to  quickly explore the space of  parameters. Therefore,  our chemical model will be a \textit{static} one simulating in a simple fashion 
the formation of the galaxy  by the collapse of primordial gas in presence of DM. In other words, the model galaxy is conceived  as a mass point \citep{Chiosi1980} 
for which no information about the spatial distribution of stars and gas is available. These latter  will be distributed according to suitable prescriptions 
(see below where the distribution laws and the normalization of the physical quantities are explained in detail for the different morphological types). 
Finally it is worth recalling that the chemical history of spheroidal systems is best described by models in which galactic winds can occur.

\textbf{Disc galaxies}.
In the case of  pure disc galaxies or  the disc component of intermediate type galaxies, it is reasonable to
 suppose that  discs are the result of accumulation of primordial or partially enriched gas at a suitable rate
\citep[as originally suggested by the dynamical studies of][]{Larson1976,Burkert1992}. If so the
formalism presented above can be extended to model galactic discs, provided we identify the baryonic mass $M_{BM}(t)$ 
with the surface mass density and consider the disc as made by a number of isolated and independent concentric rings in which
 the mass grows as a function of time \citep{Portinari1999,Portinari2000} or as a number of rings in mutual communication thanks 
to  radial flows of gas and dust \citep{Portinari2000}. However, for the purposes of this study the simple  one-zone formulation 
is fully adequate also for disc galaxies: the radial dependence is left aside  and the disc is modelled as an dimensionless object 
as in the classical paper by  \citet{Talbot1971}.
This simple models well reproduces the results of dynamical models \citep{Larson1976,Burkert1992,Carraro1998b}, with the exception 
of the radial flows of gas. As for the spheroidal objects above, the spatial distribution of stars and gas in the disc will be introduced 
by hand later on (see below). The chemical history and dynamical structure of galactic discs are fully compatible with absence of galactic winds.

\textbf{Gravitational potential of a spherical system and galactic wind}. In order to 
determine the physical 
conditions under which galactic winds may occur, 
we need to evaluate the gravitational potential of spherical systems made of DM 
and BM with different radial distributions. 
In the following we adopt the formalism developed by \citet[][]{Tantalo1996} that is 
shortly summarized here for the sake of clarity. 
The spatial distribution of DM with respect to 
BM follows the dynamical models of \citet{Bertin1992} and \citet{Saglia1992}: the mass and 
radius of the DM (${\rm M_{DM}}$ and ${\rm R_{DM}}$) are linked to those of the BM  
(${\rm M_{BM}}$ and ${\rm R_{BM}}$) by

\begin{equation}
{\rm  {M_{BM}(t) \over M_{DM} } \geq {1\over 2\pi} ({R_{BM}(t)\over R_{DM}})
       [1 + 1.37 ( {R_{BM}(t) \over R_{DM} } )]   }
\label{dark}
\end{equation}

\noindent The mass of the dark component is supposed to be constant in time and equal to ${\rm M_{DM}= \beta M_{BM}(t_{G}) }$, 
where $M_{BM}(t_G)$ the asymptotic  content of BM of a galaxy, and $\beta \simeq 6$
 is given by the baryon ratio in the $\Lambda$CDM Universe we have 
adopted \citep{Hinshaw2009}. With the assumptions, the binding gravitational 
energy of the gas is given by

\begin{equation}
{\rm \Omega_{g}(t)=-{\alpha}_{BM} G {M_{g}(t) M_{BM}(t)\over R_{BM}(t) } -
G {M_{g}(t) M_{DM} \over  R_{BM}(t) } \Omega'_{BD}  }
\label{gas_pot}
\end{equation}

\noindent where ${\rm M_g(t)}$ is the current value of the gas mass, 
$\alpha_{BM}$ is a numerical factor $\simeq 0.5$, while

\begin{equation}
{ \rm \Omega'_{BD}= {1\over 2\pi} ({R_{BM}(t)\over R_{DM}}) [1 + 1.37
            ( {R_{BM}(t) \over R_{DM} } )] }
\label{dark_pot}
\end{equation}
\noindent is the contribution to the gravitational energy given by the presence of dark matter. 
Adapting the original assumptions by  \citet{Bertin1992} and \citet{Saglia1992} to the present 
situation, we assume ${\rm M_{BM}/M_{DM}}=0.16$ and ${\rm R_{BM}/R_{DM}}=0.16$. 
Using these values for ${\rm M_{BM}/M_{DM}}$ and ${\rm R_{BM}/R_{DM}}$, the  contribution to 
gravitational energy by the DM  is ${\rm \Omega'_{BD}=0.03}$.
Assuming that at each stage of the infall process the amount of luminous mass 
that has already accumulated gets soon virialized and turned into stars,  
the total gravitational energy and radius of the material already settled onto the 
equilibrium  configuration can be approximated with the relations
for the total gravitational energy and radius as function of the mass \citep{Saito1979a,Saito1979b} 
for elliptical galaxies whose spatial distribution of  stars is such that 
the global luminosity profile follows the ${\rm R^{1/4} }$ law. 
The relation between ${\rm R_{BM}(t)}$ and ${\rm M_{BM}(t)}$ became:

\begin{equation}
{\rm R_{BM}(t) = 26.1\times (M_{BM}(t)/10^{12} M_{\odot})^{(2-\eta)} }
 ~~{\rm kpc}
\label{lum_rag}
\end{equation}
\noindent with $\eta=1.45$  \citep[see][]{Arimoto1987,Tantalo1998}.

\noindent
If DM and BM are supposed to have the same spatial distribution,
 Eqs.~(\ref{gas_pot}) and (\ref{dark_pot}) are no needed: the  binding energy becomes

\begin{equation}
{\rm \Omega_{g}(t) = \Omega_{L+D}(t) M_{g}(t) [2 - M_{g}(t)] }.
\label{gas_new}
\end{equation}

\noindent
In the spheroidal systems, when the thermal energy of the gas in the galaxy, 
heated by the SN{e} explosions, stellar winds and UV radiation from massive stars and 
cooled down by radiative  processes, equals or exceeds the gravitational potential energy of this

\begin{equation}
E_{th,g}(t) \geq \Omega_{g}(t)
\label{wind}
\end{equation}

\noindent the  \textit{galactic wind} is supposed to occur, star formation is quenched and all the remaining gas is expelled. 
In contrast, no galactic winds are supposed to occur in disc galaxies.

\textbf{Basic chemical equations}.
The complete formalism of the chemical evolution models providing the backbone of the photometric history of galaxies
can be found in \citet{Tantalo1996} for a spherical system and in \citet{Portinari2000} for disc galaxies with radial 
flows (see also \citet[][]{Cassara2012} for an exhaustive review  of the various models in literature). 
Here we show only the final equations for the chemical evolution of the ISM made of  dust and gas lumped 
together in a single component.

Indicating with  $M_{g}\left( t\right)$ the mass of gas at the time $t$, the corresponding gas fraction is

$$G\left( t\right) =\frac{M_{g}\left( t\right) }{M_{B}\left( t_{G}\right) }.$$

\noindent Denoting with  $X_{i}\left( t\right) $ the  mass abundance of the $i$-th chemical species, we may write

$$G_{i}\left( t\right) =X_{i}\left(t\right) G\left( t\right)$$

\noindent where by definition $\sum_{i}X_{i}=1$.

 The evolution of the normalized masses $G_{i}$ and abundances $X(i)$ of the i-th element, the system of equations is   given by:

\begin{eqnarray} \label{chem_ev}
\frac{d}{dt}G_{i}\left(t \right) &=& -X_{i}\left(t
\right)\psi\left(t \right)+                               \nonumber\\
&+&\int_{0}^{t-\tau_{M_{b,l}}}\psi\left(t^{\prime}\right)
\left[ R_{\phi(M)} \right]_{M\left(t-t^{\prime}\right)}
dt^{\prime}+                                              \nonumber\\
&+& \left(1-A
\right)\int_{t-\tau_{M_{b,l}}}^{t-\tau_{M_{b,u}}}\psi\left(t^{\prime}\right)
\left[ R_{\phi(M)} \right]_{M\left(t-t^{\prime}\right)}
dt^{\prime}+                                              \nonumber\\
&+&
\int_{t-\tau_{M_{b,u}}}^{t-\tau_{M_{u}}}\psi\left(t^{\prime}\right)
\left[ R_{\phi(M)} \right]_{M\left(t-t^{\prime}\right)}
dt^{\prime}+                                              \nonumber\\
&+&
A\int_{t-\tau_{M_{1,min}}}^{t-\tau_{M_{1,max}}}\psi\left(t^{\prime}\right)
\left[ R_{f(M_1)} \right]_{M_1\left(t-t^{\prime}\right)}
dt^{\prime}+                                              \nonumber\\
&+& R_{SNI}E_{SNI,i}+                                     \nonumber\\
&+& \left[\frac{d}{dt}G_{i}\left(t \right)
\right]_{inf}-\left[\frac{d}{dt}G_{i}\left(t \right) \right]_{out}.
\end{eqnarray}

\noindent where
$$\left[ R_{\phi(M)} \right]_{M\left(t-t^{\prime}\right)}=\left[\phi\left( M \right)
R_{i}\left(M\right)\left(-\frac{dM}{d\tau_{M}}\right)
\right]_{M\left(t-t^{\prime}\right)}$$
and
$$\left[ R_{f(M_1)} \right]_{M_1\left(t-t^{\prime}\right)}=\left[f\left( M_1 \right)
R_{i}\left(M_1\right)\left(-\frac{dM_1}{d\tau_{M_1}}\right)
\right]_{M_1\left(t-t^{\prime}\right)}$$

\noindent The first term at the r.h.s. is the depletion of the ISM because of the star formation process that consumes the interstellar matter; 
the following three terms at r.h.s. are the contributions of single stars to the enrichment of the element \textit{i}, the fifth term is the 
contribution by the primary star of a binary system, the sixth term is the contribution of type Ia SN{e}, the following term describes the infall of
 primordial material, and finally the last one takes into account the eventual outflow of matter at the onset of the galactic wind in elliptical galaxies.
$\textit{f}\left( M_{1} \right)$  is the distribution function of the primary $M_{1}$ mass in a binary system, between
$M_{1,min}=M_{b,l}/2$ and $M_{1,max}=M_{b,u}$. $R_{SNI}$ is the rate of type Ia SN{e} rate , while $E_{SNI,i}$ is the ejecta of the chemical 
element $i$ always in type Ia SN{e}. Further details on the calculation of yields and the adopted formalism can be found in \citet{Chiosi1986}, 
\citet{Matteucci1986} and \citet{Portinari1998}. For a complete presentation of the equations of chemical evolution see \citet{Cassara2012}.

\textbf{Initial Mass Function}. In literature there are several possible laws for the IMF: at least nine according to \citet{Cassara2012,Cassara2013} have derived the SSP 
photometry  in presence of dust for all of them. In  this paper, for the sake of simplicity we consider to the classical Salpeter's law

\begin{equation}
         dN = \phi(M) dM \propto M^{-2.35}dM
\end{equation}
where the proportionality constant is fixed by imposing that the fraction $\zeta$ of the IMF mass comprised between $\simeq 1 M_{\odot}$ 
(the minimum mass whose age is comparable to the age of the Universe)   to the upper limit $M_u$ ($\simeq 100\, M_\odot$), i.e. the mass interval 
effectively contribution to nucleosynthesis, is fixed. Accordingly
\begin{equation}
  \zeta = { \int_1^{M_u} M \phi(M)dM    \, / \,   \int_{M_l}^{M_u} M \phi(M)dM  }.
\end{equation}

\textbf{Star Formation Rate}.  We adopt the classical law by \citet{Schmidt1959}. The SFR, i.e. the number of stars of mass $M$ born in 
the time interval $dt$ and mass interval $dM$, is
$dN/dt=\Psi \left(t\right) \Phi \left( M\right) dM$. The rate of star formation $\Psi \left( t\right) $ is the \citet{Schmidt1959} law which, 
adopted to our model $\Psi \left( t\right) =\nu M_{g}\left( t\right) ^{k}$ and normalized to $M_{B}\left( t_{G}\right) $, becomes

\begin{equation} \label{Schmidt_law}
\Psi \left( t\right) =\nu M_{B}\left( t_{G}\right) ^{k-1}G\left(
t\right) ^{k}
\end{equation}

\noindent
The parameters $\nu $ and $k$ are crucial: $k$ yields the dependence of the star formation rate on the gas
content; current values are $k=1$ or $k=2$. The factor $\nu$ measures the efficiency of the star formation process.
In this type of model, because of the competition between gas infall, gas consumption by star formation, and gas ejection by dying stars, 
the SFR starts  very low, grows to a maximum and then declines. The time scale $\tau$ of Eq.~\ref{rate_inf} roughly corresponds to the age at which the star formation 
activity reaches the peak value.\\

The chemical models for pure spheroids and/or discs provide the mass of stars, $M_{*}\left(t\right)$, the mass of gas $M_{g}\left(t\right)$ and the metallicity $Z\left(t\right)$
that are used as entries for the population synthesis code.
In the case of composite galaxies made of a disc and a bulge, 
the mass of the galaxy is the sum of the two components. The assumption is that disc and bulge  evolve independently and  each component 
will have its $M_{*}\left(t\right)$, $M_{g}\left(t\right)$ and $Z\left(t\right)$.

\subsection{Stars and ISM: their spatial distribution}\label{spatial}

In most EPS models, those in particular that neglect the presence of an ISM in form of gas and dust,  the SED of a galaxy is simply obtained by convolving the SSP 
spectra with the SFH \citep{Arimoto1987,Bressan1994,Tantalo1998} and no effect of the spatial distribution of the various components (stars and ISM) is considered. 
For intermediate type galaxies with a disc and a bulge (and maybe even a halo), the situation is mimicked  by considering different SFHs for the various 
 components \citep{Buzzoni2002,Buzzoni2005,Piovan2006b}. This simple simple approach can no longer be used  in presence of  the ISM and the absorption and IR/sub-mm
 emission of radiation by dust. In particular, the emission requires a spatial description, whereas the sole treatment of the extinction could be simulated 
by applying a suitable extinction curve. In the general case of  dust extinction/emission \citep{Silva1998,Piovan2006b} the spatial distribution of the ISM, dust 
and stars in the galaxy must be specified. The observational evidence is that the spatial distribution of stars and ISM depends on the galaxy
 morphological type and that it can be reduced to:  pure spheroids, pure discs, and composite systems made of spheroid plus disc in different proportions 
(the irregulars are neglected here). Therefore one needs a different geometrical description for the various  components, to each of which a suitable star formation history 
is associated. Finally in the present approach, no  bursts of star formation are included into the chemical model  \citep[see][for somes examples of star-burst model galaxies]{Piovan2006b}.
In the following we adopt the same formalism proposed by \citet{Piovan2006b} which is also reported here for the sake of completeness and easy understanding by the reader.  
The formalism is presented  in order of increasing complexity.
\subsubsection{Disc Galaxies}\label{spatial_disc}

The mass density distribution of stars ($\rho_{\ast}$), diffuse ISM ($\rho_{ISM}$), and MCs ($\rho_{MC}$), inside a galactic disc is approximated to  a double 
decreasing exponential law. 
Considering a system of polar coordinates
 with origin at the galactic center [$r$, $\theta$, $\phi$], the height above the equatorial plane is $z = r cos \theta$ and the distance from the galactic center 
along the equatorial plane is $R = r sin \phi$; $\phi$ is the angle between the polar vector $r$ and the z-axis perpendicular to the galactic plane passing through the center. 
The azimuthal symmetry rules out the angle $\phi$.
The density laws for the three components are:

\begin{equation}\label{g1}
\rho^{i}=\rho^{i}_0 \exp\left(\frac{-r\sin\theta}
{R_{d}^{i}}\right)\exp\left(\frac{-r|\cos\theta|}{z_{d}^{i}}\right)
\end{equation}

\noindent
where ''$i$'' can be ''$\ast$'', ''$ISM$'', ''$MCs$'', i.e. stars, diffuse ISM and star forming MCs. ${R_{d}^{\ast}}$, ${R_{d}^{MC}}$, ${R_{d}^{ISM}}$ are the 
radial scale lengths of stars, MCs and ISM, and ${z_{d}^{\ast}}$, ${z_{d}^{MC}}$, ${z_{d}^{ISM}}$  the corresponding scale heights above the 
equatorial plane. Stars and star forming MCs have, as a first approximation, the same distribution: ${R_{d}^{\ast}}={R_{d}^{MC}}$ and ${z_{d}^{*}}={z_{d}^{MC}}$.
The scale parameters are chosen taking into account  the observations for the type of object  to model: for the disc of a typical massive galaxy like the Milky Way (MW), 
the typical assumption is  $z_d\simeq 0.3-0.4 kpc$, and $R_d$ is derived from either observations of  the gas and star distribution or from empirical relations (\citet{Im1995}, 
$\log(R_d/kpc) \sim -0.2 M_B-3.45$, where $M_B$ is the absolute blue magnitude). Typical values for $R_d$ are around $5kpc$.

\indent The constants $\rho^{i}_0$ vary with the time step. Indicating with  $t_{G}$ the age of the model galaxy 
the gaseous components ask the normalization constants  $\rho^{MC}_0(t_G)$ and $\rho^{ISM}_0(t_G)$ since both 
loose memory of their past history.  
For the stellar component, $\rho^{*}_0(t)$ is needed all over the galaxy life $0 < t < t_{G}$.
The stellar emission is calculated using the mix of stellar populations of any age $\tau^{\prime}=t_{G}-t$. 
The normalization constants come by integrating the density laws over the volume and by imposing the integrals to equal the mass obtained from the chemical model. 
The mass of each component $M_{i}(t)$ is: 

\begin{equation}\label{g2}
M^{i}=\int_V{\rho^{i}_0\exp\left(-\frac{r\sin\theta}{R_d^i}\right)
\exp\left(-\frac{r|\cos\theta|}{z_d^i}\right)}dV.
\end{equation}

\noindent The mass of stars born at the time $t$ is given by 	 $\Psi(t)$: $\rho^{*}_0(t)$ will be obtained by using $M^{*}\left(t \right)=\Psi(t)$.
$M^{ISM}\left(t \right)$ is the result of gas accretion, star formation and gas restitution by dying stars. 
The current total mass $M^{MC}(t_G)$ is a fraction of $M^{ISM}(t)$ and the remaining is the gas component $M_{g}(t)$. 
The double integral (in r and $\theta$) is numerically solved for $\rho^{i}_0(t)$  to be used in 
Eq.~(\ref{g1}). The galaxy radius $R_{gal}$ is left as a free parameter of
the model.

\indent The last point is the subdivision of the whole volume of a disc galaxy into a number of sub-volumes.
The energy source inside each of these can be approximated to a point source located in their centers, and the coordinates $[r,\theta,\phi]$ are divided in 
suitable intervals. As far as the radial coordinate,
 $n_{r}$ = 40-60 is a good approximation in securing the overall energy balance among the sub-volumes, speeding up the computational time and yielding numerically accurate results. 
The number of radial intervals come by imposing that the mass density among two adjacent sub-volumes scales by a fixed ratio
 $\rho_j/\rho_{j+1}=\zeta$, with constant $\zeta$. 
The grid for the angular coordinate $\theta$ is chosen in such a way that spacing gets thinner approaching the equatorial plane.
We split the angle $\theta$ going from 0 to $\pi$ in $n_{\theta}$ sub-values. We need an odd value for $n_{\theta}$ so that we have ($n_{\theta}$ - 1) /2 sub-angles 
per quadrant. The angular distance $\alpha_{1}$ between the two adjacent values of the angular grid is chosen following \citet{Silva1999}: $R_{gal}$ 
subtends a fraction $f < 1$ of the disc scale height ($z_{d}$). 
The grid for the angular coordinate $\phi$ is chosen to be suitably finely spaced near $\phi$ = 0 and to get progressively broader and broader moving away 
clockwise and counterclockwise from $\phi$ = 0.

\subsubsection{Early-type Galaxies and Bulges}\label{spatial_ell}

The luminosity distribution of ETGs is customarily described by King law. 
Following \citet{Fioc1997}, we use a density profile slightly different from the King law to secure a smooth behavior at the galaxy radius $R_{gal}$.
The mass density profiles for stars, MCs, and diffuse ISM are

\begin{equation}\label{g3}
\rho^i=\rho^{i}_0 \left[1+\left(\frac{r}{r_c^i}\right)^2\right]^{-\gamma_i}
\end{equation}

\noindent where as usual ''$i$'' stands for ''$\ast$'', ''$ISM$'', ''$MCs$''. 
$r^{*}_{c}$, $r^{MC}_{c}$, $r^{M}_{c}$  are the core radii of the distributions of stars, MCs, and diffuse ISM; 
the exponents  $\gamma_{*}$ and  $\gamma_{MC}$ can be 1.5 \citep{Piovan2006b} and $\gamma_{ISM}$ is not well known. 
\citet{Froehlich1982}, \citet{Witt1992}, \citet{Wise1996} suggest to adopt, for the elliptical galaxies, $\gamma_{M}\simeq 0.5-0.75$.
Here, we consider $\gamma_{M}=0.75$. The density profile has to be truncated at the galactic radius $R_{gal}$, free parameter of the model, 
to prevent the mass $M(r)\rightarrow\infty$ for $r\rightarrow\infty$. 
The constants $\rho^{i}_{0}(t)$, $\rho^{MC}_0(t_G)$, $\rho^{ISM}_0(t_G)$  can be found by integrating the density law over the volume and by equating this value 
of the mass to the correspondent one derived from the global chemical model.
The last step is to fix the spacing of the coordinate grid $(r,\theta,\phi)$. The spherical symmetry simplifies this issue.
and the spacing of the radial grid is made keeping in mind the energy conservation constrain. 
We take a sufficiently large number of grid points $n_r\simeq40-60$. 
The coordinate $\phi$ is subdivided into an equally spaced grid, with $n_{\phi}$ elements in total, and $\phi_{1} = 0$, $\phi_{1} = 0$. 
For the azimuthal coordinate $\theta$ we adopt the same grid we have presented for the discs.

\subsubsection{Intermediate-type galaxies}\label{spatial_inter}

Intermediate-type galaxies go from the
early S0 and Sa (big bulge) to the late Sc and Sd (small or negligible bulge).
Different SFHs for the disc and the bulge can reproduce this behavior.
We adopt a system of polar coordinates with origin at the galactic center $(r,\theta,\phi)$: azimuthal symmetry rules out the coordinate $\phi$.
In the disc, the density profiles for the three components are the double decreasing exponential laws of Eq.~\ref{g1} and 
the scale lengths are $R^{*}_{d,B}$, $R^{M}_{d,B}$, $R^{MC}_{d,B}$, $z^{*}_{d,B}$, $z^{M}_{d,B}$, $z^{MC}_{d,B}$ 
(suffix \textit{B} indicates the \textit{disc-bulge composite model}). 
In the bulge the three components follow the King-like profiles (Eq.~\ref{g3}) with 
the core radii $r^{*}_{c,B}$, $r^{M}_{c,B}$, $r^{MC}_{c,B}$ referred to the bulge.
The SFHs of disc and bulge evolve independently: the total content in stars, MCs and ISM  is
the sum of the disc and bulge contributions.

The composite shape of the galaxy lead the definition of  \textit{a new mixed grid sharing
the properties of both components}. 
$R_{B}$ is the bulge radius, $R_{gal}$ the galaxy radius. 
The radial grid is defined building two grids of radial coordinates, $r_{B,i}$ and $r_{D,i}$. 
The grid of the bulge is thicker toward the center of the galaxy: the coordinates $r_{i,B}$ of the bulge grid if $r_{i} < R_{B}$ are considered,
while for $r_{i} > R_{B}$ the coordinates of the disc $r_{D,i}$, until $R_{gal}$, are used. 
The angular coordinate $\theta$ follows the same pattern.
The azimuthal grid is chosen in the same way both for the disc and the bulge, both having azimuthal symmetry.

\subsubsection{The elemental volume grid}\label{vol_grid}

Given the geometrical shape of the galaxies, the density distributions of the three main components, and
the coordinate grid $(r,\theta,\phi)$,
the galaxy is subdivided into  $(n_r,n_{\theta},n_{\phi})$ small volumes $V$.  
Thereinafter the volume $V(r_{iV},\theta_{jV},\phi_{kV})$ will be indicated as $V(i,j,k)$. 
The mass of stars, MCs, and diffuse ISM in each volume are derived from the corresponding density laws,
neglecting all local gradients in ISM and MCs. 
The approximation works since the elemental volumes are small.

\section{Synthetic photometry of a galaxy}\label{spe_syn_ism}

As already said, the light emitted by a galaxy has two  main components: the light emitted by individual stars and the light emitted-reprocessed by the ISM.   
Our model of the  SED emitted by galaxies of different morphological type along any direction strictly follows the formalism and results developed by 
\citet{Piovan2006a,Piovan2006b}. 
In the following we shortly summarize the prescriptions we have adopted to describe the stellar and ISM contributions.

\subsection{The diffuse ISM: extinction and emission}

\citet{Piovan2006a}, \citet{Piovan2011_4541}, \citet{Piovan2011_4561}, and \citet{Piovan2011_4567} presented detailed studies of the extinction and emission properties of dusty
 ISMs. They took into account three dust components: graphite, silicates and PAHs and reached an excellent overall agreement  between theory and observational data for 
the extinction and emission of the ISM in the MW, Large Magellanic Cloud (LMC) and Small Magellanic Cloud (SMC). As we are now going to include their results
for dusty ISMs  in our model galaxies, it is wise to briefly summarize here the basic quantities and relationships in usage for the sake of completeness and clarity.

The total cross section of scattering, absorption and extinction is given by

\begin{equation}
\sigma_{p}\left( \lambda \right)
=\int_{a_{\min,i}}^{a_{\max,i}}\pi a^{2}Q_{p}\left( a,\lambda
\right) \frac{1}{n_{H}}\frac{dn_{i}(a)}{da} da \label{sigabs}
\end{equation}

\noindent the index $p$ stands for absorption (abs), scattering (sca), total extinction (ext), the index $i$ identifies the type of grains, 
$a_{min,i}$ and $a_{max,i}$ are the lower and upper limits of the size distribution for the i-type of grain, $n_{H}$ is the 
number density of $H$ atoms, $Q_{p}\left( a,\lambda \right)$ are the dimension-less absorption and scattering coefficients \citep{Draine1984,Laor1993,Li2001} and, 
finally $dn_{i}(a)/da$ is the distribution law of the grains \citep{Weingartner2001a}.

Using the above cross sections we calculate the optical depth $\tau_{p}(\lambda)$ along a given path

\begin{equation}
\tau_{p}\left( \lambda \right) =\sigma _{p}\left( \lambda \right)
\int_{L}n_{H}dl=\sigma_{p}\left(\lambda \right) \times N_{H}
\label{tauabs}
\end{equation}

\noindent $L$ is the optical path and all other symbols have their usual meaning.
The assumption is that the cross sections remain constant along the optical path.

$j_{\lambda}^{small}$, $j_{\lambda}^{big}$ and $j_{\lambda}^{PAH}$ are the contributions to the emission by small grains, big grains and PAHs, respectively. 
How these quantities are calculated is widely described  in \citet{Piovan2006a} to whom the reader should refer for more details. 
The key relationships are the following ones.
The contribution to the emission by very small grains of graphite and silicates is

\begin{eqnarray} \label{smallemission}
j_{\lambda }^{small}&=&\pi \int\nolimits_{a_{min
}}^{a_{flu}}\int\nolimits_{T_{min}}^{T_{max}} a^{2}Q_{abs}\left(
a, \lambda \right)B_{\lambda }\left( T\left(
a\right) \right)\nonumber \times \\
&& \times \frac{dP\left(a\right)}{dT} dT
\frac{1}{n_{H}}\frac{dn\left(a\right)}{da}da
\end{eqnarray}

\noindent where $dP \left(a\right)/dT $ is the distribution temperature from $T_{min}$ to $T_{max}$  attained by grains with generic dimension $a$ under an incident 
radiation field and $B_{\lambda }\left( T\left( a\right) \right) $ is the Planck function. $Q_{abs}\left( a, \lambda \right)$ are the absorption coefficients, 
$dn\left(a \right)/da$ is the \citet{Weingartner2001a} distribution law for the dimensions, 
$a_{flu}$ is the upper limit for thermally fluctuating grains, $a_{min}$ is the lower limit of the distribution.

The emission by big grains of graphite and silicates is evaluated assuming that they behave like  black bodies in equilibrium with the radiation field. 

\begin{equation} \label{bigemission}
j_{\lambda}^{big}=\pi \int_{a_{flu}}^{a_{max}}  a^{2} Q_{abs}
\left( a,\lambda \right) B_{\lambda} \left( T \left(a \right)
\right) \frac{1}{n_{H}}\frac{dn\left(a\right)}{da} da
\end{equation}

\noindent where $a_{max}$ is the upper limit of the distribution and the meaning of the other symbols is the same as in Eq.~\ref{smallemission}.

The emission by PAHs is given by

\begin{eqnarray} \label{PAHemission}
j_{\lambda}^{PAH} &=& \frac{\pi }{n_{H}hc}\int\nolimits_{\lambda
_{min}}^{\lambda _{max}}I \left( \lambda^{^{\prime }} \right)
\lambda^{^{\prime
 }} \int\nolimits_{a_{PAH}^{low}}^{a_{PAH}^{high}} \frac{dn\left(a\right)}{da} \times \nonumber\\
&\times & a^{2} \left[ Q_{abs}^{IPAH} \left( a,
\lambda^{^{\prime}} \right)
S_{ION} \left( \lambda^{^{\prime}} ,\lambda ,a \right) \chi_{i} + \right. \\
& +& \left. Q_{abs}^{NPAH} \left( a, \lambda^{^{\prime}} \right)
S_{NEU} \left( \lambda^{^{\prime}} ,\lambda ,a \right) \left( 1-
\chi_{i} \right) \right] da  d\lambda^{^{\prime}}  \nonumber
\end{eqnarray}

\noindent where the ionization of PAHs is taken into account \citep{Weingartner2001b} and $\chi_{i}=\chi_{i}\left(a\right)$ is the fraction of ionized PAHs with dimension $a$. 
$S_{ION} \left(
\lambda^{^{\prime}} ,\lambda,a \right)$ and $S_{NEU} \left(\lambda^{^{\prime}} ,\lambda ,a \right)$ give the  energy emitted at wavelength $\lambda$ by a molecule of dimension $a$,
 as a consequence of absorbing a single photon with energy  $hc/\lambda ^{^{\prime }}$. $a_{PAH}^{low}$ and $a_{PAH}^{high}$ are the limits of the distribution and 
$I \left( \lambda^{^{\prime }} \right)$ is the incident radiation field.

\subsection{Dust-free and dust-embedded SSPs}

In the following we group the SSPs  according to  whether or not they incorporate the effect of dust during the MC phase of the young massive stars. 
In all cases,  the effect of the self-absorbing envelopes of the AGB phase of intermediate and low mass stars is taken into into account.

(i) \textbf{Dust-free SSPs}. These SSPs describe the situation in which a generation of stars has already evaporated the parental MC in which it was embedded. 
This implies that a certain amount of time has already elapsed from initial star forming event. So these SSPs no longer need to include the effects of self obscuration and 
radiation reprocessing exerted by the parental MC or local ISM.  However, they still include these effects when caused by the  dust shells surrounding the AGB stars. 
For these SSPs we consider the recent study made by \citet{Cassara2013} in which the new models of TP-AGB stars by \citet{Weiss2009} have been used. 
The reader should refer to \citet{Cassara2013} for all details.

(ii) \textbf{SSPs-embedded in dusty MCs}.
In the early stages of their evolution we can consider stars as still heavily obscured by their parental dusty MCs.  
Libraries of these SSPs have been presented in \citet{Piovan2006a} as a function of four parameters: (1) optical depth $\tau$, (2) metallicity $Z$, 
(3) PAHs ionization state (three possibilities, that is, PAHs ionized with full calculations of the ionization state, neutral PAHs and ionization state of the PAHs 
as in a mixture of Cold Neutral Matter, Warm Neutral Matter and Warm Ionized Matter in the same relative proportions as in the MW) and, finally, (4) the abundance of 
Carbon in very small grains. These libraries are still up-to-date because no significant changes have been made to the structure and evolution of massive stars in the meantime. 
Therefore for these  SSPs we adopt the library  by \citet{Piovan2006a}, in which the absorption and emission of the radiation  emitted by the young stars embedded in MCs 
are accurately calculated with the Ray-Tracing method. The \citet{Piovan2006a} SSPs, however, neglect the effects by self-contamination in AGB stars. Although 
complete sets of SEDs with all these effects simultaneously taken into account would be desirable, or in other words the \citet{Piovan2006a} SSPs should be folded into 
those by \citet{Cassara2013},  the above approximation is fully acceptable for a number of reasons:  (1) the
spectral regions interested by  AGB stars does not coincide with the spectral regions interested by the interaction of young stars with the dust of the MCs; 
(2) the details  of the SEDs caused by the transfer of energy from FIR to NIR are found to play a marginal role; 
(3) even if the self-absorption by the AGB dusty envelopes is included, the effect is found to be marginal unless very high optical depths are chosen for the cloud.

To conclude in the following we adopt the SSPs by \citet{Piovan2006a} up to the evaporation of the parental MC and switch to those by \citet{Cassara2013} afterwards.

\subsection{Molecular Clouds and their evaporation}\label{evapor_MC}

According to the current view of star formation stars are born and live for part of their life inside MCs. 
As already recalled the radiation emitted by these stars in the UV region of the spectrum is absorbed by  MCs and re-emitted in the infrared. 
Therefore the radiation emitted by a galaxy can be severely altered by the presence of the MCs. The ideal approach would be  to be able to follow the evolution of MCs 
that are gradually consumed by the star forming process
and swept away by the radiation emitted by the underneath stars (SN{e} explosions and stellar winds of massive stars). 
A task that goes beyond the aims of the present study and that is simplified to evaluating the time scale for the MC evaporation. In the real case,
however, the clouds are destroyed on a timescale of the order of 10 Myr on average, typical lifetime of the molecular clouds \citep{Dwek1998,Zhukovska2008}. 
 A simple way to simulate the process above is to assume that as the time goes on, the SSPs fluxes reprocessed by dust decrease, while  the amount of flux left unprocessed  
increases.
The time scale, $t_{0}$,  for the evaporation of the MC will depend on
the properties of the ISM,  the efficiency  of the star formation  and the energy injection by young stars inside.  
We expect $t_0$  to be of the same order of the lifetime of massive stars (the age range going from 3  to 50 Myr). 
In a low density environment with a moderate rate of star formation, $t_{0}$ is likely close to  the lowest value (lifetime of the most massive stars of the population, 
case of a typical spiral galaxy), while $t_{0}$ will be close  to the upper value in a high-density environment,
(star-burst galaxies) where very obscured star formation can occur in a high density ISM, more difficult to destroy.

Finally, there is an important effect due to the the metallicity. See \citet{Piovan2006a} for a more details.

\subsection{The spectral energy distribution of a galaxy}\label{SED}

The total SED emerging from the galaxy is simulated once the main physical components, their spatial distribution, the coordinate system and 
the grid of elemental volumes are known, and the interaction among stars, dusty ISM and MCs is modelled.
A generic volume ${V^{\prime}}=V(i^{\prime},j^{\prime},k^{\prime})$ of the galaxy will receive the 
radiation coming from all other elemental volumes $V=V(i,j,k)$ and the radiation traveling from one volume to another 
interacts with the ISM comprised between them. The energy is both absorbed and emitted by the ISM under the interaction 
with the radiation field. 
Two simplifying hypotheses are followed: 

\begin{enumerate}
\item The dust of a generic volume $V$ does not contribute to the radiation field impinging on the volume $V^{'}$. 
This is due to the low optical depths of the diffuse ISM in the MIR/FIR: dust can not effectively absorb in significant amount the radiation it emits,
except for high density regions \citep{Piovan2006a}. 
The incoming radiation depends only on stars and MCs. 

\item The radiative transfer from a generic volume $V$ to $V^{'}$ is calculated by means of effective optical depth:

\begin{equation}\label{g6}
        \tau_{eff}=\sqrt{\tau_{abs}(\tau_{abs}+\tau_{sca})}
    \end{equation}
For all details see \citet{Piovan2006a} and  \citet{Silva1998}.
\end{enumerate}

\noindent The total radiation field incident on $V^{\prime}$ is:

\begin{eqnarray}\label{g7}
J(\lambda,V^{\prime}) &=&\sum^{n_{r}}_{i=1}\sum^{n_{\theta}}_{j=1}\sum^{n_{\phi}}_{k=1}\frac{V[j^{MC}
(\lambda,V)+j^{*}(\lambda,V)]}{r^{2}(V,V^{\prime})} \times  \nonumber \\
 &\times&  e{^{[-\tau_{eff}(\lambda,V,{V^{\prime}})]}}
\end{eqnarray}

\noindent the summations are carried over the whole ranges of $i$, $j$, $k$ with $i\neq i^{\prime}$ $j\neq j^{\prime}$ $k\neq k^{\prime}$; 
$r^{2}(V,V^{\prime})$ is the value averaged over the volume of the square of the distance between the volumes $V$ and $V^{\prime}$. 
The effective optical depth  $\tau_{eff}$ of  Eq.~\ref{g7} is given by :

\begin{eqnarray}\label{g8}
 \tau_{eff}(\lambda,V,V^{\prime})&=&\sqrt{\sigma_{abs}(\lambda)
 [\sigma_{abs}(\lambda)+\sigma_{sca}(\lambda)]}\times \nonumber \\
 &\times&\int^{V(i^{\prime},j^{\prime}
 k^{\prime})}_{V(i,j,k)}n_{H}(l)dl
\end{eqnarray}

\noindent The integral represents the number of H atoms contained in the cylinder between $V$ and $V^{\prime}$.
The two terms  $j^{MC}(\lambda,V)$ and  $j^{*}(\lambda,V)$ are the emission by MCs and stars per unit volume of $V(i,j,k)$ and they are calculated 
at the center of the volume element.

\noindent To calculate $j^{MC}(\lambda,V)$ and  $j^{*}(\lambda,V)$ the fraction $f_{d}$ of the  SSP luminosity that is reprocessed by dust and  
and the time scale $t_0$ for this to occur are requested.
The fraction is

\begin{eqnarray}
\label{g9}
f_d=
\begin{cases}
1           & t\leq t_0 \\
2-t/t_0     & t_0 <t \leq 2t_0\\
0           & t\geq 2t_0\\
\end{cases}
\end{eqnarray}
The fraction of SSP luminosity that escapes without interacting with dust is  $f_{f}=1-f_{d}$. 
See \citet{Piovan2006a} for a detailed description of the calculations for the monochromatic luminosity
of a dust free and dust enshrouded SSPs: here we just report, for sake
of clarity, the emission of stars and MCs per unit volume, $j^{*}(\lambda,V)$ and $j^{MC}(\lambda,V)$:

\begin{eqnarray}\label{g11}
j^{*}(\lambda,V) &=& \int_{2t_{0}}^{t_{G}}\rho_{*}(t)L_{\lambda}^f(\tau^{\prime},Z)dt+   \nonumber \\
                 &+&\int^{2t_{0}}_{t_{0}} \left(\frac{t}{t_{0}}-1\right) \rho_{*}(t){L_{\lambda}^f}(\tau^{\prime},Z)dt
\end{eqnarray}
and

\begin{eqnarray}\label{g12}
j^{MC}(\lambda,V) &=& \int_{0}^{t_{0}}\rho_{*}(t)L_{\lambda}^d(\tau^{\prime},Z)dt+ \nonumber \\
                  &+&\int^{2t_{0}}_{t_{0}}\left(2-\frac{t}{t_{0}}\right) \rho_{*}(t){L_{\lambda}^d}(\tau^{'},Z)dt
\end{eqnarray}

Once the incident radiation field $J(\lambda, V^{\prime})$ is known, we can obtain the emission per unit volume from the dusty ISM. 
The azimuthal and spherical symmetries of the galaxy models become very important, and it allows to
calculate the dust emission at $\phi=0$ for all the possible values of $r$ and $\theta$ on this ''galaxy slice''. 
The total radiation field for unit volume emitted by a single element is:

\begin{equation}\label{g13}
j^{TOT}(\lambda,V)=j^{MC}(\lambda,V)+j^{*}(\lambda,V)+j^{ISM}(\lambda,V),
\end{equation}

\noindent $j^{ISM}(\lambda,V)$ is the radiation outgoing from a unit volume of the dusty diffuse ISM. 
The total outgoing emission from the volume $V$, $j^{TOT}(\lambda,V)\times V$, 
is of course different from volume to volume.

The monochromatic luminosity measured by an external observer is calculated considering that the radiation emitted by each elemental volume $(n_r, n_{\theta},n_{\phi})$  
has to travel across a certain volume of the galaxy itself before reaching the edge, escaping from the galaxy, and being detected.

The radiation is absorbed and diffused by the ISM along this path, and the external observer will see 
the galaxy along a direction fixed by the angle $\Theta$ ($\Theta=0$: galaxy seen \textit{face-on}, 
$\Theta=\pi/2$ galaxy seen \textit{edge-on}). 
Hence:

\begin{equation}\label{g14}
L(\lambda,\Theta)=4\pi\sum_{i=1}^{n_r}\sum_{j=1}^{n_{\theta}}
\sum_{k=1}^{n_{\phi}}Vj^{TOT}(\lambda,V)e^{[-\tau_{eff}(\lambda,V,\Theta)]},
\end{equation}

\noindent and $\tau_{eff}(\lambda,V,\Theta)$ is the effective optical depth between $V(i,j,k)$ and the galactic edge along the direction. 
The detailed description can be found in \citet{Piovan2006b}.

\section{The composition of dust}\label{compo_dust}

To introduce this topic it is worth a quick summary about the dust properties and the mixture adopted in the models: for a much more extended analysis
of this issue and all the details about the calculations of the emission/extinction effects, see \citet{Piovan2006a}.

\indent The physical properties of the interstellar grains are derived from the dust extinction curves in the UV/optical region 
of the spectra and the emission spectra in the infrared bands (from the near up to the far infrared), in different physical environments. 
From the amount of information that can be obtained looking at the effects of extinction and emission related to dust grains, 
it is possible to derive the features useful to constrain and define a quasi-standard model of interstellar dust, made up of three components.

\noindent The characteristic broad bump of the extinction curve in the UV at 2175 $\AA$ and the absorption features 
at 9.7 $\umu$m and 18 $\umu$m \citep{Draine2003a} require a two components model, made of graphite and silicates while a 
population of very small grains (VSGs), is necessary to reproduce the emission  observed by IRAS 
at 12 $\umu$m and 25 $\umu$m. 

\indent VSGs are not exclusively made of silicates (the 10 $\umu$m emission feature of silicates
is not detected in diffuse clouds \citet{Mattila1996,Onaka1996}): more likely they are composed by
carbonaceous material with broad ranges of shapes, dimensions, and chemical structures (\citet{Desert1986a} and
\citet{Li2002a} present a detailed discussion of this topic).

The last contributors to dust are the PAH molecules, that originated the emission lines at 3.3, 6.2, 7.7, 8.6, and 11.3 $\umu$m
which have been first observed in luminous reflection nebulae, planetary HII regions and nebulae  \citep{Sellgren1983,Mathis1990} 
and in the diffuse ISM with IRTS \citep{Onaka1996,Tanaka1996} and ISO \citep{Mattila1996}.
These spectral features are referred to as the aromatic IR bands (AIBs).

\indent It appears clear that any realistic model of a dusty ISM, able to explain the UV-optical extinction and the IR emission of galaxies,
need the inclusion of at least three components: graphite, silicates, and PAHs. 
Furthermore, the big grains can be treated as in thermal equilibrium with the radiation field, while the VSGs 
could have temperatures above the mean equilibrium value.
The properties of a mixture of grains are obtained once their cross sections, their dimensions, and the kind of interaction with the local radiation field
are known. 
For more information see \citet{Piovan2006a}. 
We only summarize here the models that inspired our three dust components ISM, 
with graphite, silicates and PAHs. The cross sections for graphite are from \citet{Draine1984}, 
for silicates are from \citet{Laor1993} and PAHs are from \citet{Li2001}, taking the latest releases from the B. T. Draine webpage. 
The extinction curves and the distribution of dust grains as a function of their dimension are taken from \citet{Weingartner2001a}. 
The emission of graphite and silicates, both for thermally fluctuating VSGs and big grains in thermal equilibrium, is based upon the
 classical paper \citet{Guhathakurta1989}, while for PAHs we adapted \citet{Puget1985}. Finally, the ionization state of PAHs is 
calculated with the physical models by \citet{Draine1987,Bakes1994,Weingartner2001b}.

\begin{table*}
\begin{center}
\caption{ Baryonic masses of galaxies or galaxy components. Masses are in units of $10^{12}M_{\odot}$  }
\label{tab_masses}
\vspace{1mm}
\begin{tabular*}{50.0mm}{l l l}
\hline
   Galaxy Type &  Bulge       & Disc  \\
     
\hline
   Sd-Irr      &              & 0.09450  \\
   Sd          & 0.01765      & 0.08700  \\
   Sc          & 0.03510      & 0.07360  \\
   Sbc         & 0.04500      & 0.06650  \\
   Sb          & 0.05500      & 0.05660  \\
   Sab         & 0.06450      & 0.04716  \\
   Sa          & 0.72600      & 0.04600  \\
   S0          & 0.08160      & 0.03400  \\
   E-S0        & 0.08540      & 0.02690  \\
   Elliptical  & 0.11800      &          \\
\hline
\end{tabular*}
\end{center}
\end{table*}

\section{Parameters for chemical and spectro-photometric models}\label{model_param}

We summarize and shortly comment here the main parameters  of the chemical and companion 
spectro-photometric models justifying the choice we have made for each type of model galaxies.

\subsection{Chemical parameters}

\begin{itemize}
\item[-] The galactic mass ${\rm M_{BM}(t_{G})}$.
    In the infall models it represents the asymptotic value reached by the 
		baryonic component of a galaxy at the present time, the galaxy age $t_{G}$.  
		This asymptotic mass is used to normalize the gas and star masses of the galaxies. The procedure is straightforward and ${\rm M_{BM}(t_{G})}$ strictly coincides with 
		real baryonic mass of the galaxy at the present time in the case of disc galaxies, because their evolution is calculated in absence of galactic wind. 
		Conversely in the case of spheroidal systems (bulges and/or ETGs), the occurrence of galactic  winds by gas heating due to SN explosion and stellar winds, requires
		some some cautionary remarks.
    Galactic winds takes place when the gas thermal energy exceeds the gravitational binding energy of and consequently gas is expelled and star formation is halted. 
		No subsequent revival of star formation is possible in these models.  Therefore the real mass of the baryonic component coincides with the total mass 
		of the stellar populations built up to the wind stage. This value is lower than ${\rm M_{BM}(t_{G})}$. Keeping this in mind, 
		also in this case the normalization mass is ${\rm M_{BM}(t_{G})}$.  For a detailed description of the galactic wind process and its effects 
		on the masses of gas and stars, see \citet{Tantalo1996,Tantalo1998,Cassara2012}. 
		Given these premises, galaxy models for bulges and discs of different masses have been calculated using  the  values of ${\rm M_{BM}(t_{G})}$
		listed in  columns (2) and (3) of Table \ref{tab_masses}.  Remember that the real present day mass of spheroidal systems is somewhat smaller 
		than the listed value and also model dependent because of the occurrence of galactic winds
    that halt star formation and expel an sizable fraction of the initial BM.

\item[-] The ratios ${\rm R_{BM}/R_{DM}}$ and ${\rm M_{BM}/M_{DM}}$, gravitational potential and the effect of dark matter, are both set equal to 0.16;

\item[-] The exponent $k$ of the \citet{Schmidt1959} star formation rate; all models are calculated using $k=1$;

\item[-] The efficiency $\nu$ of the star formation rate. All bulge models have $\nu=5$. Even if different values of $\nu$ at 
    varying the galactic mass might be more appropriate to match real situations
    \citep{Tantalo1996,Tantalo1998}, we  keep it constant at varying the mass. 
		It is worth noticing that in some way the effect of the $\nu$  is not expected to be as strong as in \citet{Tantalo1996,Tantalo1998}, 
		because the mass range  we are considering is much narrower than in those studies. The value we adopt can be considered typical of a 
		system with mass of 10$^{11}$ $M_{\odot}$. The efficiency  $\nu$  for the discs is significantly lower: all disc models have 
		$\nu=0.50$ \citep{Portinari1998}, however adjusted the value to $\nu=0.35$  in order to match the mean metallicity in a typical disc like the
		Milky Way  \citep{Buzzoni2005}. If more complicated star formation laws are 
		used to reproduce the star formation history in disc galaxies \citep{Portinari2000}, the efficiency $\nu$ should be 
		adjusted to match the observational value of the mean metallicity.

 \item[-] The initial mass function (slope and $\zeta$). The slope is kept constant at the classical value of Salpeter, 
whereas the fraction $\zeta$ of the IMF containing stars able to enrich the ISM in chemical elements during the Hubble time is $\zeta=0.5$. 
As pointed in \citet{Tantalo1996}, this is a good choice in order to get models with ${\rm M/L_{B}}$ ratios in agreement with the observational
 data and it allows also to be consistent with the super-solar metallicities suggested in \citet{Buzzoni2005} for the bulges in
 intermediate type galaxies. In the case of disc galaxies a lower value of $\zeta$ is adopted, i.e. $\zeta=0.17$. 
The choice  is suggested by the typical mean metallicities in discs, i.e. $Z\lesssim 0.006 - 0.008$ \citep{Buzzoni2005}.

\item[-] The infall time scale $\tau$. Although the time scale of mass accretion is 
often considered as a free parameter of the models, we assume $\tau=0.3$ for all the models in order 
to reduce the number of free parameters and to mimic results from NB-TSPH numerical simulations by \citet{Merlin2012} 
on the time scale of collapse of baryonic matter and formation of the stellar content of an elliptical like object. 
The ever continuing, less intense star formation in disc galaxies suggest $\tau=3$ \citep{Portinari1999,Portinari2000}.

\item[-] Age of galaxies $t_{G}$: considering the $\Lambda$CDM model of the Universe and 
 redshift of galaxy formation $z=z_{form}=20$, we get $t_{G}=13.30$ Gyr. 
All galaxies are supposed to begin their star formation history at the same time, i.e. redshift.
\end{itemize}

\subsection{Spectro-Photometric Parameters}

Here we summarize and discuss only the most important parameters, distinguishing, as usual, 
between the bulge component of an intermediate type galaxy (or the elliptical galaxy) 
and the disc component (or the disc galaxy):

\begin{itemize}

\item[-] \textbf{Spatial structure}: \textbf{$r_c^M$} is scale radius of the ISM in the King law (Eq.~\ref{g1}); we assume $r_c^M$=0.5 kpc. In general, 
the gas is made up of molecular clouds with active star formation \textit{and} diffuse interstellar medium. Consequently, we need the length scales of MCs, 
diffuse ISM, and stars for which  we assume the same value $r_c^M$=0.5 Kpc. For the effects due to  variations of scale radii see \citet{Piovan2006a}.\\
The ratio \textbf{$r_c^*/r_c^M$} allows the distribution of gas and stars according to different scale radii. 
For the sake of simplicity we assume here that both components have the same spatial distribution  
\citep[see][for more details]{Chiosi2002,Cassara2008,Cassara2012}.\\
\textsf{Radial and vertical mass distributions in discs:} the parameters  are  $z_d^*$=0.4 kpc,  $z_d^{gas}$=0.4 kpc,  $R_d^*$=5 kpc, $R_d^{gas}$=5 kpc (see Eq.~\ref{g2}).\\
\textsf{Mass distribution in bulges:}  in Eq.~(\ref{g3}) we assume $\gamma_{*}$= 1.5, $\gamma_{MC}$= 1.5,  and $\gamma_{gas}$=0.75 as appropriate.

\item[-] \textbf{"fgasmol"}. This parameter fixes the amount of molecular gas present in the galaxy at the age with the peak of star formation, with respect to the total gas.
 This value is then used to scale proportionally the amount of MCs at different ages of the evolution of the galaxy. Indeed, we assume that star formation occurs in the 
cold MCs and therefore they should be dominant in the ISM at the peak of SFR. For elliptical galaxies or bulges it has been used \textit{only} before the onset of the 
galactic wind, after which the star formation process halts. In ETGs and bulges fgasmol=0.8. In disc galaxies we assume  fgasmol=0.6 as suggested by  \citet{Piovan2006b}, 
who took into account estimates of the masses of H$_{2}$ and HI/HII obtained from observations of late-type galaxies in the local universe;

\item[-] \textbf{ Evaporation time $t_0$}: In bulges and ETGs we adopt $t_0= 30 \times 10^6 yr$: MCs are embedded in a primordial environment 
of relatively high density. Consequently a longer time scale  is required  to dissolve these MCs compared to those in a normal environment such as the solar vicinity. 
In discs, most likely of lower density, we adopt $t_0=6 \times 10^6 yr$. This value is taken  from \citet{Piovan2006b}. 
Furthermore we assume  that the evaporation time scale of MCs in disc galaxies of the local universe is the same during the whole evolutionary history as 
suggested by their nearly constant  star formation rates.
\end{itemize}


\section{SEDs of galaxies of different morphological types}\label{SED_gal}

With the aid of the chemical models calculated for pure bulges and discs, 
we build now composite galaxy models going from pure bulges (spheroids) 
to pure discs passing through a number of composite systems with different combinations
 of the two components somehow mimicking the Hubble Sequence of galaxy morphological types. 
The combinations of bulge and disc masses are listed in columns (2) and (3) of Table \ref{tab_masses}. 
These values are estimated taking into to account for the different values of the $L_{Bulge}/L_{Tot}$ 
ratio that are needed to  compare theoretical galaxy colours with the observed ones. 
In all models, the total mass is  $\sim 1\times10^{11}M_{\odot}$ that according to  \citet{Buzzoni2005} 
is typical of intermediate type galaxies (made by bulge + disc).

\subsection{SEDs at the age $t_{Gal}=13.30$ Gyr}\label{monoeta}

In this section we present theoretical SEDs for galaxies of various morphological types, 
taking into account the contribution of the different physical components to the whole galaxy emission. 
The analyzed age is the final age of the models, that it, $t=t_{Gal}=13.30$ Gyr, calculated considering as redshift of formation
$z_{form}=20$, in the current cosmological framework.

\begin{figure*}
\centerline{
\includegraphics[width=0.33\textwidth]{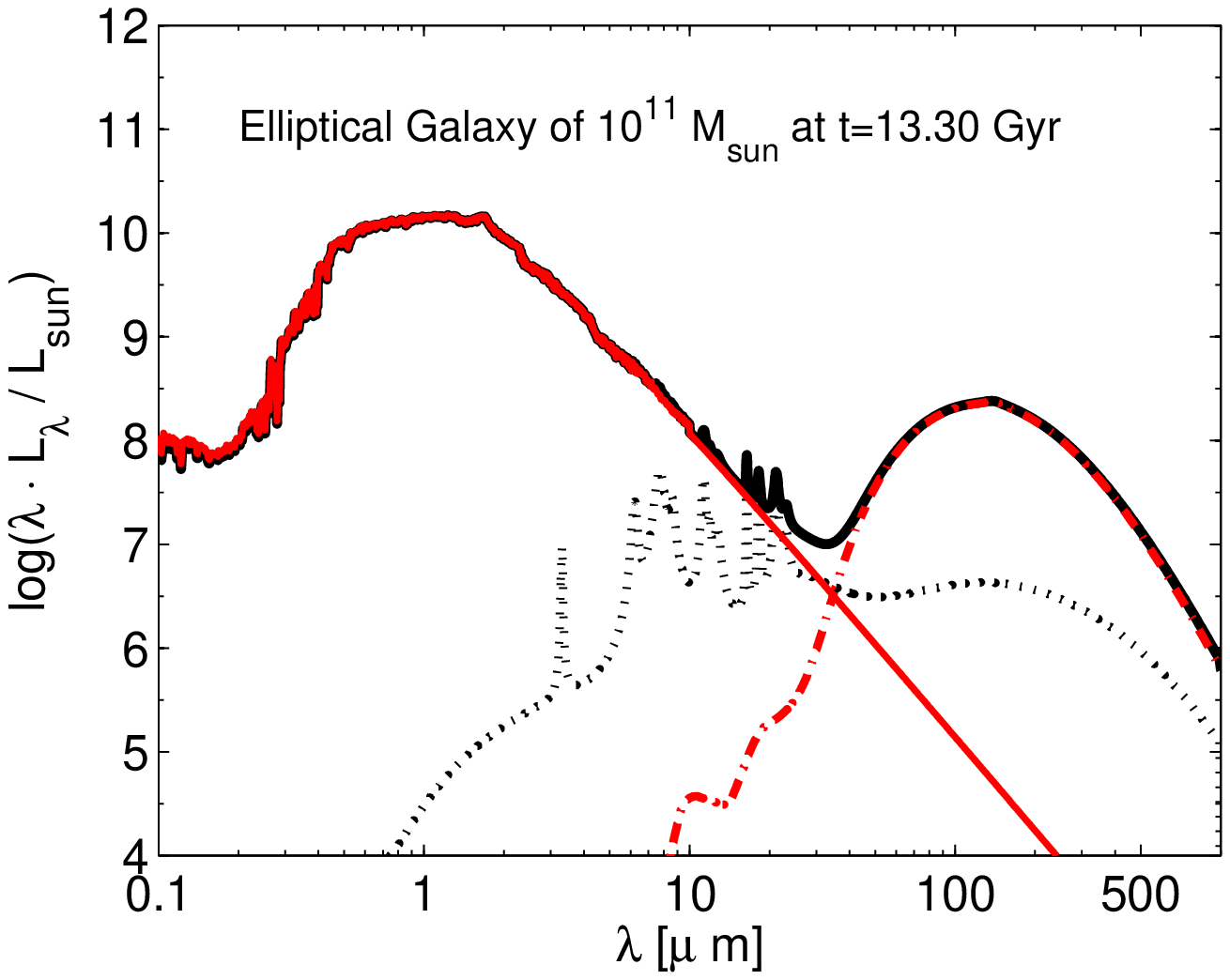}
\includegraphics[width=0.33\textwidth]{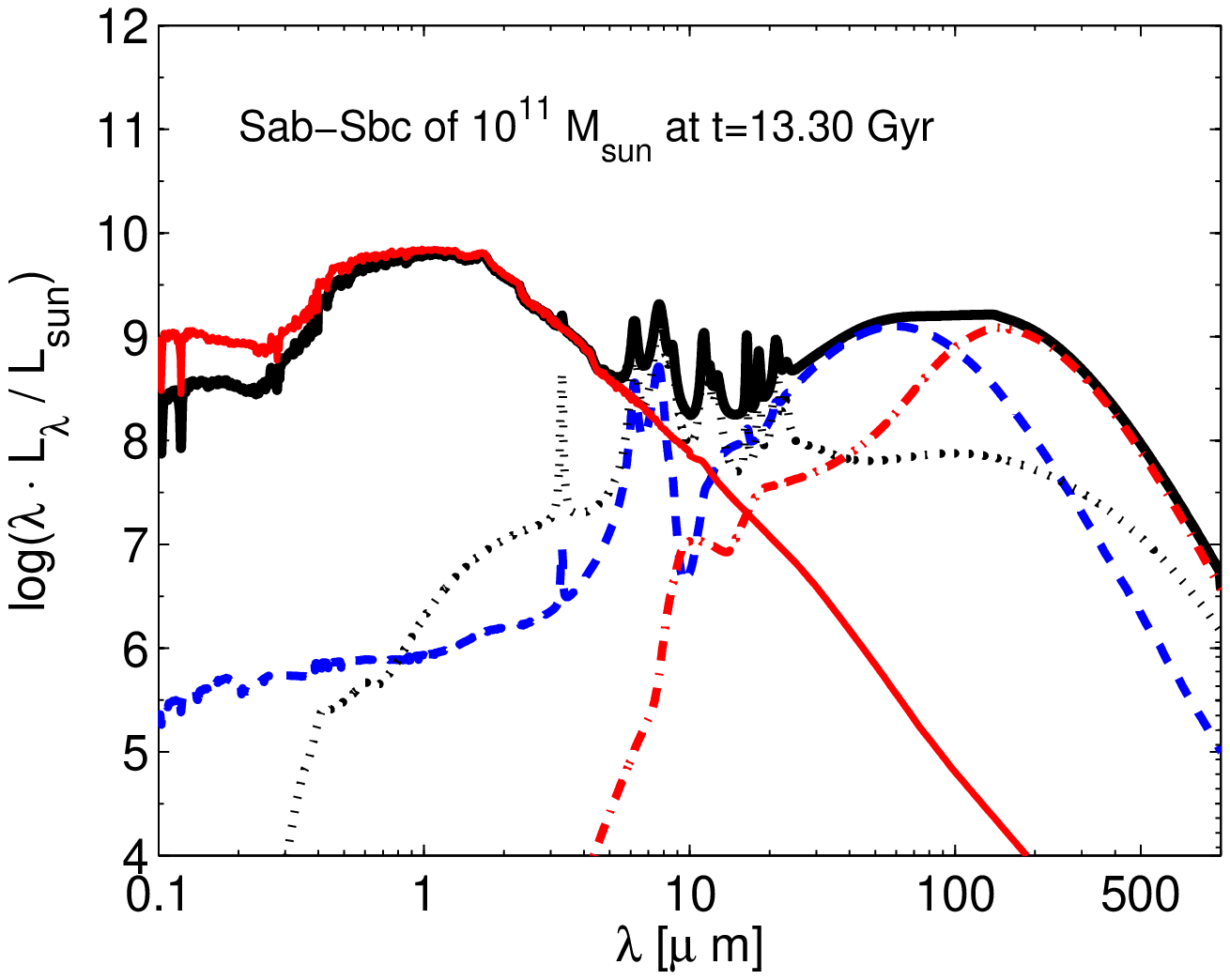}
\includegraphics[width=0.33\textwidth]{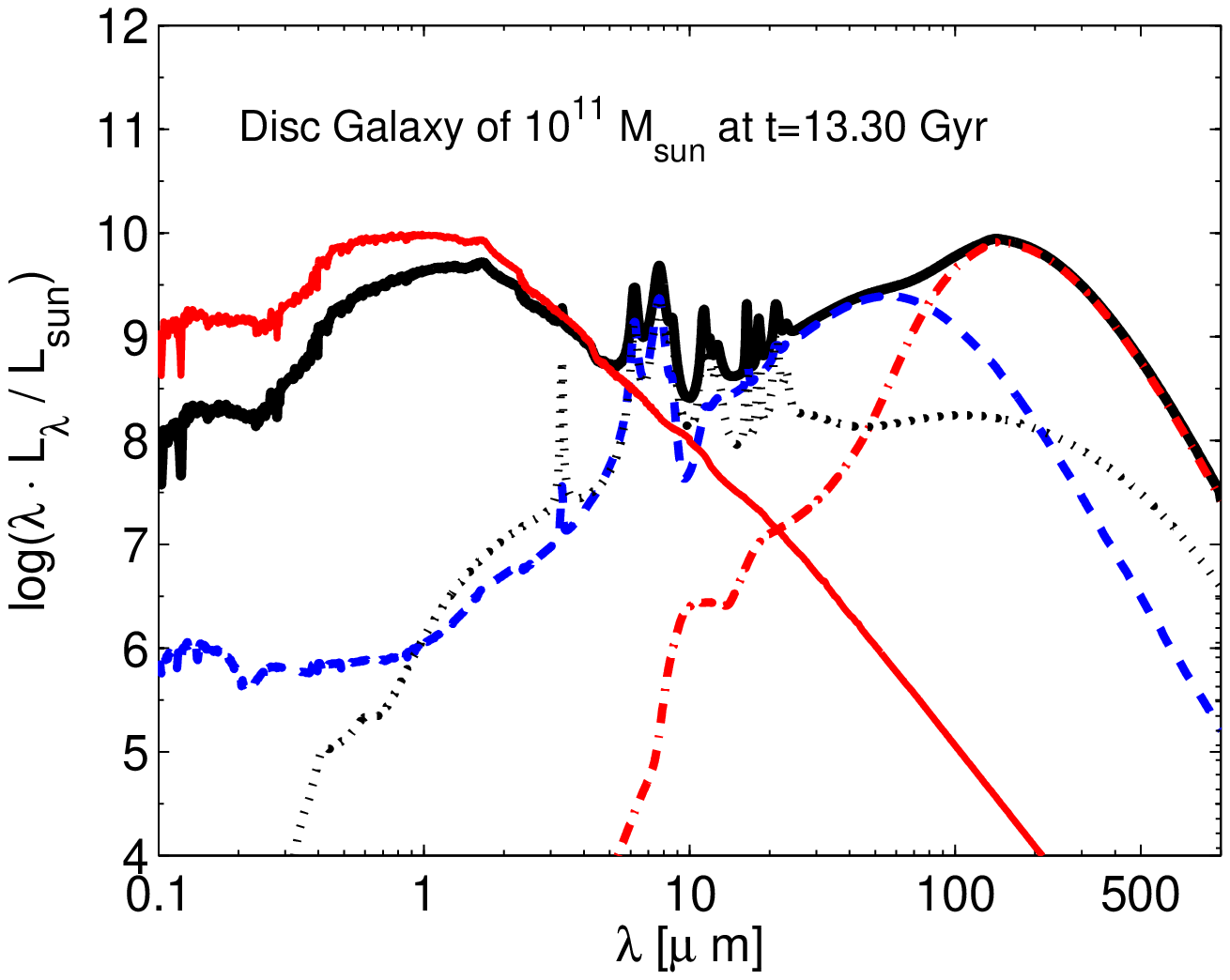} }
\caption{Left Panel: SED of the model \textit{elliptical galaxy} of $M=10^{11} M_{\odot}$ (black solid line).
We represent also the emission of both graphite and silicate grains (red dot-dashed line), the emission of PAHs (black dotted line), 
and the SED where only the extinction effect of the MCs is included (red solid line).
Middle Panel: The same as in the left panel, but for a \textit{Sbc-Sab galaxy} of $M_{tot}=10^{11} M_{\odot}$. 
The meaning of the lines is the same as in left panel. The blue dashed line highlights the contribution of the emission of MCs. 
Right Panel: The same as the left and middle panels, but for a pure disc galaxy of $M=10^{11} M_{\odot}$. The meaning of the lines  is always the same as before. }
\label{bulge_inter_disc}
\end{figure*}

In the panels of Fig.\ref{bulge_inter_disc} we show the SEDs of galaxies of different bulge to disc ratios and at the same age of 13.30 Gyr. 
The morphological classification of the models have been made looking at the theoretical $[\textrm{B}-\textrm{V}]$ and $[\textrm{U}-\textrm{B}]$ 
colours (see below). In more detail: the left panel of Fig. \ref{bulge_inter_disc} shows the SED of a pure elliptical galaxy of $10^{11} M_{\odot}$ 
(black solid line). We represent also the emission of both graphite and silicate grains (red dot-dashed line), the emission of PAHs 
(black dotted line), and the old stellar population whose light is dimmed  by the parent MCs  (red solid line). 
The middle panel displays the  SED of the model Sbc-Sab galaxy, with $M_{Bulge}=0.351 \times 10^{11} M_{\odot}$ and $M_{Disc}=0.736\times 10^{11} M_{\odot}$. 
The luminosity of the same physical components as in the left panel is represented;   we can observe the contribution of the emission of MCs (blue dashed line)
 due to the star formation still active in the disc. This effect can not be appreciate in the case of an elliptical galaxy since the galactic wind swept 
off the ISM hence stopped  star formation. Together with the global EPS with the contribution of dust, we observe the emission of graphite and silicate
 grains, the emission of PAHs, the old stellar population extinguished \textit{only} for the MCs effect (and not for the effect of extinction of the 
diffuse interstellar dust) and, finally, the emission of MCs. Finally,  the right panel gives the SED of a pure disc galaxy of $10^{11} M_{\odot}$
with the contribution by the different physical components to the whole galaxy emission.

It is evident that the shape of the SEDs of the various components gradually changes 
passing from the elliptical model to the disc model: few differences can be observed comparing the two intermediate galaxy types.\\
These considerations hold for the models at 13.30 Gyr (z=0):

\begin{itemize}
\item[-] For the elliptical model: the global emission (black solid line) shows  a peak in the UV (thus
reproducing the ultraviolet excess observed in elliptical galaxies) and a weak IR peak, which is clearly due 
to a poor amount of dust grains in the diffuse medium (there is no contribution of the molecular clouds at the 
emission in the IR region since star formation has stopped). 
The extinction effect is weak since at the final age of the galaxy small amounts of gas and dust are present.

\item[-] For the disc model: both the environments with the presence of dust, 
namely the diffuse interstellar medium and the star forming regions, 
contribute to the IR luminosity in significant amount and both play a role in the extinction of the UV/optical radiation;

\item[-] For the intermediate types (disc plus bulge): the emission of the different components is 
quite similar to those of the disc model. However, in the UV region, the SED where the stellar population is extinguished \textit{only} 
for the MCs effect (red lines - taking into account only the effect of obscuration of young stars) and the total emission (black lines) 
are in practice indistinguishable for $\lambda\gtrsim 0.5$ $\umu$ m, while for the disc model they are clearly separate (right panel of Fig. \ref{bulge_inter_disc}). 
This is ultimately due to the smaller amount of gas still present in the intermediate model 
compared to the disc only model. In the former case the galactic wind has pushed away out all the bulge gas, 
whereas in the disc the galactic wind does never occur.
\end{itemize}

The contribution of the PAHs, silicates and graphite grains grows going from the elliptical model to the disc model; 
this effect is due, as already pointed out, to the small amount of dust and gas present in the elliptical galaxy at the 
final age, whereas in disc galaxies, star formation continues until the present age. As a consequence of this,
at any  age, spectro-photometric models will contain \textit{all} the physical components, i.e.  newly born stars still embedded in their parental molecular cloud, 
stars of various ages and metallicities free from molecular clouds and, finally, a substantial contribution of the diffuse ISM.

\subsection{Disc galaxy: the effect of inclinations}\label{dis_inc}

Disc galaxies can be observed along different inclinations toward the observer. 
In our models four inclinations are considered and obviously, we expect the final SED to be different 
according to the viewing angle, going from  face-on to edge-on galaxies. 
The angles under consideration are : $0^0$ (face-on), $\pi/6$ with respect to the z-axis perpendicular to the equatorial plane, 
$\pi/3$ with respect to the z-axis, and $\pi/2$ (edge-on).

\begin{figure}
\begin{center}
  {\includegraphics[width=0.5\textwidth]{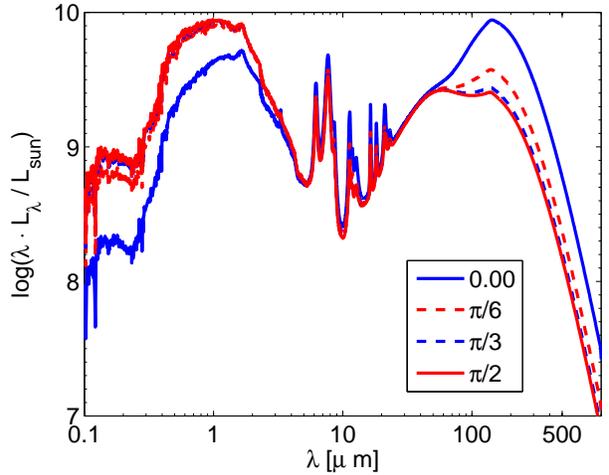} }
  \caption{SED of the model disc galaxies of  $M=10^{11} M_{\odot}$ for different viewing angles.}  \label{inclinazioni}
  \end{center}
\end{figure}

\noindent Differently from what happens for the models of elliptical galaxies because of their nearly spherical symmetry, the luminosity emitted from an edge-on spiral 
galaxy will be heavily absorbed by  the equatorial dust lane between the stars and the observer, and will present a pronounced peak in the IR region of the spectrum. 
The same galaxy seen face-on will show a less intense FIR emission and a more intense emission in the UV/optical region
compared to the edge-on model. Fig. \ref{inclinazioni} shows the total emission of the model disc galaxy of $10^{11} M_{\odot}$ according to different inclinations. 
The SEDs for this galaxy show, as expected, the opposite trend of the IR emission and extinction in the UV-optical region. 
For $\lambda \geq 150$ $\umu$m, the
edge-on emission of the dust is greater than the emission at all the other inclinations. The opposite in the UV-optical region: as expected, for the edge-on galaxy, 
the emission is lower than for the other inclinations. It is clear that every time that we consider SEDs, colours or magnitudes of dust-rich galaxies that are not 
spherically symmetric, thus introducing the dependence on the viewing angle, the results are significantly different depending on the angle. For the same model they 
span a range of possible SEDs and magnitudes.

\subsection{Evolutionary models of different ages}\label{Evol_models}

\begin{figure}
\begin{center}
\subfigure{\includegraphics[width=0.5\textwidth]{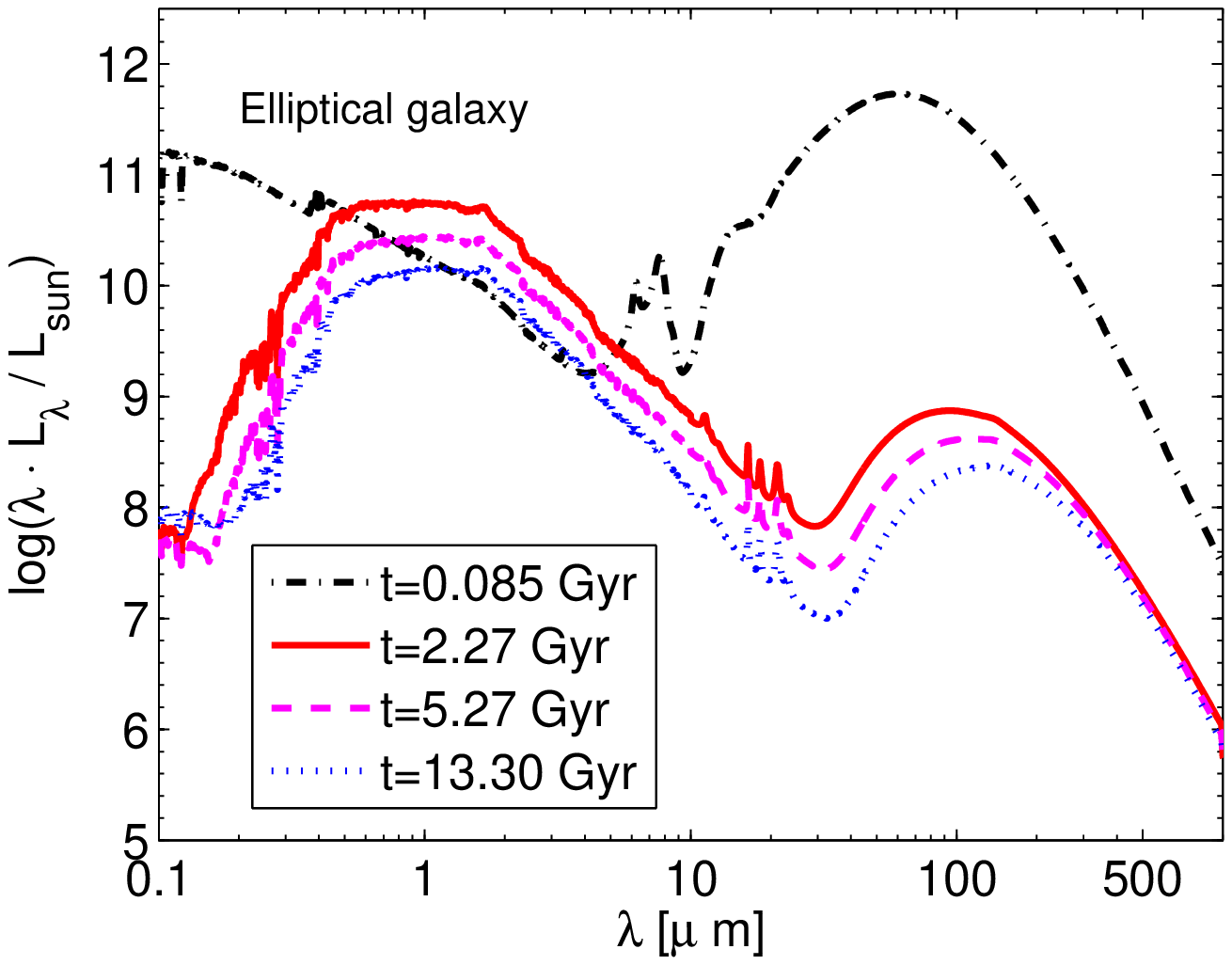}}
\subfigure{\includegraphics[width=0.5\textwidth]{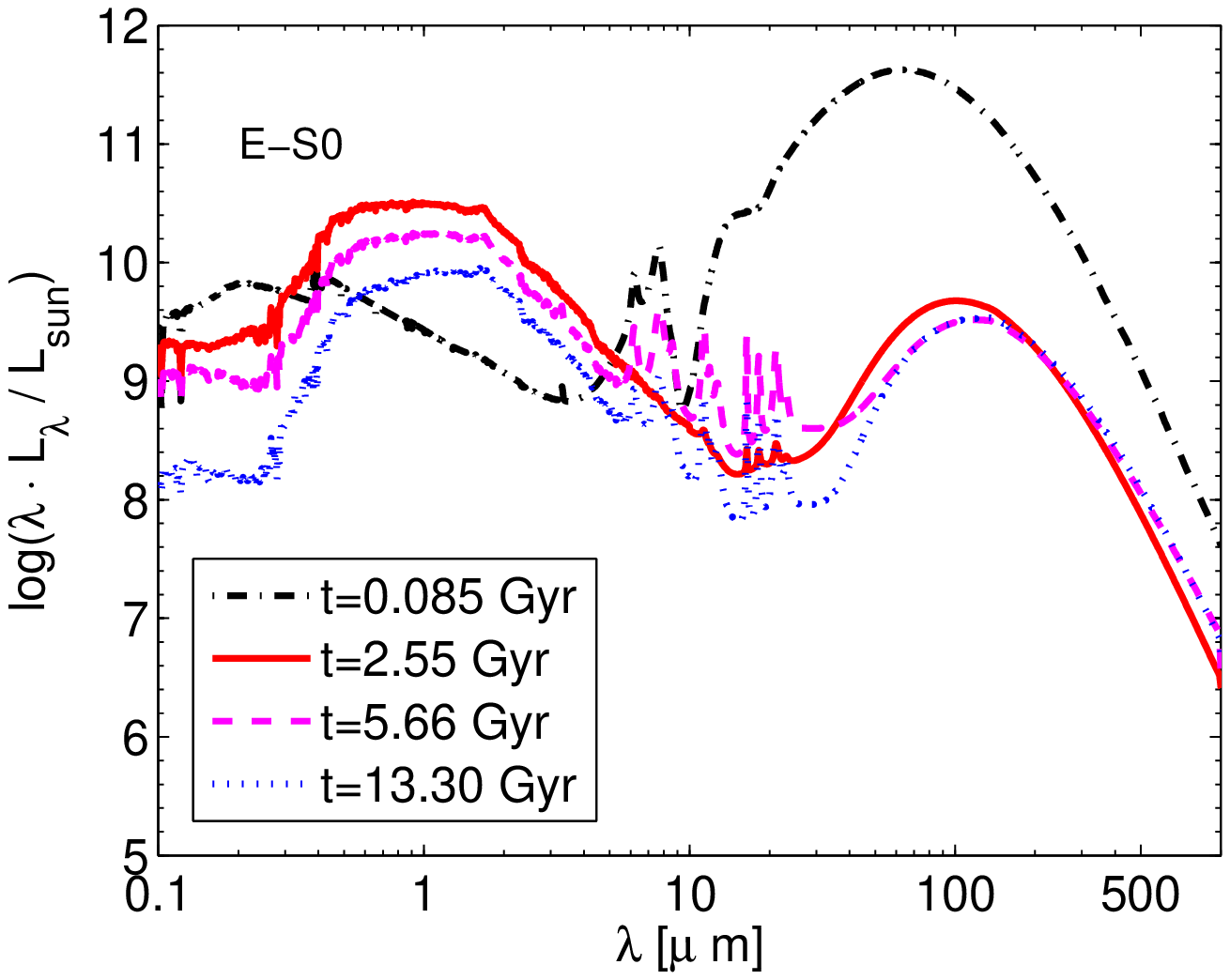} }
\subfigure{\includegraphics[width=0.5\textwidth]{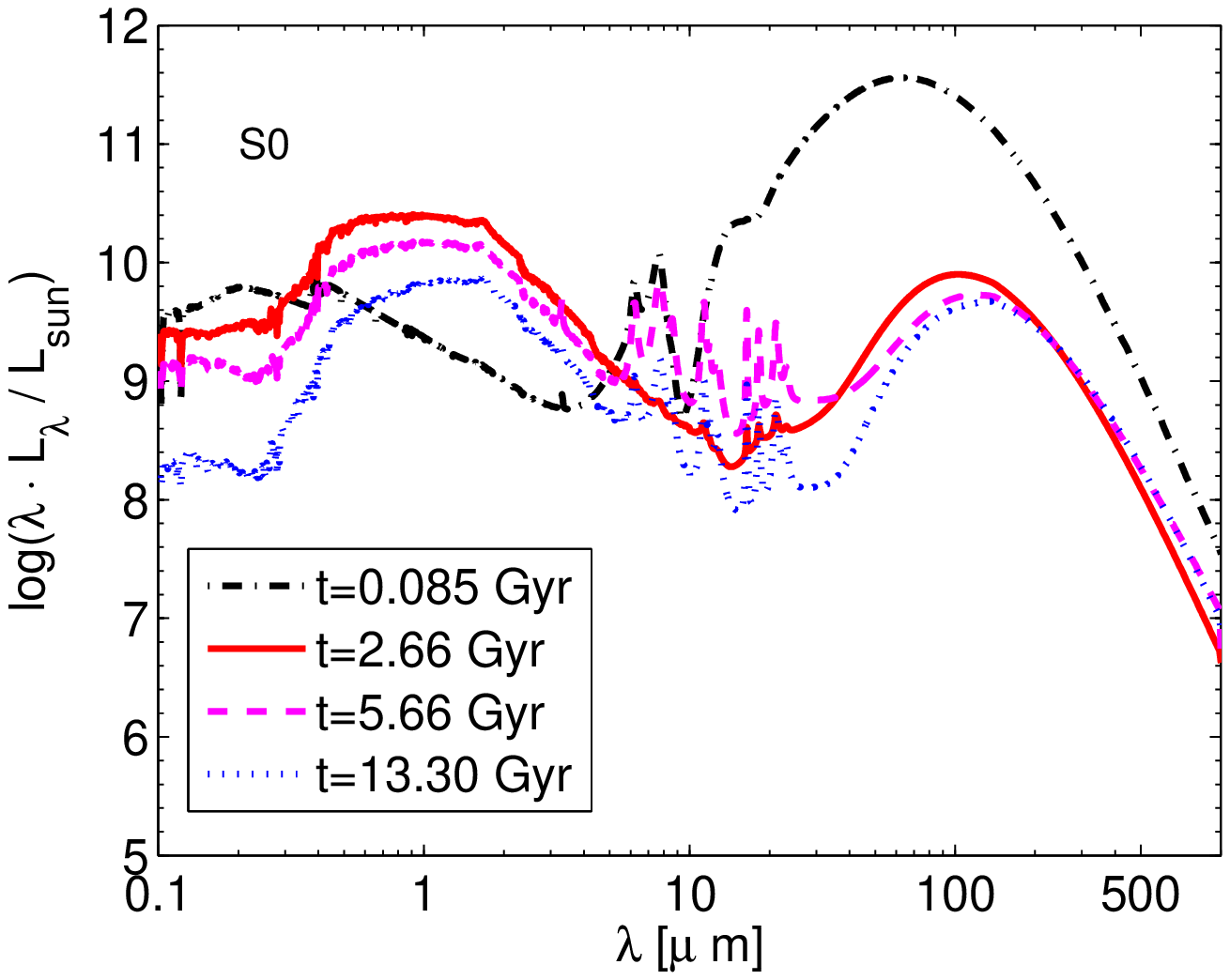} }
  \caption{Time evolution of the SED of modelled galaxies of early morphological types - 
	upper panel: elliptical; middle panel: E-S0; lower panel: 
	S0 - of $M=10^{11} M_{\odot}$ for four significative ages, as the legend indicates.}
  \label{evol}
  \end{center}
  \end{figure}

\begin{figure}
\begin{center}
\subfigure{\includegraphics[width=0.5\textwidth]{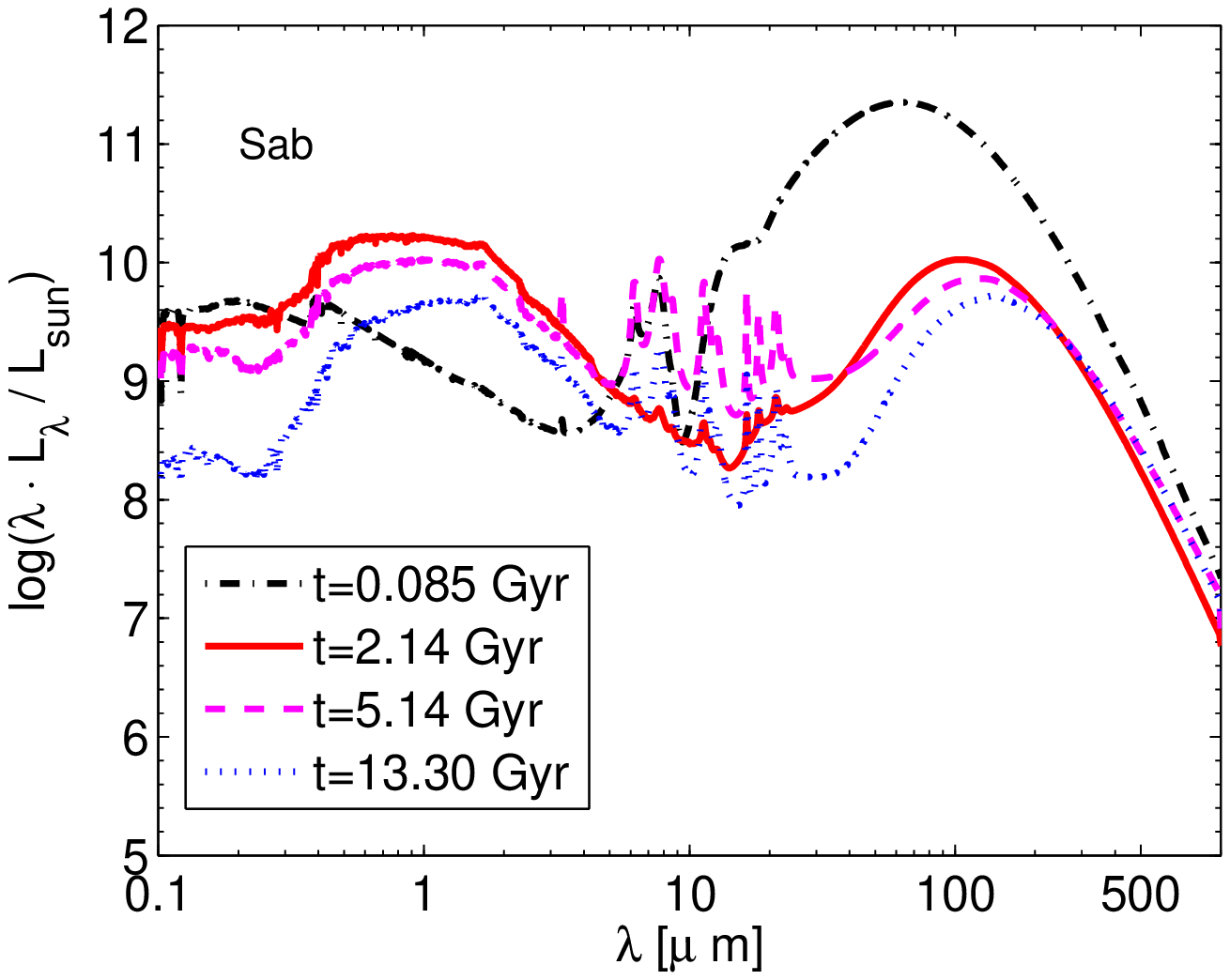}}
\subfigure{\includegraphics[width=0.5\textwidth]{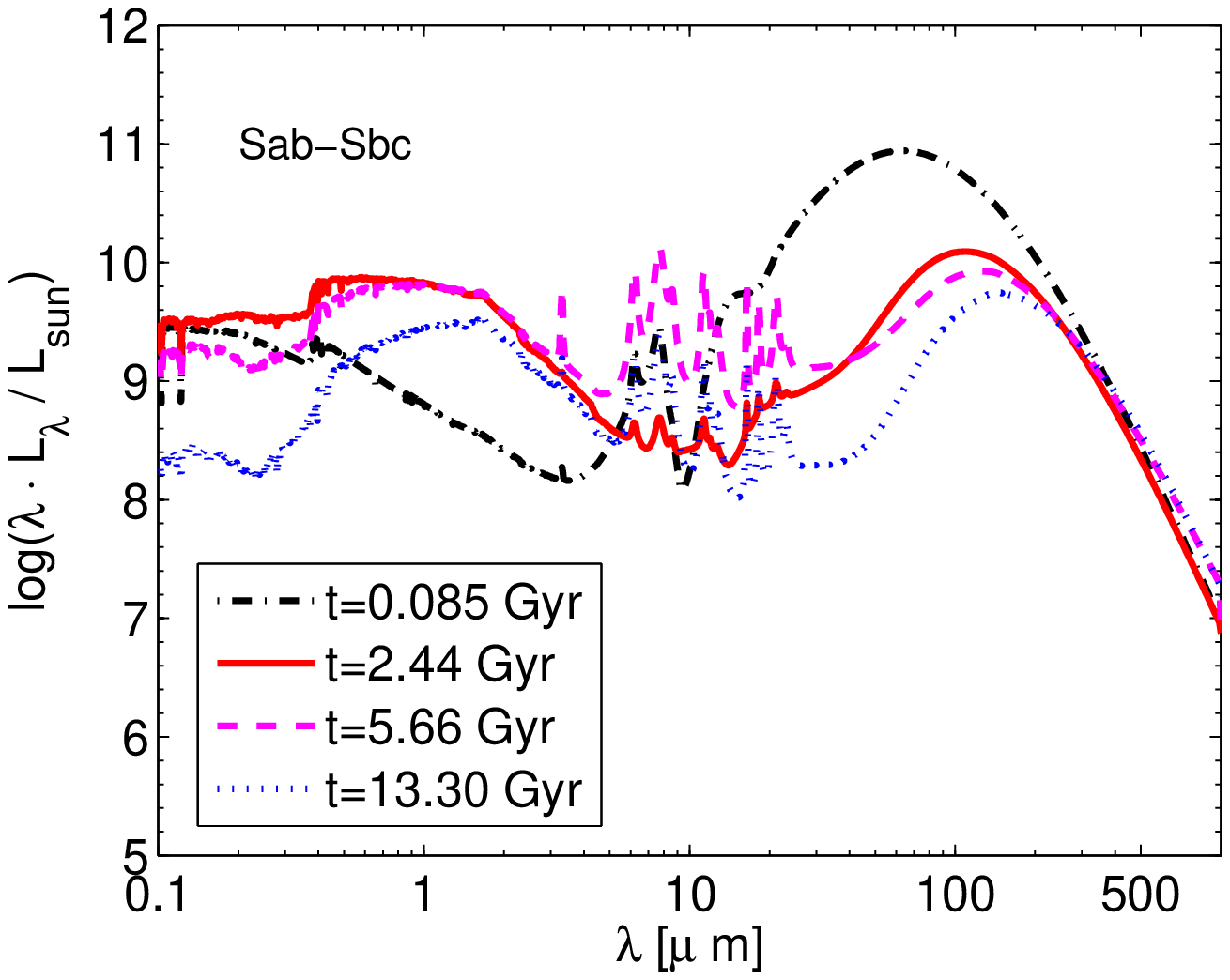}}
\subfigure{\includegraphics[width=0.5\textwidth]{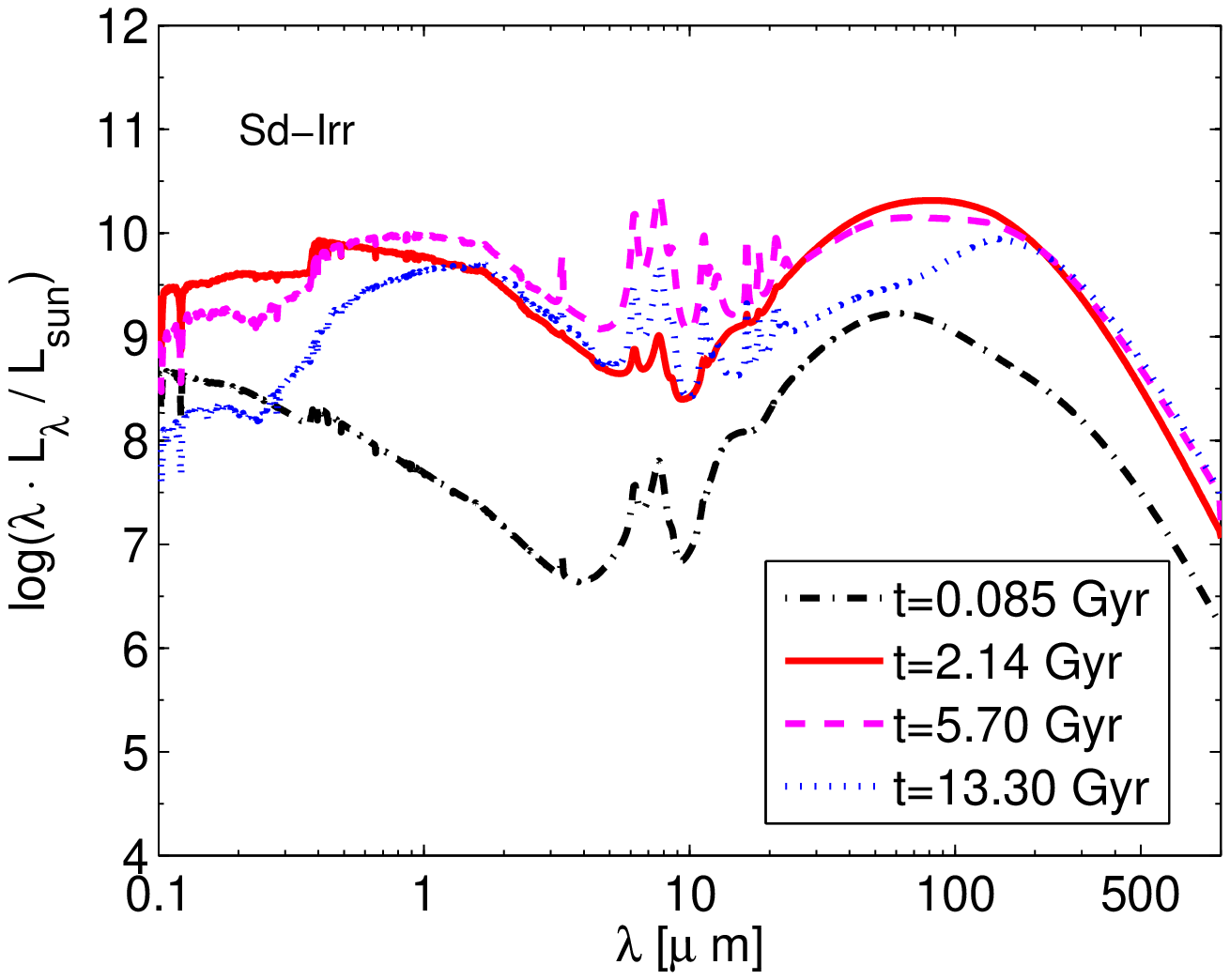}}
  \caption{Time evolution of the SED of modelled galaxies of late morphological types - upper panel: Sab;
  middle panel: Sab-Sbc; lower panel:  Sd-Irr (disc), of $M=10^{11} M_{\odot}$ for four significative ages, as the legend indicates.}
  \label{evol2}
  \end{center}
  \end{figure}

While in Sect. \ref{monoeta} we analyzed the SEDs of galaxies at the age of 13.30 Gyr (present day age), now we examine  how the SEDs of the 
same models vary along their  \textit{evolutionary history}. The situation is illustrated  in the various panels of  Fig. \ref{evol} (upper panel: elliptical galaxy;
middle panel: E-S0 galaxy; lower panel: S0 galaxy)  and Fig. \ref{evol2} (upper panel: Sab galaxy; middle panel: Sab-Sbc galaxy; lower panel: disc galaxy (Sd)) 
which show how the total EPS emission of the models changes at varying their age. Grouping of the models follows the morphological type, 
and all galaxies are supposed to start their SFH at redshift $z_{for}=20$.

\noindent Looking at the SEDs displayed in various panels we  can make the following remarks: for the pure elliptical model at the age of t=0.085 Gyr, 
the emission is strongly concentrated in the IR region of the spectrum: the large amount of dust present in the galaxy at this age absorbs 
the radiation emitted by the young stars  the UV/optical region, and it re-emits it in the IR. During the early stage,  star formation occurs in medium  
highly obscured by dust and the young stars play the dominant role in the total SED. Immediately after the onset of the galactic wind
(which is supposed to occur \textit{simultaneously and instantaneously} for the \textit{entire} gas content of the galaxy), the gas is swept away and  star formation is halted. 
One can assume that the star formation is virtually complete when $t=t_{gw}$.
We can therefore explain the SEDs of the elliptical galaxy for the ages $2.55$, $5.66$ and $13.09$ Gyr: they represent the aging of a stellar population becoming 
older and older with a small diffuse gas and dust content. The diffuse medium absorbs the stellar radiation in small amount, while 
the majority of the emission is due to cool stars in the NIR region.

For the S0 models, the presence of a small disc component allows the presence of an ever continuing star formation. It follows that, even if for t=0.085 Gyr the SED 
is quite similar to the elliptical galaxy (dust dominated emission), for older ages the SED is very different and the diffuse ISM significantly
contributes to the MIR/FIR emission. It is also interesting to note that the PAHs features appear only after a significant enrichment in metals: this is due to the 
choice of the \citet{Weingartner2001a} extinction curves. At low metallicity we adopt their SMC flat curve with a poor or negligible contribution of the PAHs. 
Finally, as expected, the UV emission is much stronger than in models for  elliptical galaxies; this is simply due to the small disc-like  component, in which 
 star formation never stops. The disc component could be replaced by a bulge-like component with stellar ages spanning much broader interval than in the case a
 pure spheroidal galaxy.

In the Sab, Sab-Sbc and Sd (disc) models, the disc mass,  the amounts of dust and their effects in turn grow with the morphological type. The total 
emission increases with time, reaches a maximum in correspondence of the peak of the star formation (both in the optical region and in the IR) and 
then decreases with the decrease of the star formation rate, according to the typical SFH adopted for discs \citep{Piovan2006b,Cassara2008}. 
As already pointed out, the star formation rate does not fall sharply as in the case of elliptical galaxies: it reaches a peak and then slowly declines, 
however continuing  up to  the present age. It is worth noticing that for t=0.085 Gyr, the SED is \textit{clearly} splits  in two peaks (FIR and UV); 
at increasing the age, the trend is smoothed out as intermediate-age and old stars contribute significantly to the 1 $\umu$m emission. 
For the late type models, this is   even more evident than for the S0 one, the PAHs features in the the SED correlate with with the metallicity: 
the low Z extinction  introduced in our models and based on the SMC extinction curve, produce young (high-z) galaxies with rather weak PAHs features.

Finally, for the full disc model, there is no phase during the early evolutionary stages  where the galaxy SED is 
dominated by the FIR emission. As a matter of fact, we miss in this case the strong and heavily obscured burst of star formation in the bulge. 
Along the whole evolution a more regular process of star formation unrolls.

\section{Theoretical and observational colours of galaxies}\label{colours}

In this section we examine the theoretical colours obtained from SEDs of galaxies of different morphological types and compare them with some observational 
data available in literature. \citet{Buzzoni2005} presents a set of EPS  models for template galaxies along the Hubble morphological sequence. 
These models account for the individual evolution of bulge, disc and halo and provide basic morphological features, 
along with bolometric luminosity and colour evolution, between 1 and 15 Gyr. 
The integrated colours and the morphological type are tightly related: this is due to the relative contribution of stellar populations in the bulge and 
disc \citep{Arimoto1991}. 

The  \citet{Buzzoni2005} models  deal with the stellar component, which is obviously the dominant contributor to the galaxy luminosity in the UV-optical region. 
The ISM gas has more selective effects on the SED by enhancing monochromatic emission, e.g. the Balmer lines. 
As far as galaxy broad-band colours are concerned, at the present age, the gas influence is negligible. 
Internal dust could play the dominant role, especially at short wavelength ($\lambda \lesssim 3000$~\AA).  
Metallicity and stellar birth rate are constrained by  comparing  theoretical results with observational data. 
For all other details see \citet{Buzzoni2005}.

\indent Our models  differ from those of \citet{Buzzoni2005} in several aspects among which we recall: first of all, 
they do not consider the contribution of the halo. However, this should  have a marginal effect, because the halo plays a 
secondary role in the mass and luminosity budget. More relevant here, our models consider the contribution of dust. 
In any case, we also consider the case of a dust-free ISM and hence dust-free galaxy emission (i.e. SEDs due only to stars) 
in order to compare the new dusty SEDs with those with no dust \citep{Bressan1994,Buzzoni2002,Buzzoni2005}.
Furthermore, instead of assuming a simple prescription for the star formation law as in  \citet{Buzzoni2005},
we follow the history of a galaxy with the aid of a complex model that put together suitable prescriptions for gas infall, 
the star formation rate,  initial mass function,  and stellar ejecta, all of which  determine  the total amounts of gas and 
stars present at any age together with the chemical history \citep{Chiosi1980,Tantalo1996,Tantalo1998,Portinari1998,Portinari1999,Portinari2000,Piovan2006b}.
For the star formation rate we use the classical Schmidt law, whereas  \citet{Buzzoni2005} adopts the  power law $SFR =Kt^{-\eta}$ with $\eta<1$.

\begin{figure}
\begin{center}
\subfigure{\includegraphics[width=0.5\textwidth]{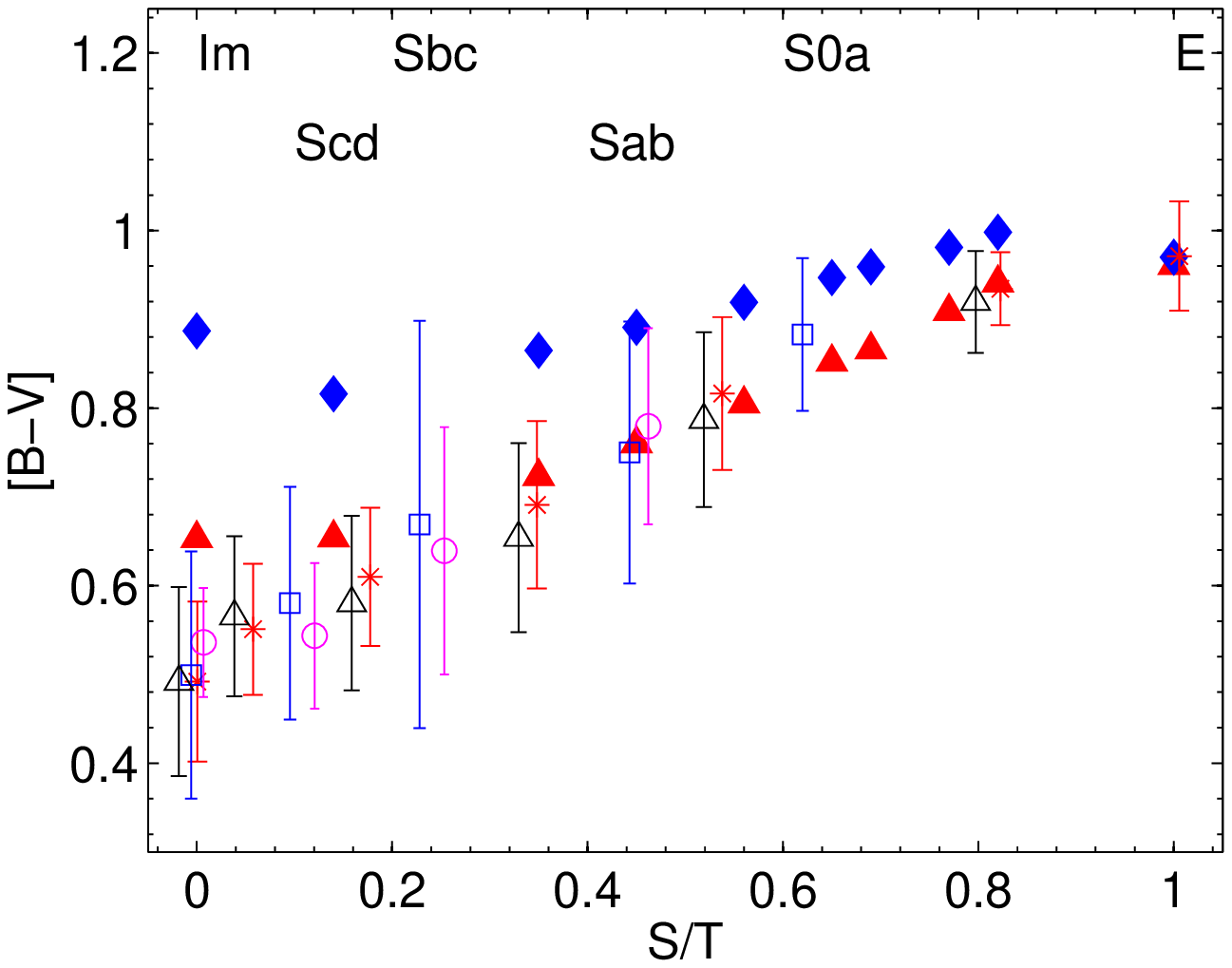}}
\subfigure{\includegraphics[width=0.5\textwidth]{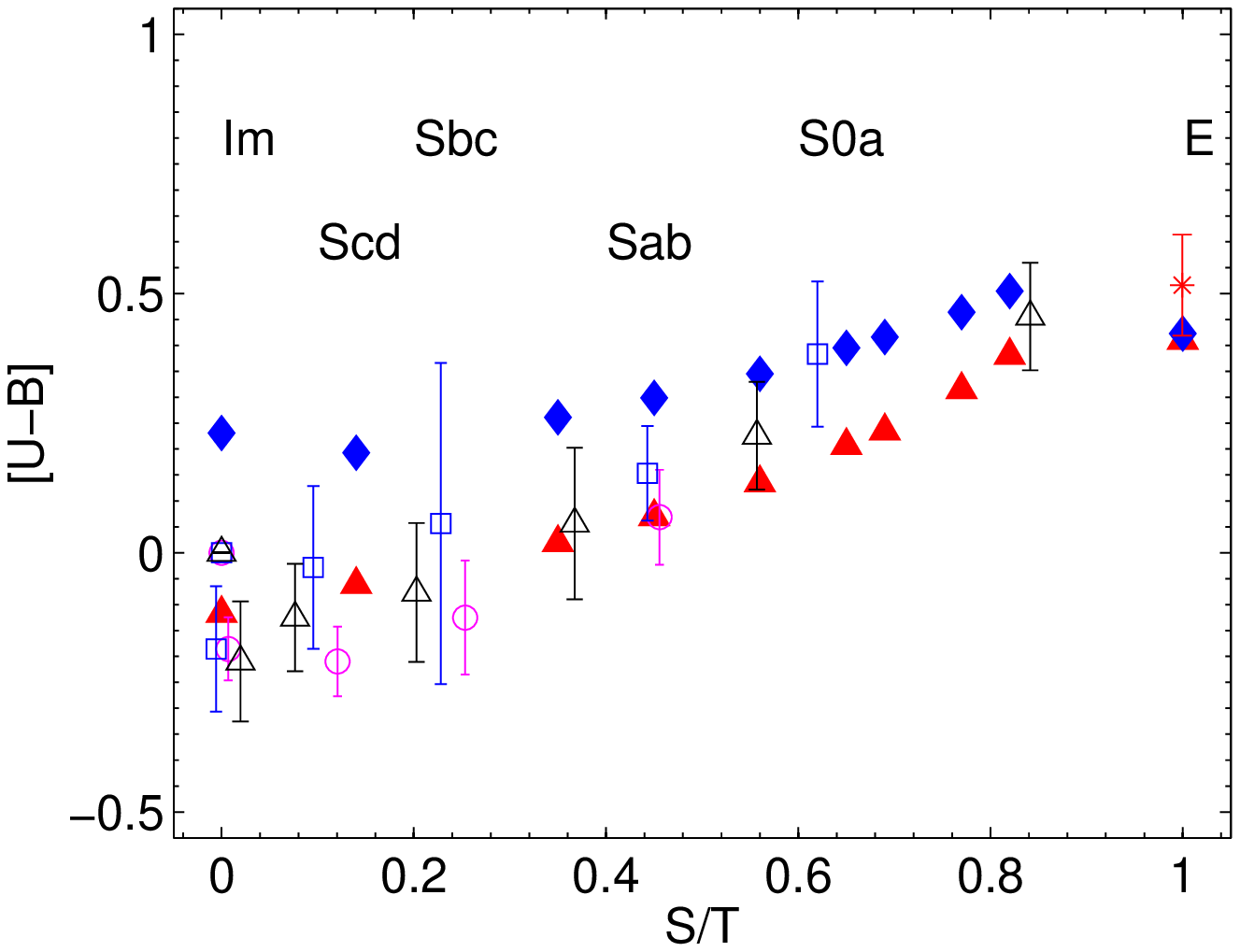}}
\caption{Galaxy colour distribution (upper panel: $[\textrm{B}-\textrm{V}]$  vs. bolometric morphological parameter
S/T=$L_{Bulge}/L_{Tot}$; lower panel: $[\textrm{U}-\textrm{B}]$  vs. bolometric morphological parameter S/T=$L_{Bulge}/L_{Tot}$). 
Data are from \protect \citet{Pence1976} (magenta circles), \protect \citet{Gavazzi1991} (blue squares), \protect \citet{Roberts1994} (red stars) and \protect \citet{Buta1994} 
(black triangles). All the data have been properly
corrected by dust extinction. The red star located at $\sim$ S/T= 1 represents the mean colour for EGs \protect
\citep{Buzzoni1995}. Our theoretical colours are indicated with blue diamonds and red triangles: they have been calculated, respectively, by means of the
 EPS with dust and EPS corrected for the contribution of dust.}
 \label{BUZZ1}
\end{center}
\end{figure}

\indent As already mentioned,  the chemical parameters for the galaxy models are chosen in such a way that the observational values for the ratio  $L_{Bulge}/L_{Tot}$ are 
reproduced (see below for more details). In contrast, for each Hubble type,  \citet{Buzzoni2005} calibrated the morphological parameter  S/T~$= L{\rm (spheroid)}/L{\rm (tot)}$. 
As the S/T calibration does not vary in the infrared range, he choose the $I$ luminosity as a reference for the model setup.
In our simulations, the S/T  ratio can not be determined \textit{a priori}; indeed, we start from the SED of a certain model galaxy, of 
which we know  \textit{in advance} the asymptotic infall mass. The  SED is fed to the photometric code, which calculates colours and magnitudes in different photometric systems.  
In our case  the disc and bulge luminosities are used \textit{a posteriori} to get clues about the asymptotic mass of the galaxy that should be used as input. 
Carefully tuning this procedure, one may eventually obtain the correct initial values for the disc and bulge mass  consistent with the S/T ratios for the different 
morphological types.
Assumed the metallicity (and hence $\zeta$, $\tau$, and $\nu$), there is an \textit{almost-linear} relation between the bulge mass 
 (the mass of the disc follows from $M_{disc,\odot}=10^{11}-M_{bulge,\odot}$ because all the intermediate types have the same total mass of $10^{11}M_{\odot}$) and the luminosity.

\begin{figure}
\begin{center}
\subfigure{\includegraphics[width=0.5\textwidth]{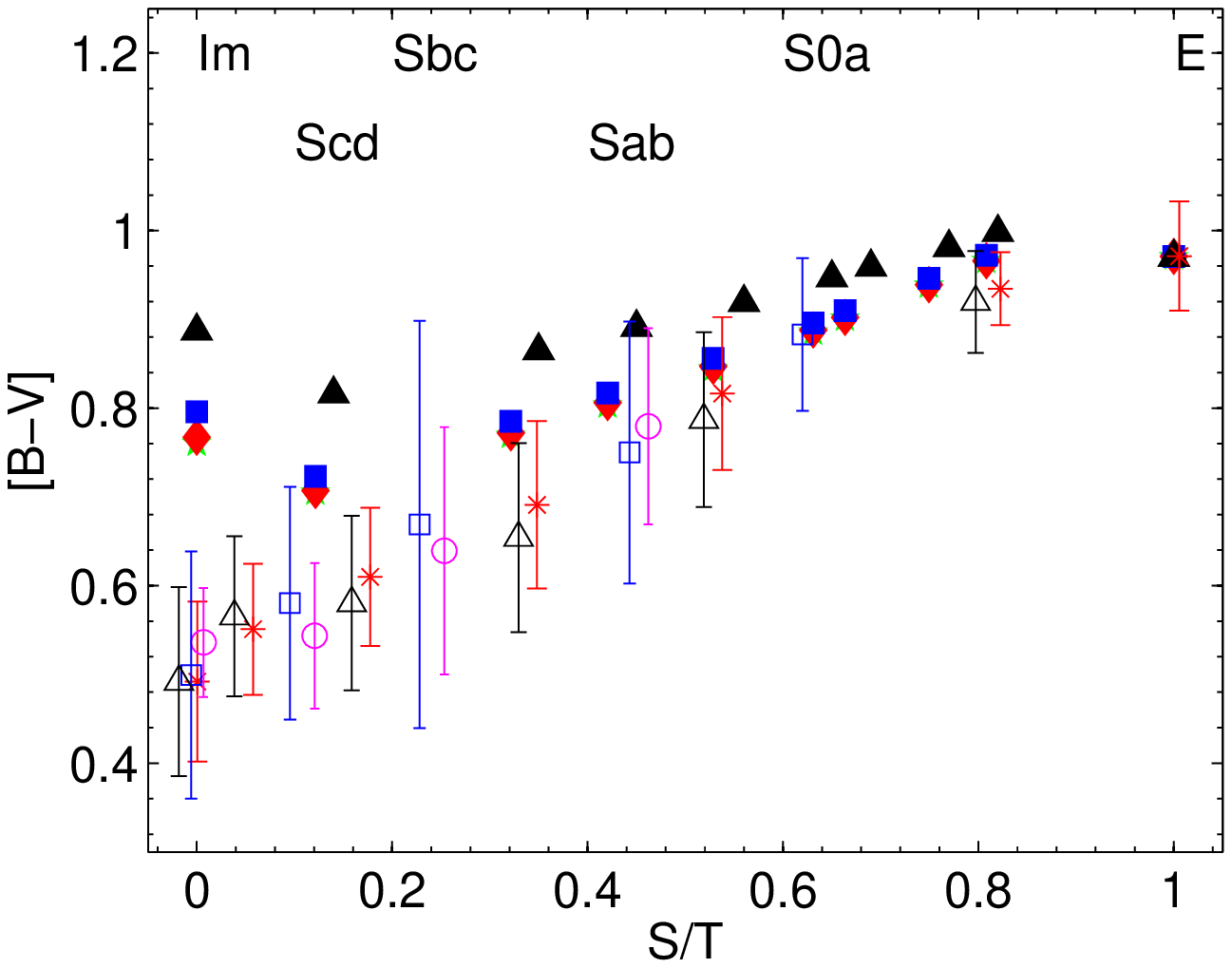}}
\subfigure{\includegraphics[width=0.5\textwidth]{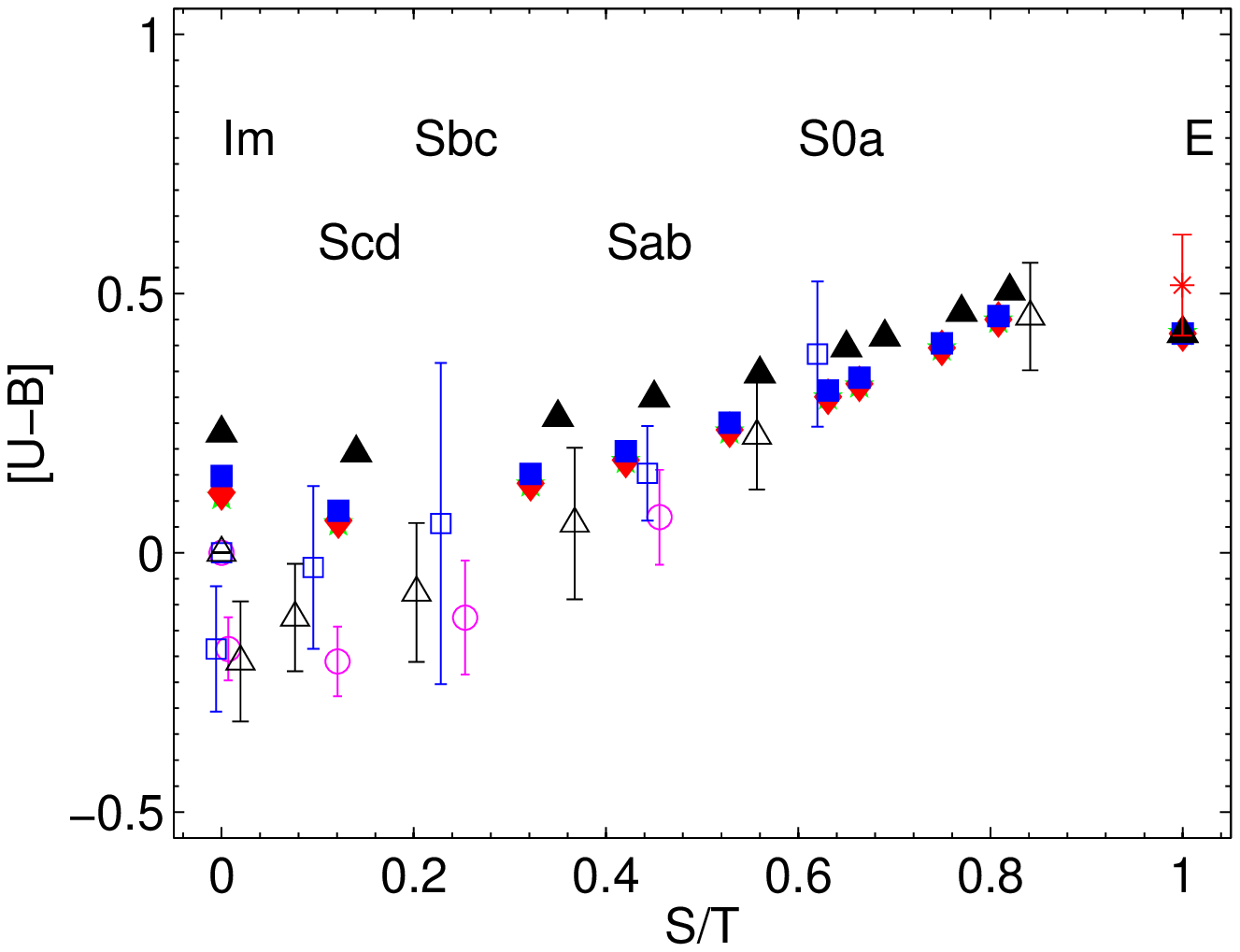}}
\caption{The same as in Fig. \protect \ref{BUZZ1}, but now only theoretical colours for EPS with dust are shown: the
different symbols and colours indicate different viewing angles. Black triangles: galaxy seen \textit{edge-on}; blue squares: 
galaxy observed at an angle of 60$^{\circ}$ measured respect to the galactic equatorial plane; red diamonds: galaxy observed at an angle of 30$^{\circ}$ 
measured respect to the galactic equatorial plane; green stars: galaxy seen \textit{face-on}.}
\label{BUZZinclinaz}
\end{center}
\end{figure}

\begin{figure}
\begin{center}
\subfigure{\includegraphics[width=0.5\textwidth]{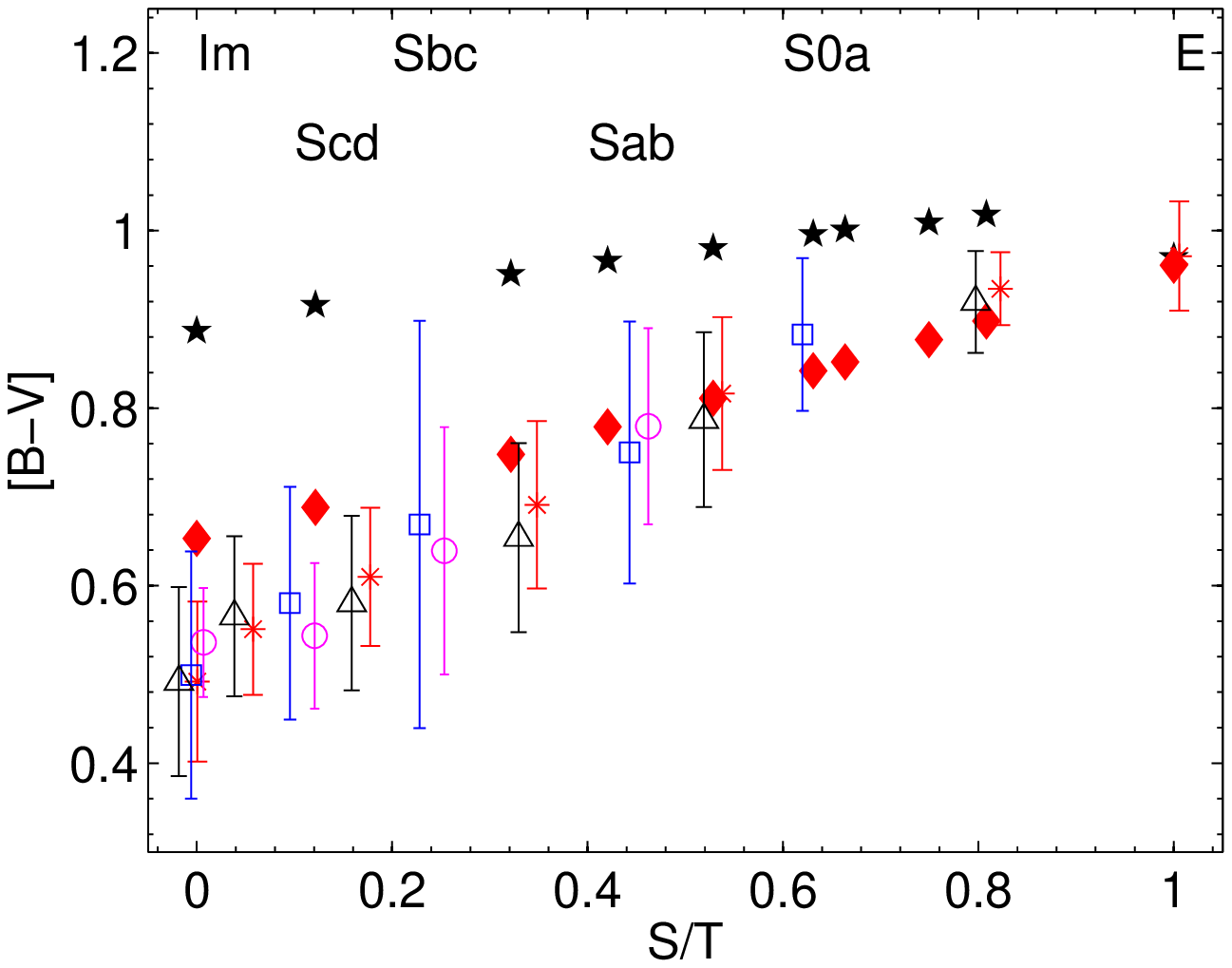}}
\subfigure{\includegraphics[width=0.5\textwidth]{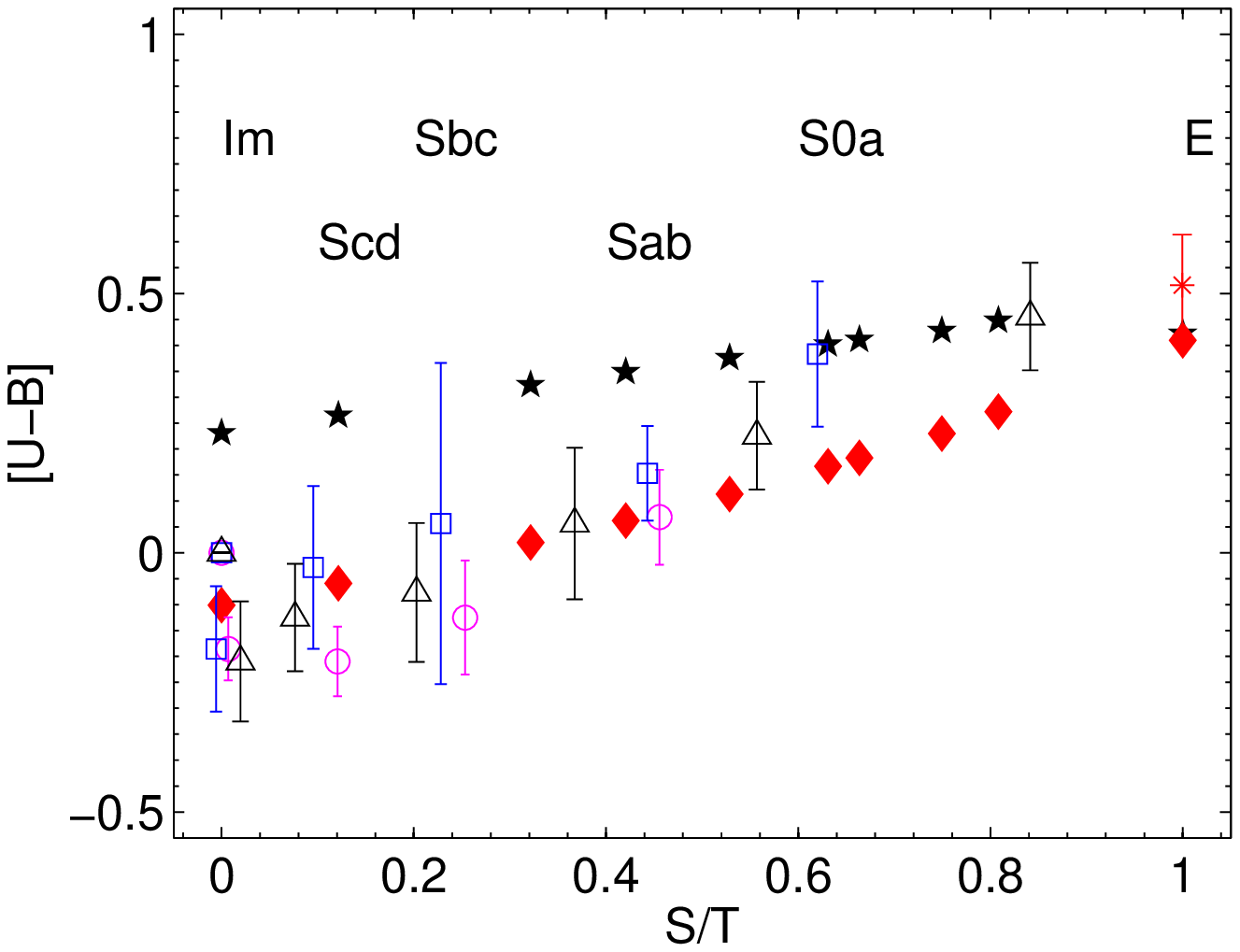}}
\caption{As in Fig. \protect \ref{BUZZ1}, but here the metallicity of the models is \textit{not fixed} as in Fig. \protect \ref{BUZZ1} 
(see the text for more details about this point).}
\label{BUZZ_Z}
\end{center}
\end{figure}

As explained in \citet{Buzzoni2005}, observations of the central bulge of the Milky Way show that it mainly contains  
metal-rich stars \citep{Frogel1988,Frogel1999} and this seems to be  quite common also among  external galaxies
\citep{Jablonka1996,Goudfrooij1999,Davidge2001}.  
The exact average value of the bulge metallicity, however, has been subject to continuously revised, 
going from  [Fe/H]~$\sim +0.2$ \citep{Whitford1983,Rich1990,Geisler1992}) to lower values, 
consistent with a standard  or even slightly sub-solar metallicity \citep{Tiede1995,Sadler1996,Zoccali2003,Origlia2005}. 
According to these arguments, for the bulge of his intermediate type galaxies \citet{Buzzoni2005} adopts  SSPs with [Fe/H]$ = +0.22$ 
in agreement  with the observations of external galaxies. 
For the disc component, relying on the observed colours of present-day galaxies, \citet{Buzzoni2005} adopts [Fe/H]$_{\rm disc} = -0.5$ dex as a 
luminosity-weighted representative value for his models. 

\indent In order to account for the prescription of \citet{Buzzoni2005}, two different sets of galactic models are calculated; 
their parameters are listed and discussed in Sect. \ref{model_param}. The only difference between the sets concerns the final metallicity.
Our chemical code does not allow (unless we force the input parameters to extreme values) to reach a super-solar metallicity for the bulge, 
in particular at the lowest galaxy masses. For realistic values of the parameters, our bulges reach solar or  slightly super-solar metallicities. 
Also for the disc, we can not easily reach the low value suggested by \citet{Buzzoni2005}: the metallicity of our discs tends to be slightly higher. The
fundamental parameter to vary  is $\zeta$. Metallicities slightly super-solar  (for the bulge) and slightly lower than the average LMC value 
(for the disc) are the best values that can be obtained for plausible value of $\zeta$ in the Salpeter IMF in model galaxies with 
total asymptotic mass of $10^{11}\,M_\odot$. We keep these values in order to maintain the general property that in any case the bulge $(Z\gtrsim 0.02)$ 
is more metal-rich than the disc $(Z\lesssim 0.008)$. In order to evaluate the effect due to different values 
of the metallicity for the two galaxy components, disc and bulge, we  calculate two different set of models:

\begin{itemize}
\item[-] The first set stands on the parameters already discussed in Sect. \ref{model_param}. 
In this case, the galactic code, performing the EPS, will interpolate between SEDs of SSPs taking for each of 
them the metallicity predicted by the chemical code at the time when that stellar population was born.

\item[-] The second set stands on the same parameters, but in this case we force the galactic code to 
generate  the same metallicities  adopted by \citet{Buzzoni2005} for the disc and bulge. In this case there is 
no interpolation on the SEDs of SSPs in metallicity, because it is fixed for both disc and bulge.
\end{itemize}

\indent The results of our simulations, together with the observed colours, are shown in the Figs. \ref{BUZZ1}, \ref{BUZZinclinaz} and \ref{BUZZ_Z}. 
They represent galaxy colour distribution, that is:  $[\textrm{B}-\textrm{V}]$  vs. bolometric morphological parameter S/T=$L_{Bulge}/L_{Tot}$ (upper panel) 
and $[\textrm{U}-\textrm{B}]$ vs. bolometric morphological parameter S/T=$L_{Bulge}/L_{Tot}$ (lower panel). 
Data are taken from \citet{Pence1976} (magenta circles), \citet{Gavazzi1991} (blue
squares), \citet{Roberts1994} (red stars) and  \citet{Buta1994} (black triangles). The red star located at $\sim$ S/T= 1 is the mean colour 
for elliptical galaxies \citep{Buzzoni1995}. All  galaxies have been corrected for reddening by the respective authors. In Fig. \ref{BUZZ1}, our 
theoretical colours are indicated with blue diamonds and red
triangles: they have been calculated, respectively, considering EPS with dust and EPS corrected for the extinction of dust, namely the classical bare EPS. 
These colours are obtained by fixing the metallicities of the disc and bulge. In Fig. \ref{BUZZ_Z}, our theoretical colours are  indicated with black 
stars and red diamonds: once more, they have been calculated considering EPS with dust and  EPS with no dust. For the models of Fig. \ref{BUZZ1} we only fix
 the chemical parameters (see Sect. \ref{model_param}), thus leaving the spectro-photometric code free to interpolate in metallicity, according to the input 
values provided by  the chemical simulations.  For the models of Fig. \ref{BUZZ_Z}, we  forced the
galactic code to adopt the metallicities of the disc and bulge according to  Buzzoni (2005). In this case there is no interpolation on the SEDs
 of SSPs in metallicity.
Finally, in Fig. \ref{BUZZinclinaz}, only theoretical colours generated by  SEDs of EPS  \textit{with dust}
are plotted, but taking into account the effect of the viewing angle.

\noindent The agreement of our simulations with the data is good and the following considerations can be made:
\begin{itemize}
\item[-] As expected, in all the figures, the theoretical colours  best fitting the extinction corrected data  are the 
dust-free ones: between the classical EPS and the EPS with dust there is a difference of $\sim$ 0.2 for both colours. 
This difference can be easily explained since the colours in the plots are all in the optical region, thus being all 
absorbed by dust, with more extinction for the bands at shorter wavelengths. A stronger difference would be observed in optical - near IR colours, 
since the near IR radiation is less absorbed by dust.

\item[-] The effect of dust is more evident in the late-type galaxies, richer in gas and dust. 
For the models of ETGs, at the present age (all the models have been calculated from z=z$_{for}$ to z=0, that is t$_{G}$ = 13.30 Gyr) 
only a small amount of dust is still present (see for instance  the  discussion in Sect. \ref{monoeta}). Colours obtained using EPS with 
or without dust are in practice  indistinguishable, while the differences increase going from early-type toward late-type galaxies.

\item[-] The effect of the metallicity is evident \textit{but} not so remarkable: the same trend is followed by the 
theoretical colours, for both cases with fixed and not fixed metallicity, Figs. \ref{BUZZ1} and \ref{BUZZ_Z}.
    This suggests that the our model well reproduces  the observations.

\item[-] The effect of the inclination of the disc  strictly follows the discussion of  Sect. \ref{dis_inc}: 
the absorption due to dust is more spectacular when the galaxies are observed edge-on.
\end{itemize}

\begin{figure*}
\begin{center}
{\includegraphics[width=0.45\textwidth]{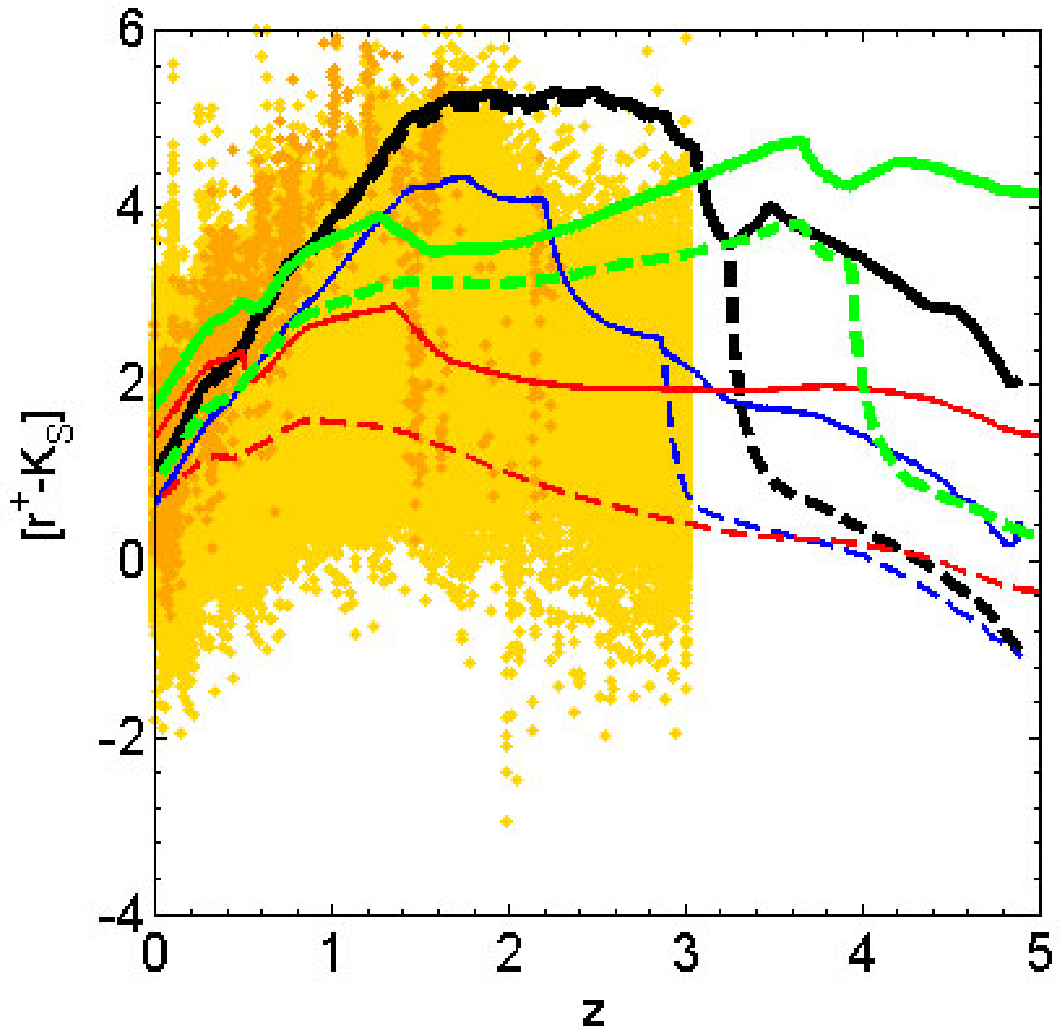}
 \includegraphics[width=0.45\textwidth]{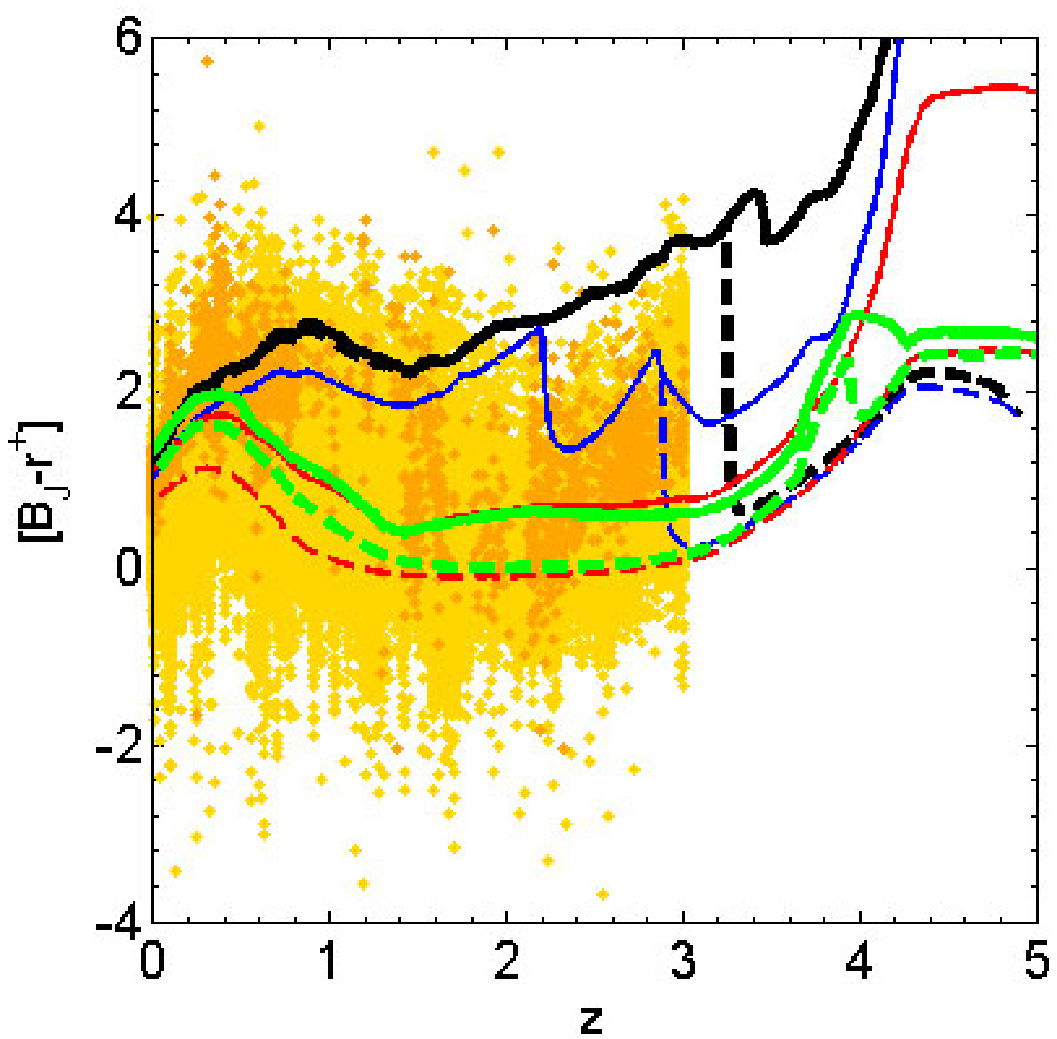} }
\caption{Left Panel: Cosmological evolution with the redshift for the  [\textrm{B$_{J}$}-\textrm{r$^{+}$}] colour of the survey COSMOS (both filters are pass-bands 
of the Subaru telescope). The sample of galaxies represented is taken from the catalogue of galaxies observed in the COSMOS survey and selected 
in \protect \citet{Tantalo2010}. The total sample of galaxies is represented in orange, while the Early Type Galaxies are represented in yellow.
Superimposed, the evolution of the colour [\textrm{B$_{J}$}-\textrm{r$^{+}$}] for three models presented 
in this work or ad-hoc calculated for this redshift evolution, namely: (1) two elliptical galaxies with masses $10^{10}M_{\odot}$ and $10^{12}M_{\odot}$ and with the same choice 
of the input parameters as in Sect. \protect \ref{Evol_models} (black and blue lines); (2) an intermediate type model Sab of $10^{11}M_{\odot}$ (green line) and (3) a 
disc galaxy (Sd) of $10^{11}M_{\odot}$ (red line). 
In all the cases we show the evolution of the colour taking into account our dusty EPS (solid lines) and classical EPS without dust (dotted lines).
Right Panel:
The same as in the left panel but for the colour  [\textrm{K$_{S}$}-\textrm{r$^{+}$}] of the survey COSMOS ($r^{+}$ is a filter of the Subaru telescope, while $K_{S}$ is from the Kitt Peak national Observatory). }
\label{BjmenoRKsmenoR}
\end{center}
\end{figure*}

\section{Cosmological Evolution of Galaxy colours } \label{colour_redshift}

In this section we present the photometric evolution of our model galaxies as a function of the redshift in the $\Lambda$CDM Universe we have 
adopted and we compare the theoretical magnitudes and colours  with some available observational data.\\
For a source observed at redshift $z$, the relation between photons observed at a wavelength $\lambda_0$ 
and emitted at  wavelength $\lambda_e$ is $\lambda_e = \lambda_0 / (1+z)$.
Furthermore, if the source has an apparent magnitude $m$ when observed in a certain photometric pass-band, its   absolute magnitude $M$ in the rest-frame 
pass-band satisfy the relation

\begin{equation}
\label{mR}
m = M + DM + K_{corr},
\end{equation}

\noindent where $DM$ is the distance modulus and $K_{corr}$ is the so-called K-correction. The distance modulus is  defined by

\begin{equation}
DM =5 log_{10} \left( \frac{D_L(z)}{10 pc} \right),
\end{equation}

\noindent where $D_L(z)$ is the luminosity distance and $1pc =3.086 \times 10^{18}  cm$.

\indent The luminosity of a source  at redshift $z$ is related to its spectral 
density flux (energy per unit time per unit area per unit wavelength) by

\begin{equation}
L(\lambda_e) = 4 \pi (1+z) D_L^2 f(\lambda_0),
\end{equation}

\noindent where $f(\lambda_0)$ is the monochromatic flux of the source at the observer. The $K$-correction is defined
as:

\begin{equation}
K_{corr} = 2.5 log_{10} (1+z) + 2.5 log_{10} \left[ \frac{
L(\lambda_0)} { L(\lambda_e) } \right].
\end{equation}

\noindent  In order to compare sources  at different redshifts, we must convert 
the apparent photometric data (magnitudes,  etc.) to  rest-frame quantities by 
applying the K-corrections and also  correct the  rest-frame quantities for 
the expected evolutionary changes during the time interval corresponding to 
the redshift difference, the so-called evolutionary correction $E(z)$.  
The $K$- and $E$-corrections are usually derived from  the theoretical SEDs 
calculated with the stellar EPS technique. Given the above definitions, 
$K(z)$ and  $E(z)$  can be expressed as magnitude differences in the following way:

\begin{eqnarray}
  K(z) &=& M(z,t_0) - M(0,t_0) \nonumber \\
  E(z) &=& M(z,t_z) - M(z,t_0)
\end{eqnarray}

\noindent where $M(0,t_0)$ is the absolute magnitude in a pass-band derived from the \textit{rest frame} spectrum of the source at the current time, $M(z,t_0)$ 
is the absolute magnitude derived from the spectrum of the source \textit{at the current time but redshifted at} $z$, and $M(z,t_z)$ is the absolute magnitude 
obtained from the spectrum of the source \textit{at time} $t_z$ \textit{and redshifted at} $z$. To summarize the absolute magnitude, $M(z)$, in some broadband 
filter and  at redshift $z$ and its apparent magnitude $m(z)$ are expressed  by

\begin{equation}
M(z) = -2.5 \log L(z,t(z)),
\end{equation}

\noindent and, passing to apparent magnitudes,

\begin{equation}
m(z) = M(z) + E(z) +  K(z) + DM(z).
\end{equation}

The relation between the cosmic time $t$ and redshift $z$, $t(z)$, of a stellar population formed at a given initial redshift $z_f$, depends on the adopted 
cosmological model of the Universe (and its parameters).

In the next section we will compare the SEDs of our model galaxies with the luminosities of  galaxies from the  \citet{Takeuchi2010} database kindly provided to 
us by Takeuchi (2012, private communication).  To this aim,  it is  useful to  remind here the procedure  to get the luminosities back from the apparent AB, 
ST and Vega magnitudes. The luminosity in a pass-band satisfies the relations

\begin{eqnarray}\label{Lum_lambda0}
\small
L\left(\nu_0\right)\Delta \nu_0 &=& L\left(\lambda_0\right)\Delta\lambda_0  \nonumber\\
&=&10^{0.4K\left(z\right)}4\pi D_L^{2}\left(z\right)\cdot  \nonumber \\
&&  10^{-0.4\left(m_{AB}+48.60\right)}\Delta \nu_0
\end{eqnarray}

\noindent
where $\Delta \nu_0$ and $\Delta \lambda_0$ are the integrals of the filter over the pass-band.
Similar equations holds for the Vega and ST systems, provided that  the  corresponding photometric  constants are used  and that the monochromatic 
flux for the ST and Vega systems is expressed per  Angstrom. The  \citet{Takeuchi2010} database yields monochromatic fluxes normalized to the pivotal 
wavelength that are corrected for E(z) and K(z), i.e. simply

\begin{equation}
\nu_{e}L\left(\nu_e\right)=4\pi
D_L^{2}\left(z\right)f\left(\nu_{0}\right)\nu_{0} .
\end{equation}

\subsection{Comparison with the observations}\label{redshift_data}

Deep and large scale surveys, from earth and space, allow nowadays to obtain extremely rich samples   of data at different redshifts 
and in different wavelengths, from the UV to the FIR. The main characteristic of these deep photometric surveys detecting a large 
number of galaxies is that a significant fraction of the detected objects appear as point sources. They can neither  be easily 
distinguished from single stars, nor easily classified from a morphological point of view. It follows that the photometric study of 
their properties is crucial, also in order to produce some morphological classification.\\ 
In this paper,  we take into account the Cosmic Evolution Survey – COSMOS official photometric redshift catalogue \citep{Scoville2007}, 
designed to probe the evolution of galaxies in the context of their large scale structure out to moderate redshift \citep[see also][]{Capak2007,Mobasher2007}. 
\citet{Tantalo2010} selected an extended sample of ETGs in the COSMOS catalogue: the 
morphological selection is made  with an automatic pipeline able to separate the objects by means of their bi-dimensional distribution of the light.

\begin{figure*}
\centerline{
\includegraphics[height=5.2cm,width=5.2truecm]{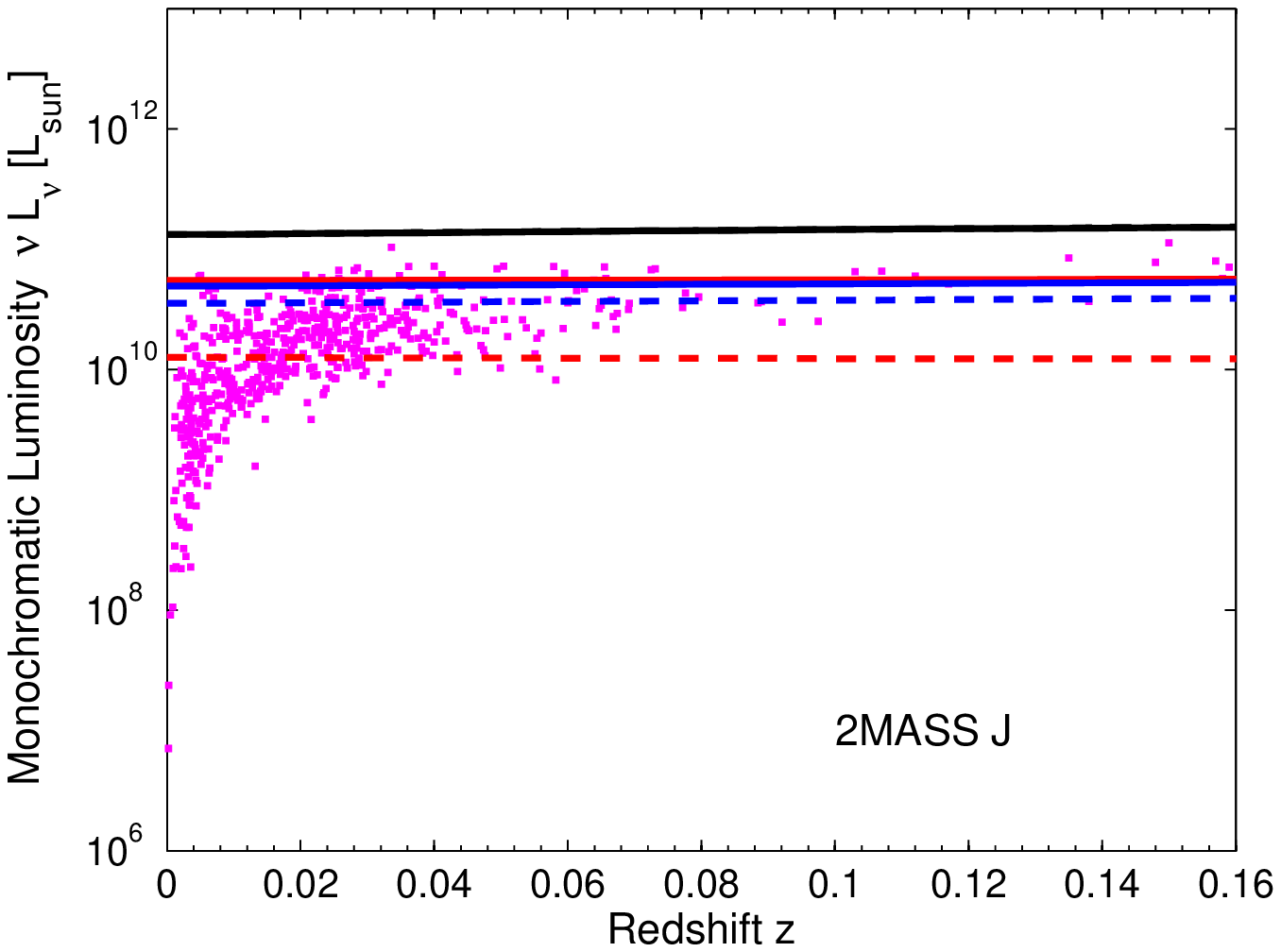}
\includegraphics[height=5.2cm,width=5.2truecm]{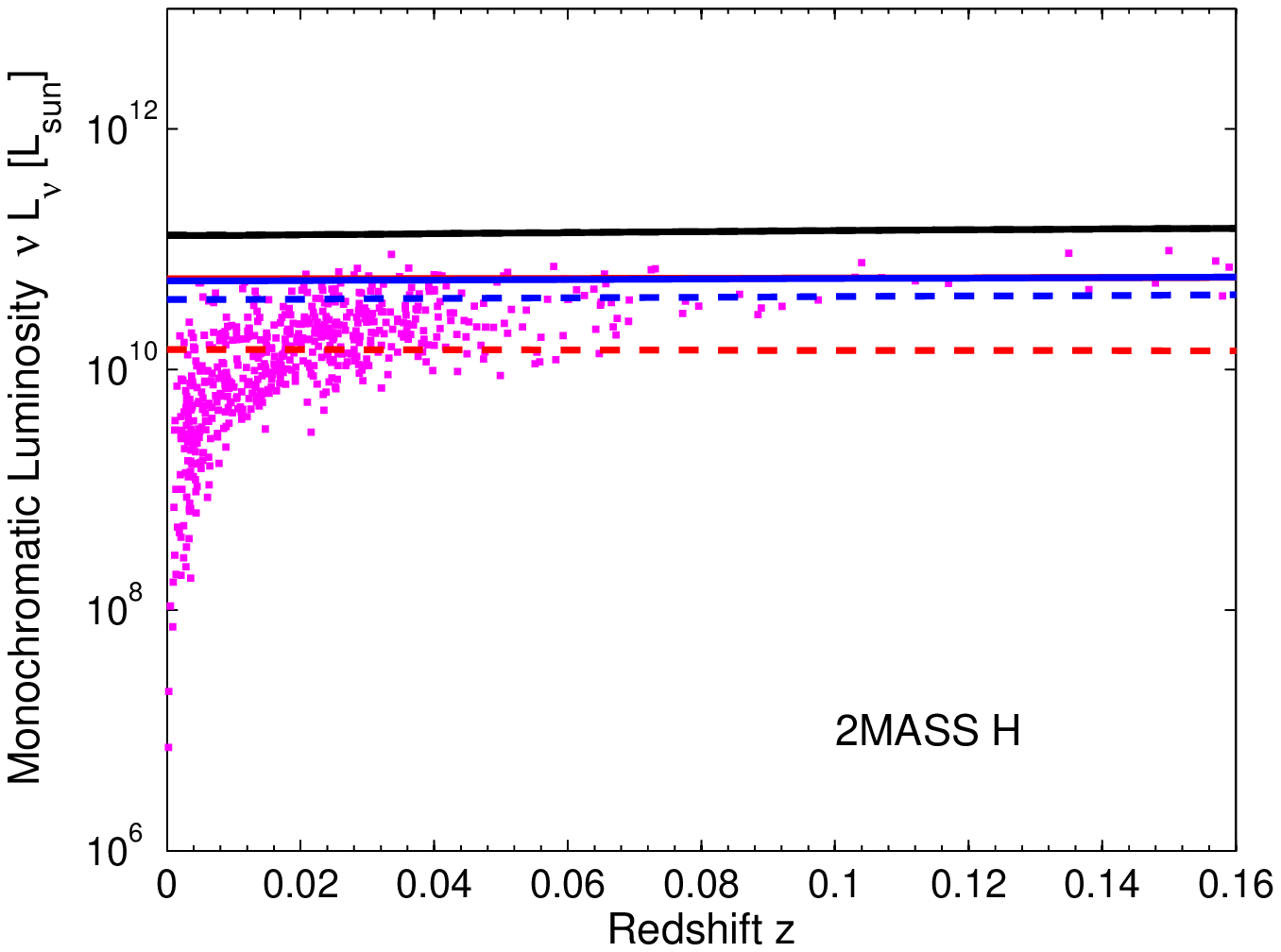}
\includegraphics[height=5.2cm,width=5.2truecm]{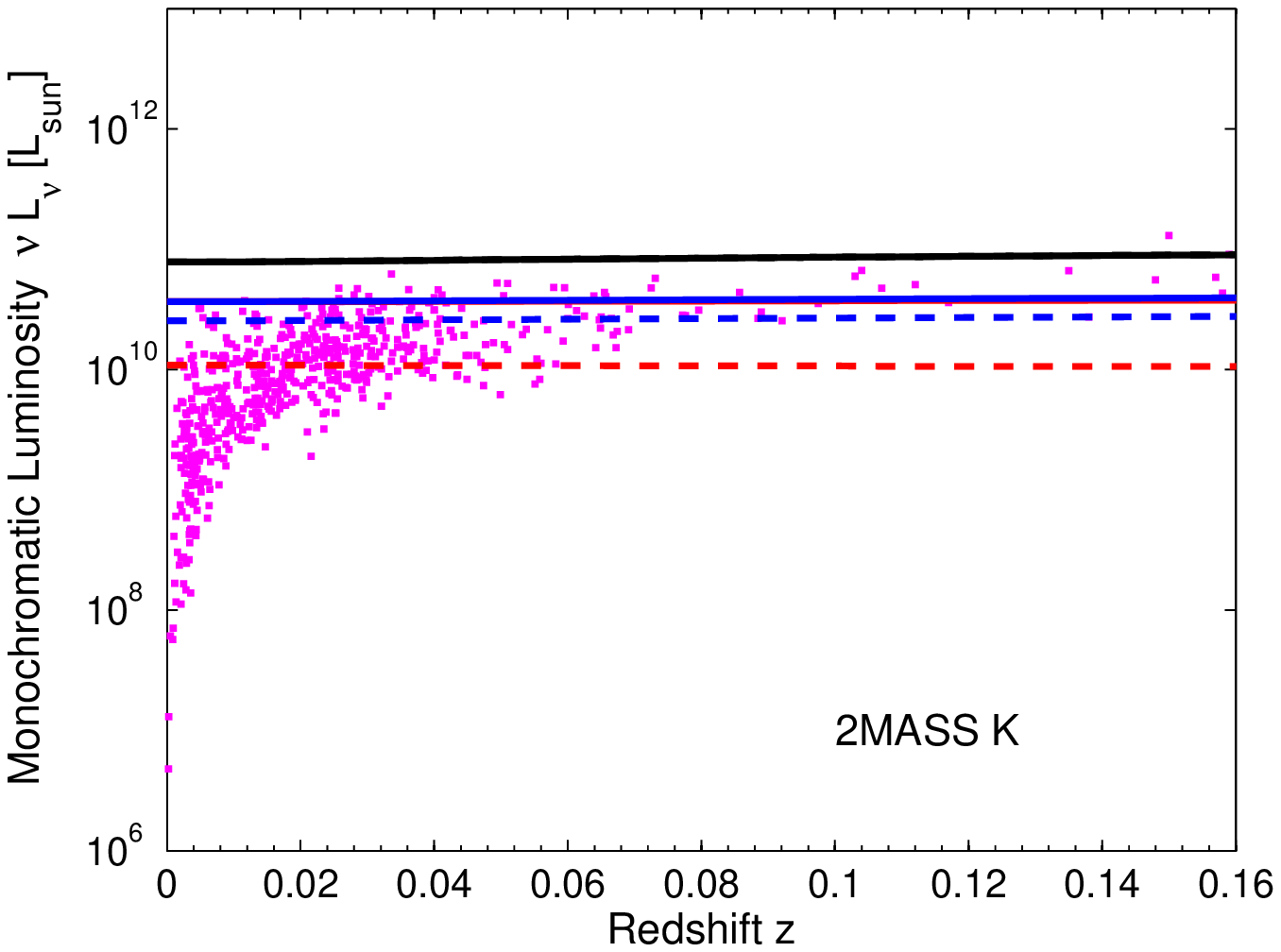} }
\centerline{
\includegraphics[height=5.2cm,width=5.2truecm]{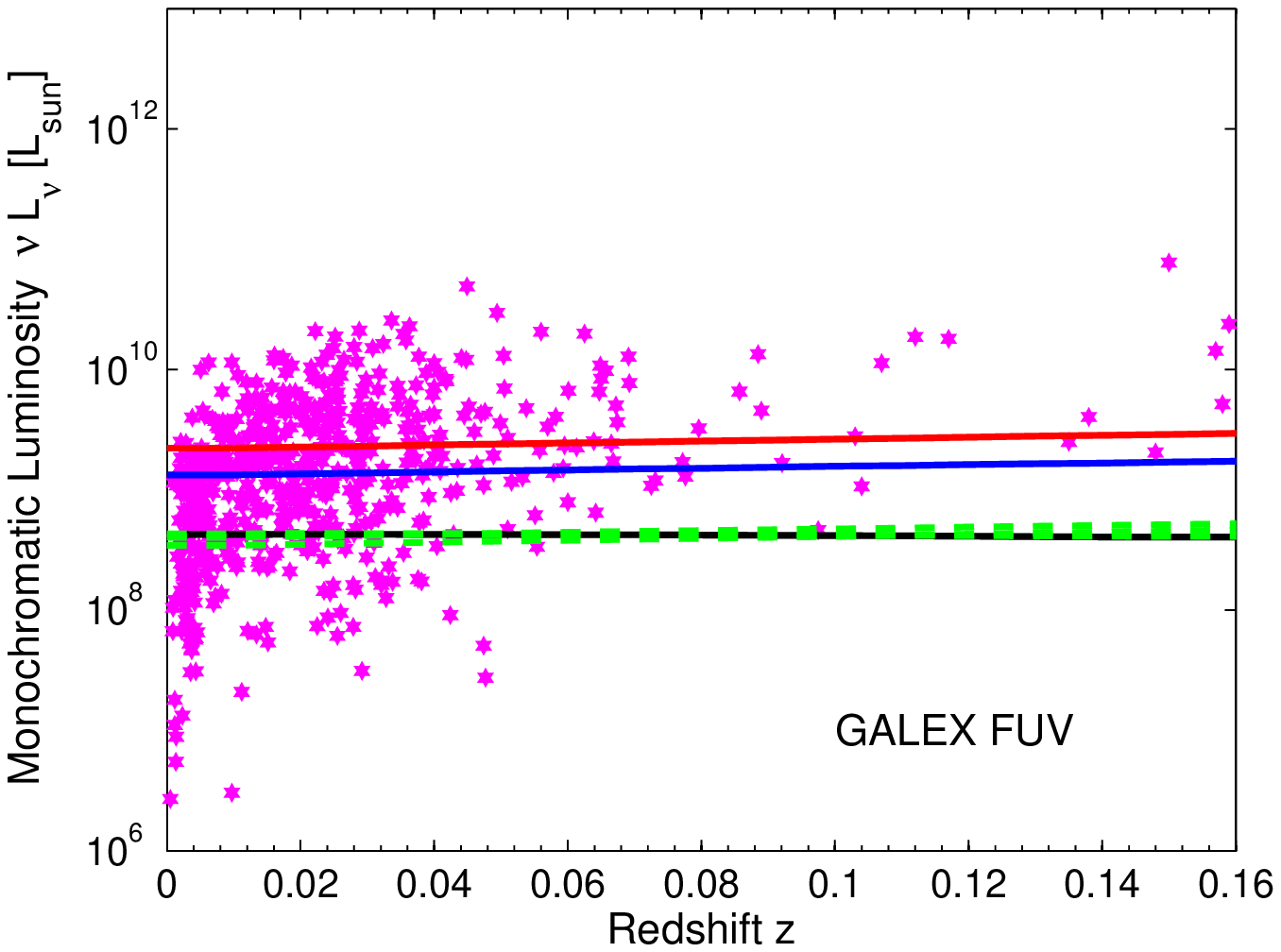}
\includegraphics[height=5.2cm,width=5.2truecm]{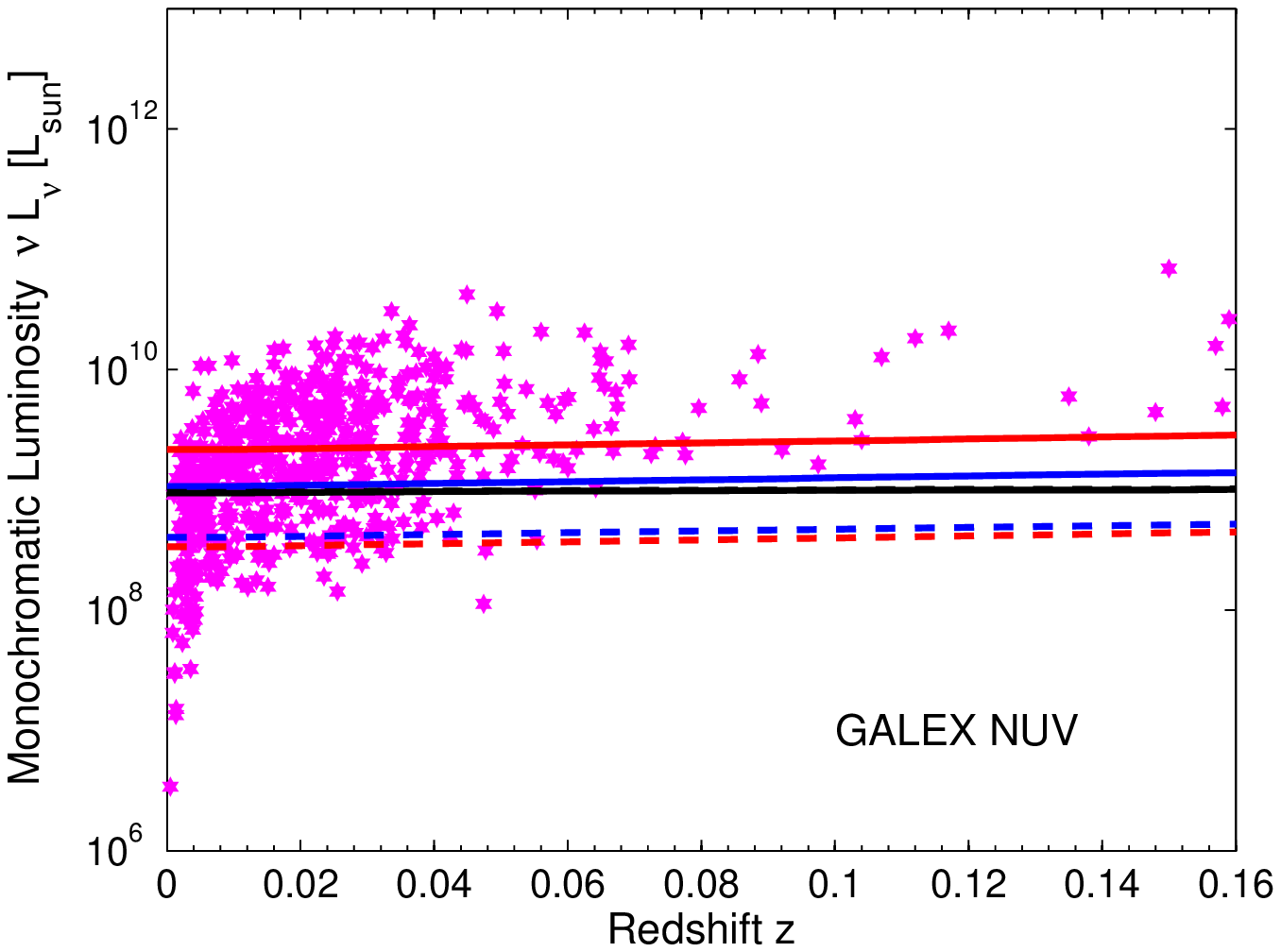}
\includegraphics[height=5.2cm,width=5.2truecm]{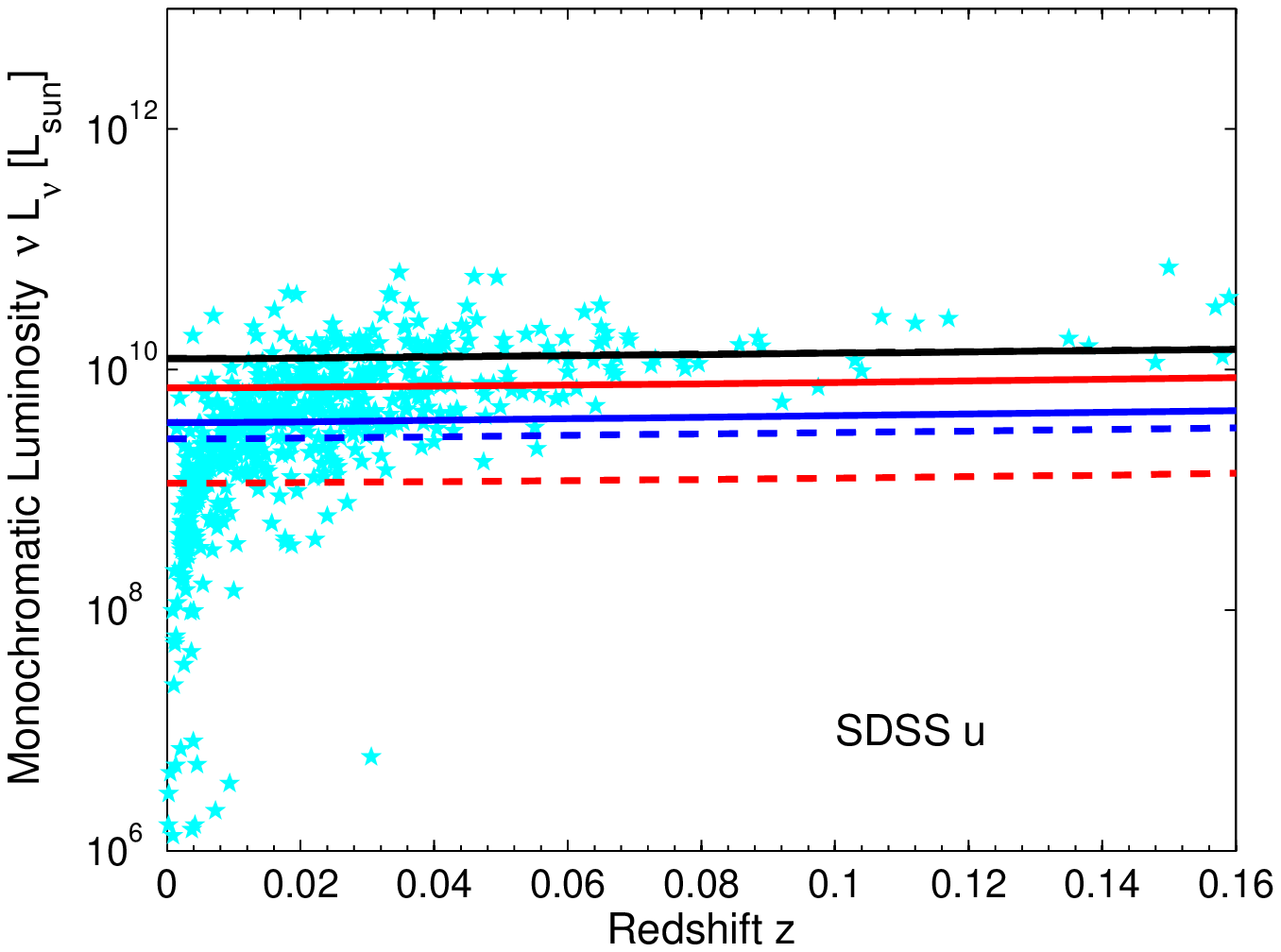} }
\caption{A comparison of the monochromatic luminosities $\nu\cdot L \left(\nu\right)$ of 3 models with asymptotic mass 10$^{12}$ M$_{\odot}$ 
with a sample of of galaxies of various morphological types and masses by \protect \citet{Takeuchi2010}.
 We represent an elliptical galaxy (black lines), an intermediate type galaxy (blue lines) and a disc galaxy (red lines). 
Solid lines represent the edge-on model, more affected by the ISM extinction, while dashed lines represent the face-on model. 
The pass-bands on display are  J, H and K bands of 2MASS, FUV and NUV of  Galex, and u of  SDSS.}
 \label{TakeuchiPRIMA}
\end{figure*}

\begin{figure*}
\centerline{
\includegraphics[height=5.2cm,width=5.2truecm]{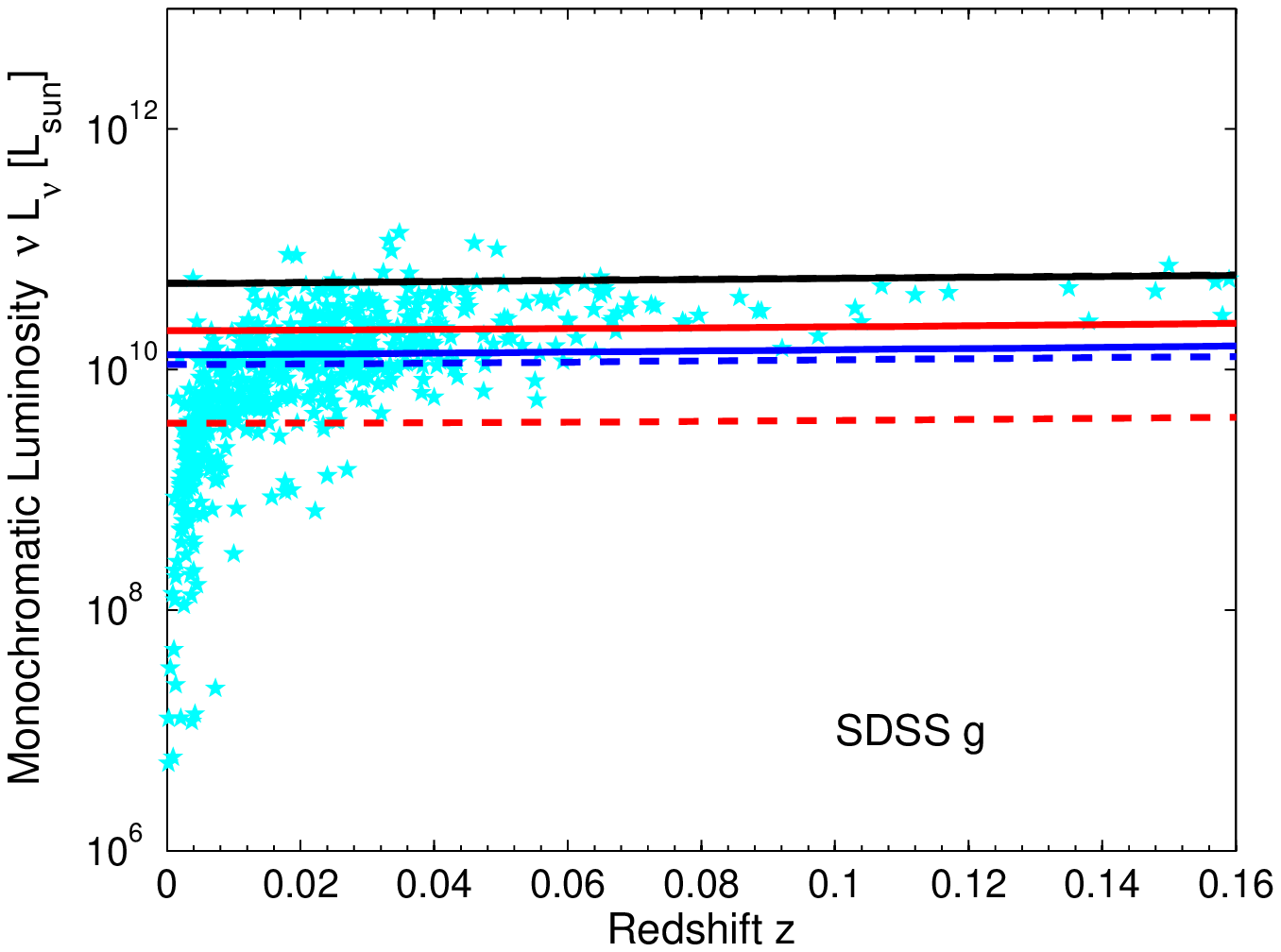}
\includegraphics[height=5.2cm,width=5.2truecm]{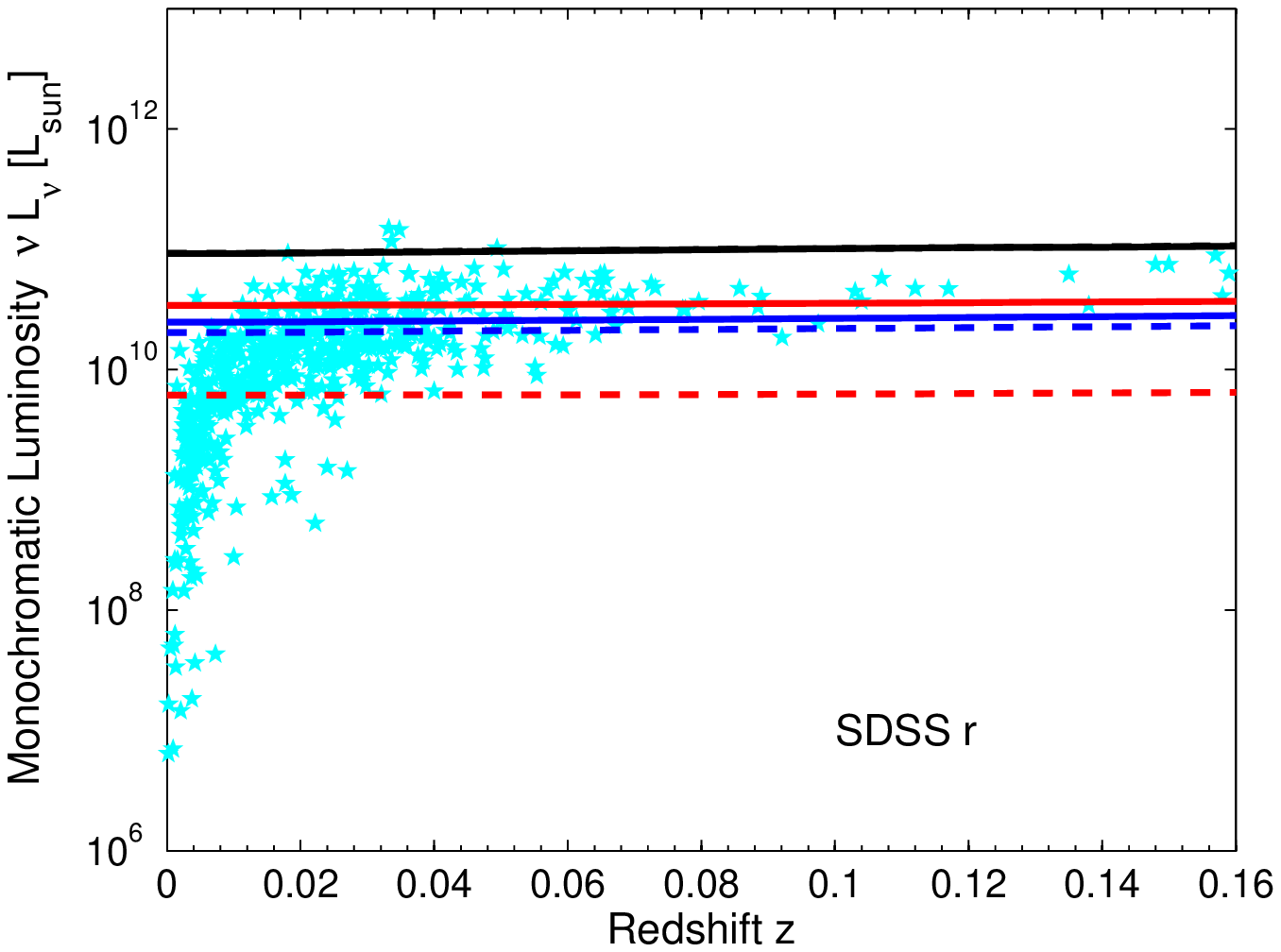}
\includegraphics[height=5.2cm,width=5.2truecm]{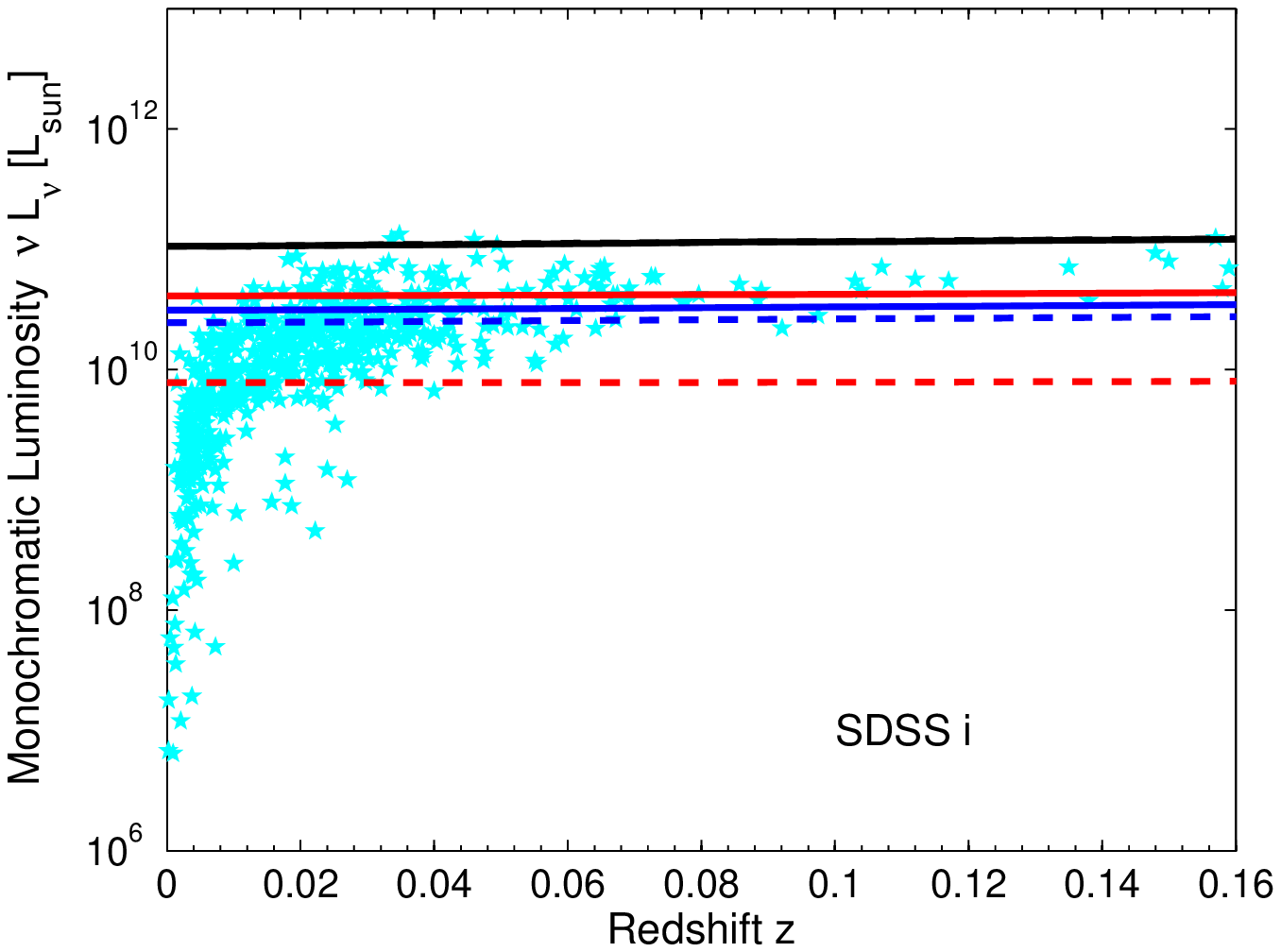}}
\centerline{
\includegraphics[height=5.2cm,width=5.2truecm]{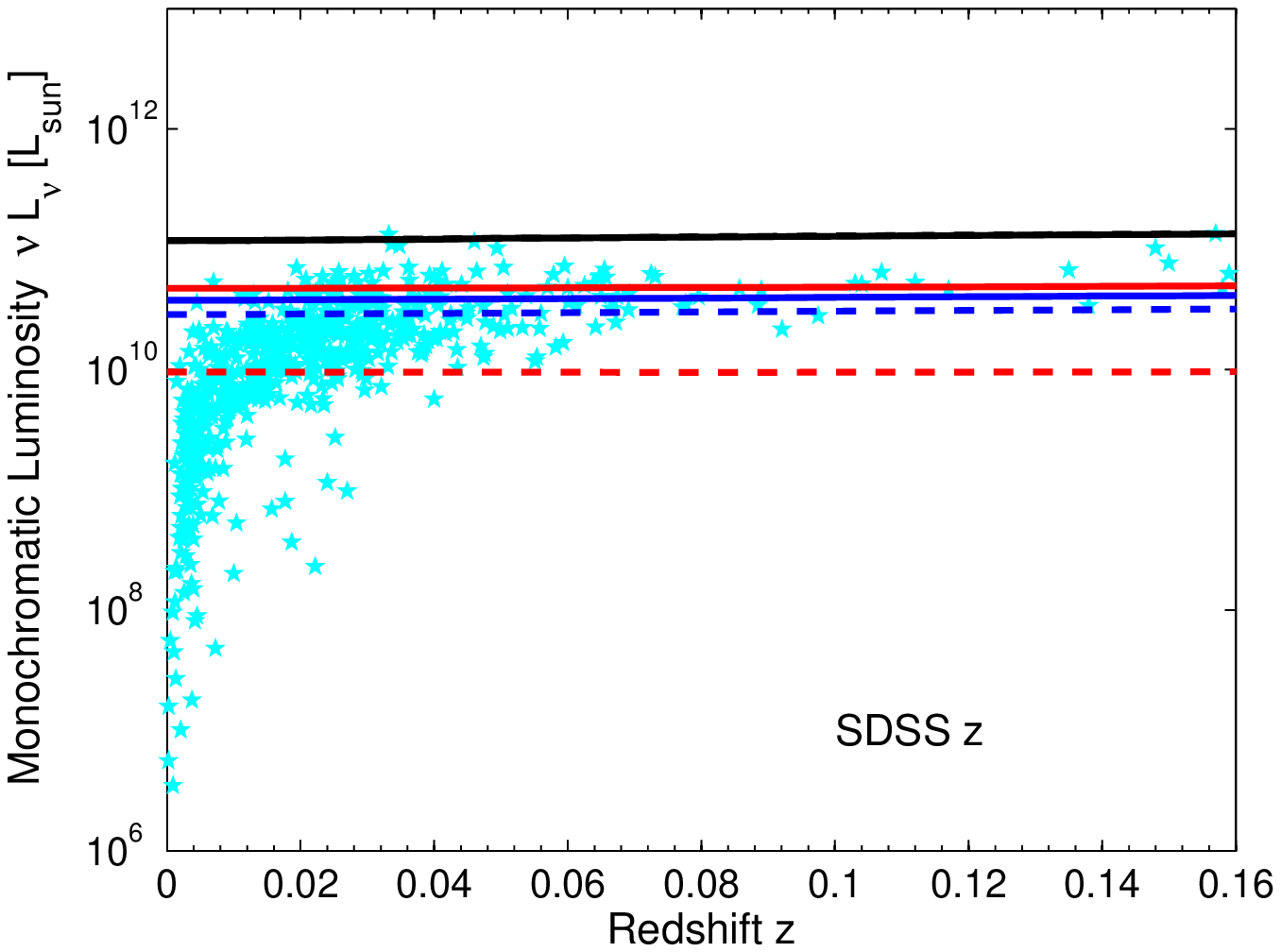}
\includegraphics[height=5.2cm,width=5.2truecm]{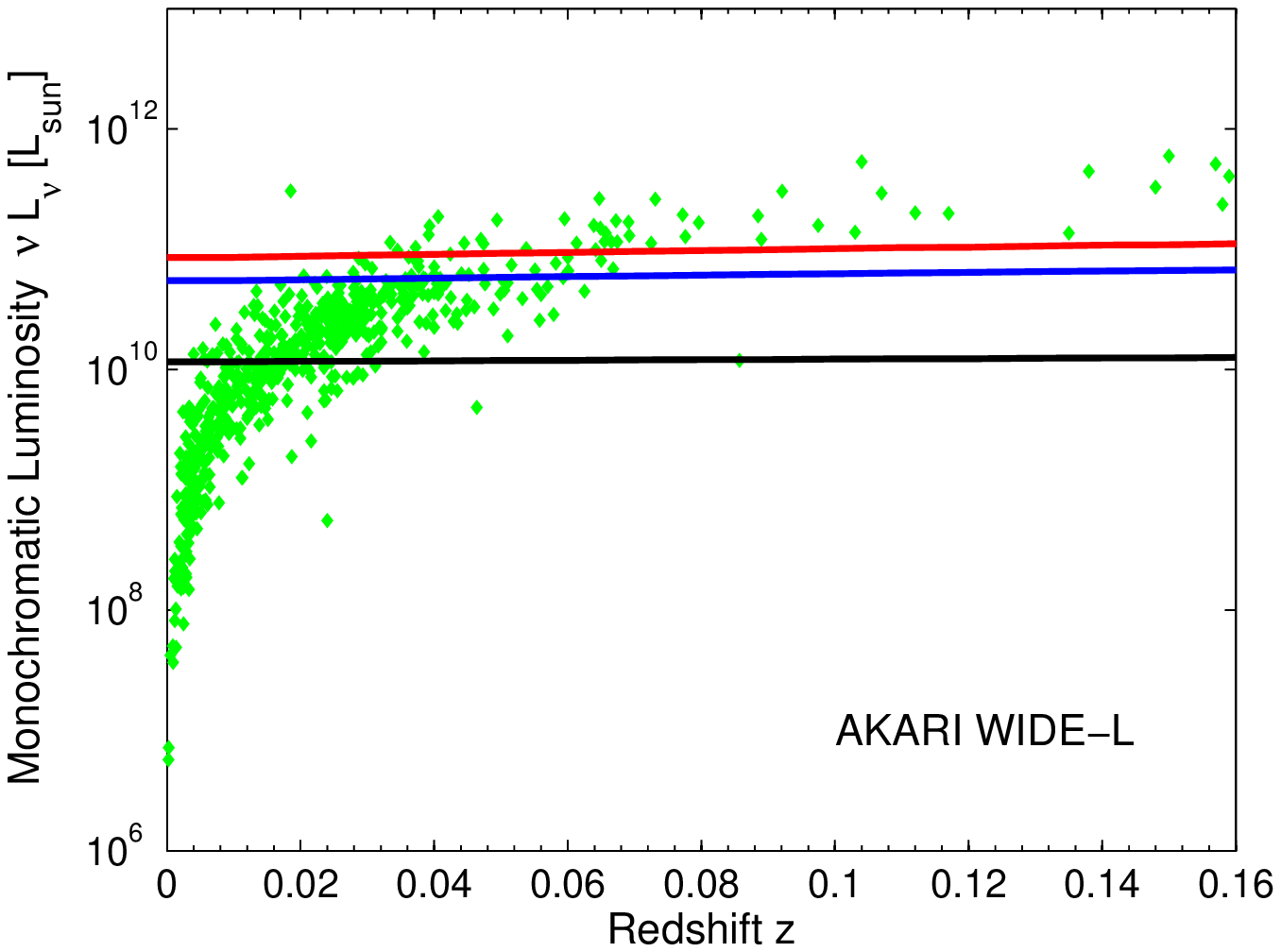}
\includegraphics[height=5.2cm,width=5.2truecm]{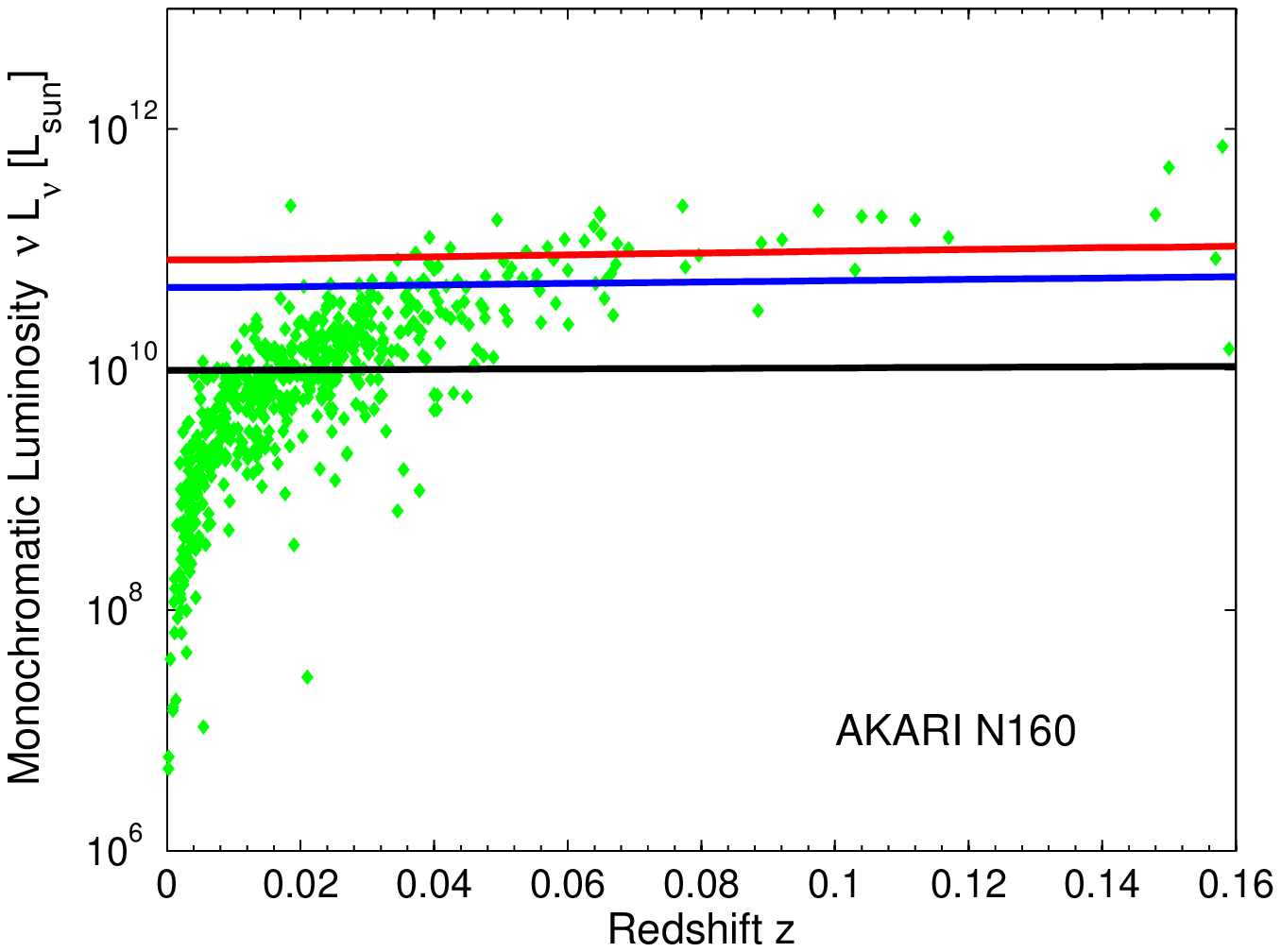} }
\caption{The same as in Fig. \protect \ref{TakeuchiPRIMA}, but for the  g, r, i, z  pass-bands of SDSS
and the AKARI pass-bands WIDE-L and N160 in the far infrared.}
\label{TakeuchiSECONDA}
\end{figure*}

\textsf{COSMOS}.
The panels of  Fig. \ref{BjmenoRKsmenoR} show the evolution with the redshift of the COSMOS colours [\textrm{B$_{J}$}-\textrm{r$^{+}$}] (left panel)
 and  [\textrm{K$_{S}$}-\textrm{r$^{+}$}]  (right panel) for our  model galaxies of different morphological types. We represent three cases:
 (i) a pure spheroidal system, for which two masses $10^{10}M_{\odot}$ and $10^{12}M_{\odot}$ are considered (the selected  sample is indeed made by ETGs);  
(ii) a galaxy of  intermediate type Sab, characterized by a disc and a bulge, with total mass $10^{11}M_{\odot}$;   
(iii) finally a pure disc galaxy with no bulge (representing a Sd type). For each galaxy type,  we also consider two cases, i.e. 
with (solid lines) and without (dashed lines) dust in the derivation of their SEDs.
The colours [\textrm{B$_{J}$}-\textrm{r$^{+}$}](left panels) and 
[\textrm{K$_{S}$}-\textrm{r$^{+}$}] (right panels) of galaxies of the COSMOS sample are also shown. 
The  ETGs are identified by the parameter  $T_{phot} \leq 1.1$ \citep[see][for more details]{Tantalo2010}. 
The orange diamonds show the total sample of galaxies, whereas the sample of ETGs  selected by \citet{Tantalo2010} is represented by the yellow diamonds. 
For the pure spheroidal galaxy the agreement with the data  is good, at least up to redshift  $z \thicksim 1-1.5$.
In the redshift interval $0<z<1$, where most of the ETGs is concentrated, colours are  in better agreement with the observational data.
In our simulations, the galactic wind stops abruptly the process of star formation so that  the galaxy evolves almost passively from the redshift of the wind $z=z_{twind}$
 to the present $z=0$. We can notice as our colours, in particular for the most massive galaxy $10^{12}M_{\odot}$ extend toward the region with the yellow circles
representing ETGs. 
This interval, however, is delicate for our models because in the chemical simulations supporting the EPS code, the galactic wind starting at $z\sim 3$ is an instantaneous
 process emptying the galaxy of gas; a more gradual process as we expect to happen in real galaxies would be more suitable allowing to avoid fluctuations in the calculated
 colours due to the discontinuity in the evolution of the gas mass. For redshift higher than $z\sim 3$, about corresponding to the onset of the galactic wind, we have no data 
to test the agreement between observations and theoretical colours. We can notice, however, the effect of the dust, by comparing the dashed (without dust) and solid lines 
(with dust). Dust absorbs stellar radiation stronger in the band $B_{J}$ than $r^{+}$. Both magnitudes grow, but, since the $B_{J}$ band is more absorbed, the colour becomes redder. 
Finally we briefly comment on  the colours of galaxies of intermediate type and pure disc. The results for COSMOS are quite interesting: the colours tend 
to stay in the region occupied by the yellow points, exactly where there are no ETGs. In particular for the [\textrm{B$_{J}$}-\textrm{r$^{+}$}] the result is good with a clearly 
different path in the colour-redshift plane followed by the different morphological types. In the [\textrm{K$_{S}$}-\textrm{r$^{+}$}]-redshift plane,
again the models of disc galaxies tend to populate   the region of the orange diamonds, whereas those for the intermediate type ones fall  in between the two extreme cases.

\textsf{Galaxy luminosities} To conclude this section, in Figs. \ref{TakeuchiPRIMA} and \ref{TakeuchiSECONDA} we present a simple comparison of the luminosities
 of our models with the data for 607 galaxies of various morphological type by \citet{Takeuchi2010} observed in different photometric systems. Of course this sample 
contains  objects spanning  wide ranges of masses and morphological types so that much narrower grids of theoretical models would be required. This is beyond the
 purposes of this study and we leave it to future work. For now we limit ourselves  to simply check that our models are consistent with the luminosity range indicated by  
 the observations. To this aim, we plot the evolution of the monochromatic luminosity of our models for three massive galaxies (elliptical, intermediate and disc) of 
about 10$^{12}$ $M_{\odot}$. Since the redshift range spanned by  the data (from $z$=0  to $z$=0.16) is rather small we do not expect our models to evolve 
significantly in luminosity. This is what we see in   Figs. \ref{TakeuchiPRIMA} and \ref{TakeuchiSECONDA}. However, our the average the models fall in the range of
the observations at all pass-bands, with some dispersion due to different inclinations of the disc. This effect is in particular relevant for the UV luminosities. 
For Akari, since dust does not absorb its own radiation there is no difference between different inclinations and the two lines, solid and dashed one, are coincident. 
As expected the model for elliptical galaxy model, present a low luminosity due to the low content of dust, whereas  the dust-rich morphological types better agree with 
the observations.

\section{Discussion and conclusion}\label{conclusions}

In this paper, improving upon the standard EPS technique, we have developed theoretical  SEDs of galaxies, whose morphology goes from disc to spherical structures, 
in presence of dust in the ISM. Properly accounting for the effects of dust  on the SED of a galaxy increases the complexity of the problem with respect to
  the standard EPS theory because it is necessary to consider the distribution of the energy sources (the stars) inside the ISM absorbing and re-emitting the 
	stellar flux. This means that the geometry and morphological type of the galaxy become important and unavoidable ingredients of the whole problem, together 
	with the transfer of radiation from one region to another. The emergent SEDs of our model galaxies have  been shown to reproduce very well, even in  details,
	the observational data for a few test galaxies of different morphological type. The model is versatile and applicable to a large range of objects of astrophysical
	interest at varying the star formation and chemical enrichment histories, the geometrical shape or morphology of the galaxies and the amounts of gas and dust contained 
	in their ISM.

Before concluding, it is worth mentioning a few points of weakness that could be improved. First,  the chemical models we have adopted are from
 \citet{Tantalo1996,Tantalo1998}, whereas the chemical yields from \citet{Portinari1998}. These models are state-of-the-art in the study of the chemical
 evolution of galaxies. However, they  do not include a proper description of the formation/destruction of dust as for instance  recently developed  by 
\citet{Dwek1998,Dwek2005} and \citet{Piovan2011_4541,Piovan2011_4561,Piovan2011_4567}. Even if the dust content can be related to the metallicity of the galaxy, 
the relative proportions of the various components of the dust would require the detailed  study of the evolution of the dusty environment and the complete
 information on the dust yields \citep{Dwek1998,Dwek2005,Galliano2005,Piovan2011_4541,Piovan2011_4561,Piovan2011_4567}. This would lead to a better and more 
physically sounded correlation between the composition of dust,  the star formation activity and rate, and the chemical enrichment history of the galaxy itself. 
All this is missing in most galaxy models in which dust is considered. The problem may be particularly severe for high metallicity environments. Second, the
 models for the disc galaxies with a central bulge need to be tested against SEDs of local galaxies of intermediate type going from $S0$ to $Sb/Sc$, trying 
to match some observational constraints, like the UV-optical average colours \citep{Buzzoni2002,Buzzoni2005}. Finally, many other physical ingredients can 
be improved and/or considered. Just to mention two of these, the extension of the SED to the radio range and the simulation of the nebular emission. Work
 is in progress to this aim.

\vskip 3mm
Extensive tabulations of the SEDs, magnitudes and colours  of  the SSPs and model galaxies are available from L. P. Cassar\`{a} upon request.

\section*{Acknowledgments}
We are grateful to Alberto Buzzoni for fruitful discussions. This study makes use of data products from 2MASS, 
which is a joint project of the University of Massachusetts and the Infrared Processing and Analysis Centre/California Institute of Technology, 
founded by the National Aeronautics and Space Administration and the National Science Foundation.

\begin{small}
\bibliographystyle{mn2e}                             
\bibliography{mnemonic,biblio_Cassara2014}              
\end{small}
\end{document}